\documentclass[fleqn,10pt]{article}
\usepackage{latexsym, graphicx, epsfig, amsmath, amssymb,amsfonts}
\usepackage{natbib,amsthm,version}
\usepackage{amsbsy,bm,multirow,enumerate}
\usepackage[titletoc,page]{appendix}
\usepackage[mathscr]{eucal}
\usepackage{mathtools}
\usepackage{color}
\usepackage[utf8]{inputenc}
\usepackage[english]{babel}
\usepackage{amsthm}
\usepackage{enumerate}
\usepackage[hidelinks]{hyperref}
\usepackage{url}
\usepackage{subfigure}
\usepackage[lined,boxed]{algorithm2e}

\newcommand{\mbs}[1]{\mathbf{#1}}

\newtheorem{theorem}{Theorem}[section]

\newtheorem{remark}{Remark}

\theoremstyle{definition}

\graphicspath{
{ ./Figures/Conv/ }
}

\title{
  An Energy-Stable Scheme for Incompressible Navier-Stokes 
  Equations with Periodically Updated Coefficient Matrix 
} 
\author{
  Lianlei Lin$^{1}$, \ Naxian Ni$^2$, \ Zhiguo Yang$^2$, \
  Suchuan Dong$^2$\thanks{Author of correspondence.
    Email: sdong@purdue.edu} \\
  $^1$School of Electronics and Information Engineering \\
  Harbin Institute of Technology, China \\
  $^2$Center for Computational and Applied Mathematics \\
  Department of Mathematics \\
  Purdue University, USA \\
 } 

\date{(September 10, 2019)}
\begin{document}
\maketitle



\begin{abstract}

  We present an energy-stable scheme for
  simulating the incompressible Navier-Stokes equations
  based on the generalized Positive Auxiliary Variable
  (gPAV) framework. In the gPAV-reformulated system
  the original nonlinear term is replaced by
  a linear term plus a correction term, where the correction
  term is put under control by an auxiliary variable.
  The proposed scheme incorporates a pressure-correction
  type  strategy into the gPAV procedure, and it satisfies a 
  discrete energy stability property.
  The scheme entails the computation of two copies of
  the velocity and pressure within a time step,
  by solving an individual de-coupled linear
  equation for each of these field variables.
  Upon discretization
  the pressure
  linear system involves a constant coefficient matrix that
  can be pre-computed, while the velocity linear system involves
  a coefficient matrix that is updated periodically,
  once every $k_0$ time steps in the current work, where $k_0$ is a user-specified
  integer.
  The auxiliary variable, being a scalar-valued number,
  is computed by a well-defined explicit formula, which guarantees
  the positivity of its computed values.
  It is observed that the current method can produce accurate
  simulation results at large (or fairly large) time step sizes
  for the incompressible Navier-Stokes equations.
  The impact of the periodic coefficient-matrix update
  on the overall cost of the method is observed to be small
  in typical numerical simulations.
  Several flow problems have been simulated to demonstrate the accuracy and
  performance of the method developed herein.

\end{abstract}


\vspace{0.05cm}

\section{Introduction}
\label{sec:intro}

%
%
%

This work concerns the numerical approximation of the incompressible
Navier-Stokes equations in an energy-stable fashion.
Energy-stable approximations are attractive in  that
they not only preserve the
dissipative nature of the underlying continuous Navier-Stokes system,
but more practically can potentially allow the use of larger time steps
in computer simulations. This type of schemes
are the focus of a number of previous works in the literature;
see e.g.~\cite{Shen1992,SimoA1994,VerstappenV2003,GuermondMS2005,LabovskyLMNR2009,DongS2010,Sanderse2013,JiangMRT2016,ChenSZ2018}.
These schemes typically treat the nonlinear term fully implicitly
or in a linearized fashion. Upon discretization, they
would typically entail the solution of nonlinear
algebraic systems within a time step, or when only a linear system
needs to be solved, would involve  time-dependent coefficient matrices
and entail frequent re-computations (every time step)
of these coefficient matrices~\cite{DongS2010}.
This is a main drawback of traditional energy-stable schemes.
Their computational cost per time step is typically high compared
with that of semi-implicit type
schemes~\cite{Chorin1968,Temam1969,KimM1985,KarniadakisIO1991,BrownCM2001,XuP2001,LiuLP2007,HyoungsuK2011,Dong2015clesobc,SersonMS2016},
which, albeit only conditionally stable, are more commonly-used in production
simulations.

%


An interesting recent development in this area is~\cite{LinYD2019},
which describes a discretely energy-stable scheme employing 
an auxiliary energy variable in its formulation. 
The Navier-Stokes equations are reformulated and augmented
by a dynamic equation for the auxiliary variable, which
is a scalar-valued number rather than a field function.
A prominent feature of this scheme lies in the reformulation of
the nonlinear term,
\begin{equation}
\frac{R(t)}{\sqrt{E(t)}}\mbs u\cdot\nabla\mbs u,
\end{equation}
where $\mbs u$ is the velocity, $R(t)$ is the auxiliary variable
and $E(t)$ is the shifted total kinetic energy of the system.
The numerical scheme proposed in \cite{LinYD2019} treats the
$\mbs u\cdot\nabla\mbs u$ component
in an explicit fashion, but controls this explicit component by
an implicit treatment of $\frac{R(t)}{\sqrt{E(t)}}$.
The scheme is shown to satisfy a
discrete energy stability property, which is
also demonstrated
by numerical experiments. 
The scheme has an interesting property that makes it computationally
attractive and competitive.
Within each time step
it requires only the solution of linear algebraic systems with
constant coefficient matrices, which can be pre-computed,
for the field functions.
One does need to additionally solve a nonlinear algebraic equation
about a scalar-valued number. But since
this nonlinear equation is about a scalar number, not a field function,
its cost is very low, accounting for about a few percent of the total
cost per time step~\cite{LinYD2019}.
A further development of this approach is discussed very recently 
in~\cite{LinLD2019}, which presents a method for treating 
the so-called energy-stable open boundary
conditions~\cite{DongKC2014,Dong2015clesobc,DongS2015,Dong2014obc} in an
energy-stable fashion on the discrete level.

%
%
%

While the auxiliary-variable approach and the numerical scheme
from~\cite{LinYD2019} possess a number of  attractive properties,
certain aspects of the method are less favorable
and leave much to be desired.
We list some of the issues here: 
\begin{itemize}
\item
  The need for solving a nonlinear algebraic equation for
  the auxiliary variable is highly undesirable.
  While its computational cost can be negligible,
  the nonlinear equation causes two complications.
  First, the existence and uniqueness of the solution for
  the auxiliary variable from the discrete scheme becomes unknown.
  Second, the positivity of the computed values for the auxiliary variable,
  as is physically required by its definition, is uncertain.

\item
  The numerical scheme of~\cite{LinYD2019} is formulated in
  a setting where the velocity and the pressure are fully coupled,
  and the discrete energy stability is proven in this coupled setting.
  When implementing the scheme, 
  the authors have made a further approximation
  about the boundary vorticity, which  
  de-couples the pressure/velocity computations in actual simulations.
  The stability proof, however, does not hold if this further approximation
  is taken into account.

\item
  It is observed in~\cite{LinYD2019} that the accuracy of the method
  deteriorates when the time step size becomes large (or fairly large).
  How to improve the accuracy of the method for large (or fairly
  large) time step sizes, while simultaneously preserving the favorable
  properties that keep the computational cost relatively low,
  is an open issue.

\item
  In the definition of the auxiliary variable, a biased total energy
  (shifted by an energy constant $C_0$) has been used in~\cite{LinYD2019}.
  It is observed that the
   $C_0$ value seems to have an influence
  on the accuracy of the simulation results (see
  the Kovasznay flow test of~\cite{LinYD2019}), which is an undesirable
  aspect. 

\end{itemize}

The first and the second issues in the above list have been addressed
by~\cite{LinLD2019}.
In the method presented in~\cite{LinLD2019}, the nonlinear algebraic
equation has been eliminated, and the auxiliary variable
at each time step is given by an explicit formula,
which ensures that its computed values are always positive.
The method is formulated in a setting in which the pressure and
 velocity are de-coupled (barring the auxiliary variable)
by a velocity-correction strategy.
This scheme retains the attractive properties found in
\cite{LinYD2019}, such as the discrete energy stability and
the need to only solve linear algebraic systems with constant pre-computable
coefficient matrices.

The numerical scheme of~\cite{LinLD2019} is able to
achieve these important properties, in large part,
thanks to the adoption of
the generalized Positive Auxiliary Variable (gPAV) approach,
which was originally developed in~\cite{YangD2019diss} for
general dissipative systems.
gPAV provides a means to use a general class of functions
in defining the auxiliary variable,
and a systematic procedure for treating
dissipative partial differential equations (PDE).
The gPAV procedure endows
energy stability to the
resultant scheme, and also can ensure the positivity of the computed
values of the generalized auxiliary variable~\cite{YangD2019diss}.
Compared with 
related works~\cite{LinYD2019,YangLD2019,YangD2019twop,ShenXY2018,Yang2016},
the gPAV framework provides a more favorable way for
treating the auxiliary variables, and it applies to
very general dissipative systems.


In the current work we focus on the accuracy issue of
the auxiliary-variable method as listed above.
We would like to explore the possibility to expand its accuracy range,
and aim to achieve accuracy at large (or fairly large)
time step sizes, without seriously sacrificing
the computational cost 
for incompressible Navier-Stokes equations.
Summarized in this paper is our effort
in this respect and a numerical scheme that largely
achieves this goal.

In the current paper we present an energy-stable
scheme for the incompressible Navier-Stokes equations
employing the gPAV strategy. The salient feature of
the scheme lies in the reformulation and numerical treatment of
the nonlinear term. In the gPAV-reformulated system
we replace the nonlinear term  by a linear term plus
a correction term, and put the correction term 
under control by an auxiliary variable (a scalar-valued number).
Upon discretization,
this leads to a velocity linear algebraic system with
a coefficient matrix that can be updated
periodically, in particular once every $k_0$ time
steps in the current work,
where $k_0$ is a user-specified integer parameter.
The proposed scheme is observed to produce accurate results at large
or fairly large time step sizes (depending on the Reynolds number).
It substantially expands the accuracy range for the time step size 
compared with the scheme of~\cite{LinYD2019}
and the scheme without the current reformulation
of the nonlinear term.
Incidentally, we observe that this scheme is not sensitive to
the energy constant $C_0$ used in defining the auxiliary variable.

The current scheme incorporates the gPAV idea and
a pressure-correction type splitting strategy, and is endowed with several
attractive properties.
It is energy-stable and satisfies a discrete energy stability property.
No nonlinear solver is involved in this scheme, for
either the field functions or the auxiliary variable.
The computations for the velocity and the pressure are de-coupled.
The method requires the computation of
two copies of the velocity and pressure within a time step,
by solving an individual de-coupled linear equation for each of them.
 Upon discretization,
the pressure linear algebraic system involves a constant coefficient
matrix that can be pre-computed,
while the velocity linear system involves a coefficient matrix that
can be updated periodically (every $k_0$ time steps).
The auxiliary variable is computed by a well-defined explicit
formula, which guarantees the positivity of its computed values.


The coefficient-matrix update induces an extra cost, due to
the re-computation and factorization involved therein.
But since this is performed only occasionally (every $k_0$ time steps),
this extra cost is effectively spread over $k_0$ time steps.
In numerical simulations $k_0$ can typically range from several dozen to
several hundred, depending on the Reynolds number
($k_0=20$ in the majority of simulations reported herein).
So the impact of occasional coefficient matrix update
on the overall computational cost of the current method
is quite small, and can essentially be negligible in many cases.



The contribution of this work lies in the energy-stable
scheme for the incompressible Navier-Stokes equations developed herein.
Its favorable properties include:
(i) improved accuracy,
producing accurate simulation results at large (or fairly large)
time step sizes;
(ii) relatively low computational cost,
requiring only the solution of linear systems with coefficient matrices
that are pre-computable or only need to be updated periodically;
and (iii) low sensitivity to the energy constant $C_0$.


The rest of this paper is organized as follows.
In Section \ref{sec:method} we discuss the reformulation of the
incompressible Navier-Stokes equations based on the gPAV framework,
and present the energy-stable scheme for the reformulated system.
We prove a discrete  energy stability property of the scheme,
and discuss the solution algorithm and its implementation based
on high-order spectral
elements~\cite{SherwinK1995,BlackburnH1999,KarniadakisS2005,ChenSX2012}.
In Section \ref{sec:tests} we test the proposed method
using several  flow problems and investigate its
accuracy, the effect of algorithmic parameters, and
the computational cost.
Section \ref{sec:summary} then concludes the presentation
with comments on a number of related issues.


\section{Energy-Stable Scheme for Incompressible Flows}
\label{sec:method}

\subsection{Governing Equations and gPAV-Reformulated Equivalent System}

Consider some flow domain $\Omega$ (with boundary $\partial\Omega$)
in two or three dimensions, and an incompressible flow contained
in $\Omega$. The dynamics of the system is described by
the incompressible Navier-Stokes equations given by, in non-dimensional form,
\begin{align}
  &
  \frac{\partial\mbs u}{\partial t} + \mbs N(\mbs u)
  + \nabla p -\nu\nabla^2\mbs u = \mbs f,
  \label{equ:nse} \\
  &
  \nabla\cdot\mbs u = 0, \label{equ:cont}
\end{align}
where $\mbs u(\mbs x,t)$ is the velocity,
$p(\mbs x,t)$ is the pressure,
the nonlinear term $\mbs N(\mbs u)=\mbs u\cdot\nabla\mbs u$,
$\mbs f(\mbs x,t)$ is an external body force,
and $\mbs x$ and $t$ are the spatial coordinate and time.
$\nu$ is the non-dimensional viscosity (inverse of Reynolds number $Re$),
\begin{equation} \label{equ:def_nu}
\nu = \frac{1}{Re} = \frac{\nu_f}{U_0 L}
\end{equation}
where $U_0$ and $L$ are respectively the characteristic
velocity and length scales,
and $\nu_f$ is the kinematic viscosity of the fluid.
We assume Dirichlet boundary condition in this work,
\begin{equation}\label{equ:bc}
\mbs u|_{\partial\Omega} = \mbs w(\mbs x,t)
\end{equation}
where $\mbs w$ is the boundary velocity.
The governing equations are supplemented by the initial condition
\begin{equation}\label{equ:ic}
\mbs u(\mbs x,0) = \mbs u_{in}(\mbs x)
\end{equation}
where $\mbs u_{in}$ is the initial velocity distribution that
satisfies equation \eqref{equ:cont} and is compatible with
the boundary velocity $\mbs w(\mbs x,t)$ on $\partial\Omega$ at $t=0$.
To fix the pressure we will impose the often-used condition
\begin{equation}
  \int_{\Omega} p d\Omega = 0.
  \label{equ:p_cond}
\end{equation}

By taking the $L^2$ inner product between equation \eqref{equ:nse}
and $\mbs u$, integrating by part and using equation
\eqref{equ:cont}, 
we obtain the
energy balance equation
\begin{equation}
  \frac{\partial}{\partial t}\int_{\Omega} \frac{1}{2}|\mbs u|^2d\Omega=
  -\nu\int_{\Omega}\|\nabla\mbs u \|^2d\Omega
  + \int_{\partial\Omega}\left[
  -p\mbs n + \nu\mbs n\cdot\nabla\mbs u - \frac{1}{2}(\mbs n\cdot\mbs u)\mbs u
  \right]\cdot\mbs u dA,
  \label{equ:energy}
\end{equation}
where
$
\|\nabla\mbs u \|^2 = \sum_{i,j=1}^{d_{im}}\partial_i u_j\partial_i u_j
$
and $d_{im}$ is the dimension in space.

We define a biased energy,
\begin{equation}\label{equ:def_E}
  E(t) = E[\mbs u] = \int_{\Omega} \frac{1}{2}|\mbs u|^2d\Omega + C_0,
\end{equation}
where $C_0$ is a chosen energy constant such that
$E(t)>0$ for all $t\geqslant 0$.
Following the gPAV framework from~\cite{YangD2019diss}
and also the work~\cite{LinYD2019},
we introduce an auxiliary variable $R(t)$ by
\begin{equation}\label{equ:def_R}
  \left\{
  \begin{split}
    &
    E(t) = R^2, \\
    &
    R(t) = \sqrt{E(t)}. \\
  \end{split}
  \right.
\end{equation}
It is important to note that both $E(t)$ and $R(t)$ are scalar-valued numbers, not
field functions. Based on its definition,
$R(t)$ satisfies the following evolution equation
\begin{equation}\label{equ:R_equ}
  2R\frac{dR}{dt}
  = \int_{\Omega} \mbs u\cdot\frac{\partial\mbs u}{\partial t} d\Omega.
\end{equation}

Noting that $\frac{R^2(t)}{E(t)}=1$, we  reformulate
equation \eqref{equ:nse} into an equivalent form
\begin{equation}
  \frac{\partial\mbs u}{\partial t} + \mbs M(\mbs u)
  + \nabla p -\nu\nabla^2\mbs u
  + \frac{R^2}{E(t)}\left[
    \mbs N(\mbs u) - \mbs M(\mbs u)
    \right]
  = \mbs f,
  \label{equ:nse_1}
\end{equation}
where $\mbs M(\mbs u)$ is  defined by,
\begin{align}
  &
  \mbs M(\mbs u) = \mbs u_0\cdot\nabla\mbs u
  + \frac{1}{2}(\nabla\cdot\mbs u_0)\mbs u,
  \label{equ:def_M1} 
\end{align}
and $\mbs u_0$ is a prescribed velocity field
that is only occasionally updated in time. 
In the current paper
we choose $\mbs u_0$ to be the velocity field $\mbs u$
at every $k_0$-th time step, where $k_0$ is an integer parameter
provided by the user.
More specifically, at any time step $n$, $\mbs u_0$ is taken
to be the velocity field $\mbs u$ at time step $mk_0$, where
$m$ is the integer satisfying
$mk_0\leqslant n< (m+1)k_0$. Therefore, $\mathbf{u}_0$ is 
updated only once every $k_0$ time steps.

We reformulate equation \eqref{equ:R_equ} as follows,
\begin{equation}
  \begin{split}
    2R\frac{dR}{dt} &= \int_{\Omega} \mbs u\cdot\frac{\partial\mbs u}{\partial t}d\Omega\\
    &\quad
  + \left(\frac{R^2}{E}-1  \right)\int_{\Omega}\left[
    -\mbs M(\mbs u) - \nabla P + \nu\nabla^2\mbs u + \mbs f
    \right]\cdot\mbs u d\Omega \\
  &\quad
  + \frac{R^2}{E}\left[
    \int_{\Omega}\left(\mbs N(\mbs u) - \mbs M(\mbs u)  \right)\cdot\mbs ud\Omega
    - \int_{\Omega}\left(\mbs N(\mbs u) - \mbs M(\mbs u)  \right)\cdot\mbs ud\Omega
    \right] \\
  &\quad
  + \left(1- \frac{R^2}{E} \right)\left|\int_{\Omega}\mbs f\cdot\mbs ud\Omega  \right|
  + \left(1- \frac{R^2}{E} \right)\left|
  \int_{\partial\Omega}\left[
    -P\mbs n + \nu\mbs n\cdot\nabla\mbs u
    -\frac{1}{2}(\mbs n\cdot\mbs u)\mbs u
    \right]\cdot\mbs u dA
  \right| \\
  &= \int_{\Omega} \mbs u\cdot\frac{\partial\mbs u}{\partial t}d\Omega
  + \int_{\Omega}\left[
    \mbs M(\mbs u) + \nabla P - \nu\nabla^2\mbs u
    + \frac{R^2}{E}\left(\mbs N(\mbs u) - \mbs M(\mbs u) \right) -\mbs f
    \right]\cdot \mbs ud\Omega \\
  &\quad
  + \frac{R^2}{E}\left[ 
    -\int_{\Omega}\nabla P\cdot\mbs ud\Omega
    + \int_{\Omega}\nu\nabla^2\mbs u\cdot\mbs u d\Omega
    - \int_{\Omega} \mbs N(\mbs u)\cdot\mbs ud\Omega
    + \int_{\Omega}\mbs f\cdot\mbs ud\Omega
    \right] \\
  & \quad
  + \left(1- \frac{R^2}{E} \right)\left|\int_{\Omega}\mbs f\cdot\mbs ud\Omega  \right|
  + \left(1- \frac{R^2}{E} \right)\left|
  \int_{\partial\Omega}\left[
    -P\mbs n + \nu\mbs n\cdot\nabla\mbs u
    -\frac{1}{2}(\mbs n\cdot\mbs u)\mbs u
    \right]\cdot\mbs u dA
  \right|.
  \end{split}
  \label{equ:R_equ_1}
\end{equation}
Note that in the above equation a number of zero terms have been
incorporated into the right hand side (RHS). In
this equation $P$ is a field function related to the pressure $p$ (not $p$
itself),
and will be specified later in equation \eqref{equ:def_P}.
Note that
\begin{multline*}
  -\int_{\Omega}\nabla P\cdot\mbs ud\Omega
  + \int_{\Omega}\nu\nabla^2\mbs u\cdot\mbs u d\Omega
  - \int_{\Omega} \mbs N(\mbs u)\cdot\mbs ud\Omega
  = \\
  -\int_{\Omega}\nu\|\nabla\mbs u \|^2d\Omega
  + \int_{\partial\Omega} \left[
    -P\mbs n + \nu\mbs n\cdot\nabla\mbs u
    -\frac{1}{2}(\mbs n\cdot\mbs u)\mbs u
    \right]\cdot\mbs u dA,
\end{multline*}
where the integration by part, the divergence theorem,
and equation \eqref{equ:cont} have been used.
We can then re-write equation \eqref{equ:R_equ_1} into
\begin{equation}
  \begin{split}
    2R\frac{dR}{dt} &= \int_{\Omega} \mbs u\cdot\frac{\partial\mbs u}{\partial t}d\Omega
    + \int_{\Omega}\left[
    \mbs M(\mbs u) + \nabla P - \nu\nabla^2\mbs u
    + \frac{R^2}{E}\left(\mbs N(\mbs u) - \mbs M(\mbs u) \right) -\mbs f
    \right]\cdot \mbs ud\Omega \\
    &\quad
    +\frac{R^2}{E}\left[
      -\int_{\Omega}\nu\|\nabla\mbs u \|^2d\Omega
      + \int_{\Omega}\mbs f\cdot\mbs ud\Omega
      + \int_{\partial\Omega} \left[
        -P\mbs n + \nu\mbs n\cdot\nabla\mbs u
        -\frac{1}{2}(\mbs n\cdot\mbs w)\mbs w
        \right]\cdot\mbs w dA
      \right] \\
    & \quad
  + \left(1- \frac{R^2}{E} \right)\left|\int_{\Omega}\mbs f\cdot\mbs ud\Omega  \right|
  + \left(1- \frac{R^2}{E} \right)\left|
  \int_{\partial\Omega}\left[
    -P\mbs n + \nu\mbs n\cdot\nabla\mbs u
    -\frac{1}{2}(\mbs n\cdot\mbs w)\mbs w
    \right]\cdot\mbs w dA
  \right|.
  \end{split}
  \label{equ:R_reform}
\end{equation}
Note that the boundary condition \eqref{equ:bc} has been applied
in the above equation.
This is the reformulated equivalent form of equation \eqref{equ:R_equ}

The reformulated equivalent system of governing equations
consists of equations \eqref{equ:nse_1}, \eqref{equ:cont} and
\eqref{equ:R_reform}, the boundary condition \eqref{equ:bc},
and the initial condition \eqref{equ:ic} and the following
initial condition for $R(t)$,
\begin{equation}
  R(0) = \sqrt{\int_{\Omega}\frac{1}{2}|\mbs u_{in}|^2d\Omega + C_0}.
  \label{equ:R_ic}
\end{equation}
In this system the dynamic variables are $\mbs u$, $p$ and $R$,
and they are all coupled together. $E(t)$ is given by
equation \eqref{equ:def_E}.
Note that in the reformulated system
$R(t)$ is treated as an approximation of $\sqrt{E(t)}$ and
is  computed by solving this system of
equations, not by using equation \eqref{equ:def_R}.

\subsection{Numerical Scheme and Unconditional Energy Stability}

We next present an unconditionally energy-stable scheme for
the reformulated system consisting of
equations \eqref{equ:nse_1}, \eqref{equ:cont} and
\eqref{equ:R_reform}, and the boundary condition \eqref{equ:bc}.

Let $n\geqslant 0$ denote the time step index, and
$(\cdot)^n$ denote the variable $(\cdot)$ at time step $n$.
Define
\begin{equation}\label{equ:ic_disc}
  \mbs u^0 = \tilde{\mbs u}^0 = \mbs u_{in}, \quad
  R^0 = R(0) \ \text{defined in equation \eqref{equ:R_ic}}.
\end{equation}
We compute $p^0$ by solving equation \eqref{equ:nse}
(together with \eqref{equ:p_cond})
at $t=0$, which in weak form is given by
\begin{equation}\label{equ:p_step_1}
  \int_{\Omega} \nabla p^0\cdot\nabla q d\Omega
  = \int_{\Omega}\left[\mbs f^0 -\mbs N(\mbs u_{in})  \right]\cdot\nabla q d\Omega
  - \nu\int_{\partial\Omega} \mbs n\times(\nabla\times\mbs u_{in})\cdot\nabla qdA
  - \int_{\partial\Omega}\mbs n\cdot\left.\frac{\partial \mbs w}{\partial t}\right|^0 qdA,
  \quad \forall q\in H^1(\Omega).
\end{equation}
Note that since the boundary velocity
$\mbs w(\mbs x,t)$ is known on $\partial\Omega$,
$\left.\frac{\partial\mbs w}{\partial t} \right|^0$
is well-defined in the above equation.

Then given $(\mbs u^n, \tilde{\mbs u}^n, R^n, p^n)$ we compute
$(\tilde{\mbs u}^{n+1},\mbs u^{n+1},p^{n+1}, R^{n+1})$ together with another
auxiliary field function $\phi^{n+1}$ through the following steps: \\
\noindent\underline{For $\tilde{\mbs u}^{n+1}$:}
\begin{subequations}\label{equ:vel}
  \begin{align}
    &
    \frac{\frac32\tilde{\mbs u}^{n+1}-2\mbs u^n+\frac12\mbs u^{n-1}}{\Delta t}
    + \mbs M(\tilde{\mbs u}^{n+1}) + \nabla p^n
    -\nu\nabla^2 \tilde{\mbs u}^{n+1}
    + \xi\left[\mbs N(\tilde{\mbs u}^{*,n+1}) - \mbs M(\tilde{\mbs u}^{*,n+1})  \right]
    = \mbs f^{n+1};
    \label{equ:vel_1} \\
    &
    \xi = \frac{\left(R^{n+3/2}\right)^2}{E[\bar{\mbs u}^{n+3/2}]}; \label{equ:vel_2} \\
    &
    E[\bar{\mbs u}^{n+3/2}]
    = \int_{\Omega}\frac12\left| \bar{\mbs u}^{n+3/2}\right|^2d\Omega + C_0;
    \label{equ:vel_3} \\
    &
    \tilde{\mbs u}^{n+1} = \mbs w^{n+1}, \quad \text{on} \ \partial\Omega;
    \label{equ:vel_4}
  \end{align} 
\end{subequations}
\noindent\underline{For $\phi^{n+1}$:}
\begin{align}
  \phi^{n+1} = \nabla\cdot\tilde{\mbs u}^{n+1}; \label{equ:phi}
\end{align}
\noindent\underline{For $p^{n+1}$ and $\mbs u^{n+1}$:}
\begin{subequations}\label{equ:p}
  \begin{align}
    &
    \frac{\frac32\mbs u^{n+1}-\frac32\tilde{\mbs u}^{n+1}}{\Delta t}
    + \nabla\left(p^{n+1}-p^n + \nu\phi^{n+1}  \right) = 0;
    \label{equ:p_1} \\
    &
    \nabla\cdot\mbs u^{n+1} = 0; \label{equ:p_2} \\
    &
    \mbs n\cdot\mbs u^{n+1} = \mbs n\cdot\mbs w^{n+1}, \quad
    \text{on} \ \partial\Omega; \label{equ:p_3} \\
    &
    \int_{\Omega} p^{n+1}d\Omega = 0; \label{equ:p_4}
  \end{align}
\end{subequations}
\noindent\underline{For $R^{n+1}$:}
\begin{equation}\label{equ:R_disp}
  \begin{split}
    &
  \left(\frac32 R^{n+1} + R^n - \frac12 R^{n-1} \right)
  \frac{\frac32 R^{n+1} - 2R^n + \frac12 R^{n-1}}{\Delta t}
  = \int_{\Omega} \tilde{\mbs u}^{n+1}\cdot
  \frac{\frac32\mbs u^{n+1}-2\mbs u^n+\frac12\mbs u^{n-1}}{\Delta t} \\
  &
  +\xi \left[
    -\nu\int_{\Omega}\|\nabla\bar{\mbs u}^{n+1} \|^2d\Omega
    + \int_{\Omega}\mbs f^{n+1}\cdot\bar{\mbs u}^{n+1}d\Omega
    + \int_{\Omega}\left(-\bar{P}^{n+1}\mbs n + \nu\mbs n\cdot\nabla\bar{\mbs u}^{n+1}
    -\frac12 (\mbs n\cdot\mbs w^{n+1})\mbs w^{n+1}\right)\cdot\mbs w^{n+1}d\Omega
    \right] \\
  &
  -\int_{\Omega}\left[
    -\mbs M(\tilde{\mbs u}^{n+1}) - \nabla P^{n+1} +\nu\nabla^2\tilde{\mbs u}^{n+1}
    -\xi\left(\mbs N(\tilde{\mbs u}^{*,n+1}) - \mbs M(\tilde{\mbs u}^{*,n+1})  \right)
    + \mbs f^{n+1}
    \right]\cdot\tilde{\mbs u}^{n+1} d\Omega \\
  &
  + (1-\xi)\left[\left|\int_{\Omega}\mbs f^{n+1}\cdot\bar{\mbs u}^{n+1}d\Omega \right|
  + \left|
  \int_{\Omega}\left(-\bar{P}^{n+1}\mbs n + \nu\mbs n\cdot\nabla\bar{\mbs u}^{n+1}
    -\frac12 (\mbs n\cdot\mbs w^{n+1})\mbs w^{n+1}\right)\cdot\mbs w^{n+1}d\Omega
  \right|\right].
  \end{split}
\end{equation}

The symbols in the above equations are defined as follows.
$\Delta t$ is the time step size.
$\tilde{\mbs u}^{n+1}$ and $\mbs u^{n+1}$ are two different approximations
of the velocity $\mbs u$ at step ($n+1$).
$\tilde{\mbs u}^{*,n+1}$ is a 2nd-order explicit approximation of
$\tilde{\mbs u}^{n+1}$, given by
\begin{equation}\label{equ:def_star}
  \tilde{\mbs u}^{*,n+1} = 2\tilde{\mbs u}^n - \tilde{\mbs u}^{n-1}.
\end{equation}
$\bar{\mbs u}^{n+1}$ and $\bar{\mbs u}^{n+3/2}$ are
second-order approximations of $\mbs u$ at time steps
($n+1$) and ($n+3/2$) respectively, and are to be specified later
in equation~\eqref{equ:def_ubar}.
$R^{n+3/2}$ and $R^{n+1/2}$ are 2nd-order approximations of
$R(t)$ at time steps $(n+3/2)$ and $(n+1/2)$, defined by
\begin{equation}\label{equ:def_Rn32}
  R^{n+3/2} = \frac{3}{2} R^{n+1} - \frac{1}{2}R^n, \qquad
  R^{n+1/2} = \frac32 R^{n} - \frac12 R^{n-1}.
\end{equation}
By equation \eqref{equ:phi} we mean that
$\phi^{n+1}$ is a projection of
$\nabla\cdot\tilde{\mbs u}^{n+1}$ into the $H^1(\Omega)$ space.
In equation \eqref{equ:R_disp} $P^{n+1}$ and $\bar{P}^{n+1}$ are defined by
\begin{equation}\label{equ:def_P}
  P^{n+1} = p^{n+1} + \nu\phi^{n+1}, \qquad
  \bar{P}^{n+1} = \bar{p}^{n+1} + \nu\bar{\phi}^{n+1},
\end{equation}
where $\bar{p}^{n+1}$ and $\bar{\phi}^{n+1}$ are second-order
approximations of $p^{n+1}$ and $\phi^{n+1}$ to be specified later
in equations \eqref{equ:def_pbar} and \eqref{equ:phi_bar}.
In equation \eqref{equ:R_disp}, note that
$\frac12(\frac32 R^{n+1} + R^n - \frac12 R^{n-1})$ is a second-order
approximation of $R^{n+1}$, satisfying the following property,
\begin{equation}\label{equ:R_relation}
  \begin{split}
    &
  \left(\frac32 R^{n+1} + R^n - \frac12 R^{n-1} \right)
  \left(\frac32 R^{n+1} - 2R^n + \frac12 R^{n-1} \right) \\
  &
  = \left(R^{n+3/2} \right)^2 - \left(R^{n+1/2} \right)^2
  = \left(\frac32 R^{n+1}-\frac12 R^n \right)^2
  - \left(\frac32 R^{n}-\frac12 R^{n-1} \right)^2.
  \end{split}
\end{equation}


Equations \eqref{equ:vel_1}--\eqref{equ:p_4} are similar to
the rotational pressure correction scheme for the incompressible
Navier-Stokes equations, except for the
$\mbs M(\tilde{\mbs u}^{n+1})$ term and the term involving $\xi$,
which couples these equations together with
equation \eqref{equ:R_disp}.
Note that
all the terms are enforced at the time step ($n+1$), except the
term $\xi$, which is approximated at time step (n+3/2) according to
equation \eqref{equ:vel_2}.
This does not affect the overall second-order accuracy,
because $\xi = \frac{\left(R^{n+3/2}\right)^2}{E[\bar{\mbs u}^{n+3/2}]}$ is
a second-order approximation of
$\frac{R^2(t)}{E(t)}=1$. Here the key is to realize
that $\frac{R^2(t)}{E(t)}=1$ for any time $t$ on the continuum level.
This approximation is a key point in
the gPAV framework~\cite{YangD2019diss}.

The scheme represented by equations \eqref{equ:vel_1}--\eqref{equ:R_disp}
is energy stable due to the following property.
\begin{theorem}\label{thm:thm_1}
  In the absence of the external force ($\mbs f=0$) and with
  homogeneous boundary condition ($\mbs w=0$), the following
  relation holds with the scheme
  given by \eqref{equ:vel_1}--\eqref{equ:R_disp},
  \begin{equation}\label{equ:eng_law}
    \left|\frac32 R^{n+1}-\frac12 R^{n} \right|^2
    - \left|\frac32 R^{n}-\frac12 R^{n-1} \right|^2
    = -\frac{\left(R^{n+3/2} \right)^2}{E[\bar{\mbs u}^{n+3/2}]}
    \nu\Delta t
    \int_{\Omega} \|\nabla \bar{\mbs u}^{n+1} \|^2d\Omega
    \leqslant 0.
  \end{equation}
\end{theorem}
\begin{proof}
  Take the $L^2$ inner products between equation \eqref{equ:vel_1} and
  $\tilde{\mbs u}^{n+1}$, and between equation \eqref{equ:p_1} and
  $\tilde{\mbs u}^{n+1}$. Summing up the two resultant equations
  together with equation \eqref{equ:R_disp}, we get
  \begin{equation}\label{equ:eng_1}
    \frac{\left(R^{n+3/2}\right)^2 - \left(R^{n+1/2}\right)^2}{\Delta t}
    = \xi\left(
      -\nu\int_{\Omega}\|\nabla\bar{\mbs u}^{n+1}  \|^2d\Omega
      + A_1 + A_2
      \right)
      + (1-\xi)\left(|A_1| + |A_2|\right),
  \end{equation}
  where we have used the relation \eqref{equ:R_relation}, and
  \begin{equation}\label{equ:def_A12}
    \left\{
    \begin{split}
      &
      A_1 = \int_{\Omega}\mbs f^{n+1}\cdot\bar{\mbs u}^{n+1}d\Omega, \\
      &
      A_2 = \int_{\Omega}\left(-\bar{P}^{n+1}\mbs n + \nu\mbs n\cdot\nabla\bar{\mbs u}^{n+1}
    -\frac12 (\mbs n\cdot\mbs w^{n+1})\mbs w^{n+1}\right)\cdot\mbs w^{n+1}d\Omega.
    \end{split}
    \right.
  \end{equation}
  If $\mbs f=0$ and $\mbs w=0$, then $A_1=A_2=0$, and equation
  \eqref{equ:eng_1} leads to equation \eqref{equ:eng_law}
  in light of \eqref{equ:vel_2} and \eqref{equ:def_Rn32}.
  Noting that $E[\bar{\mbs u}^{n+3/2}]>0$, the inequality in \eqref{equ:eng_law}
  holds.
\end{proof}

\subsection{Solution Algorithm and Implementation}
\label{sec:implementation}

While the system of equations \eqref{equ:vel_1}--\eqref{equ:R_disp}
are  coupled with one another, they can be solved in a
de-coupled fashion and the scheme can be implemented in an efficient way,
thanks to the fact that the auxiliary variable
$R(t)$ is a scalar number, not a field function.
We next present such a solution algorithm.

Let
\begin{equation}\label{equ:def_gamma0}
  \gamma_0 = \frac{3}{2}; 
  \qquad
  \hat{\mbs u} = 2\mbs u^n - \frac12\mbs u^{n-1}.
\end{equation}
We re-write equation \eqref{equ:vel_1} into
\begin{equation}
  \frac{\gamma_0}{\Delta t} \tilde{\mbs u}^{n+1}
  + \mbs M(\tilde{\mbs u}^{n+1}) - \nu\nabla^2\tilde{\mbs u}^{n+1}
  = \mbs f^{n+1} + \frac{\hat{\mbs u}}{\Delta t}
  -\nabla p^n - \xi\left[
    \mbs N(\tilde{\mbs u}^{*,n+1}) - \mbs M(\tilde{\mbs u}^{*,n+1})
    \right].
\end{equation}
Barring the unknown scalar number $\xi$, this is a linear equation
with respect to $\tilde{\mbs u}^{n+1}$. We solve this equation
together with the boundary condition \eqref{equ:vel_4} as follows.
Define two field functions $\tilde{\mbs u}_1^{n+1}$
and $\tilde{\mbs u}_2^{n+1}$ as solutions to the following problems:
\begin{subequations}
  \begin{align}
    &
    \frac{\gamma_0}{\Delta t} \tilde{\mbs u}_1^{n+1}
    + \mbs M(\tilde{\mbs u}_1^{n+1}) - \nu\nabla^2\tilde{\mbs u}_1^{n+1}
    = \mbs f^{n+1} + \frac{\hat{\mbs u}}{\Delta t}
    -\nabla p^n,
    \label{equ:u1_1} \\
    &
    \tilde{\mbs u}_1^{n+1} = \mbs w^{n+1}, \quad \text{on} \ \partial\Omega;
    \label{equ:u1_2}
  \end{align}
\end{subequations}
\begin{subequations}
  \begin{align}
    &
    \frac{\gamma_0}{\Delta t} \tilde{\mbs u}_2^{n+1}
    + \mbs M(\tilde{\mbs u}_2^{n+1}) - \nu\nabla^2\tilde{\mbs u}_2^{n+1}
    = -\left[
    \mbs N(\tilde{\mbs u}^{*,n+1}) - \mbs M(\tilde{\mbs u}^{*,n+1})
    \right]
    \label{equ:u2_1} \\
    &
    \tilde{\mbs u}_2^{n+1} = 0, \quad \text{on} \ \partial\Omega.
    \label{equ:u2_2}
  \end{align}
\end{subequations}
Then for given value $\xi$
the solution to the equations \eqref{equ:vel_1} and \eqref{equ:vel_4}
are given by
\begin{equation}\label{equ:utilde_soln}
  \tilde{\mbs u}^{n+1} = \tilde{\mbs u}_1^{n+1} + \xi\tilde{\mbs u}_2^{n+1}.
\end{equation}

Let
$
H_0^1(\Omega) = \{\ v\in H^1(\Omega)\ :\ v|_{\partial\Omega}=0\  \}.
$
The weak forms for equations \eqref{equ:u1_1} and \eqref{equ:u2_1}
are given by
\begin{align}
  \begin{split}\label{equ:u1_weak}
    &
  \frac{\gamma_0}{\nu\Delta t}\int_{\Omega}\tilde{\mbs u}_1^{n+1}\varphi d\Omega
  + \int_{\Omega}\nabla\varphi\cdot\nabla\tilde{\mbs u}_1^{n+1}d\Omega
  + \frac{1}{\nu}\int_{\Omega}\mbs M(\tilde{\mbs u}_1^{n+1})\varphi d\Omega \\
  &\qquad
  = \frac{1}{\nu}\int_{\Omega}\left(
  \mbs f^{n+1} + \frac{\hat{\mbs u}}{\Delta t} - \nabla p^n
  \right)\varphi d\Omega,
  \qquad \forall \varphi \in H_0^1(\Omega);
  \end{split} \\
  \begin{split}\label{equ:u2_weak}
    &
    \frac{\gamma_0}{\nu\Delta t}\int_{\Omega}\tilde{\mbs u}_2^{n+1}\varphi d\Omega
    + \int_{\Omega}\nabla\varphi\cdot\nabla\tilde{\mbs u}_2^{n+1}d\Omega
    + \frac{1}{\nu}\int_{\Omega}\mbs M(\tilde{\mbs u}_2^{n+1})\varphi d\Omega \\
    &\qquad
    = -\frac{1}{\nu}\int_{\Omega}\left[
      \mbs N(\tilde{\mbs u}^{*,n+1}) - \mbs M(\tilde{\mbs u}^{*,n+1})
      \right]\varphi,
    \qquad \forall \varphi \in H_0^1(\Omega).
  \end{split}
\end{align}
These weak forms, together with the boundary conditions
\eqref{equ:u1_2} and \eqref{equ:u2_2},
can be implemented using high-order spectral elements
in a straightforward fashion.

Define
\begin{equation}\label{equ:def_ubar}
  \bar{\mbs u}^{n+1} = \tilde{\mbs u}_1^{n+1} + \tilde{\mbs u}_2^{n+1}, \qquad
  \bar{\mbs u}^{n+3/2} = \frac32 \bar{\mbs u}^{n+1} - \frac12 \tilde{\mbs u}^n.
\end{equation}
Note that these are second-order approximations of $\tilde{\mbs u}^{n+1}$
and $\tilde{\mbs u}^{n+3/2}$, respectively.

To solve for $\phi^{n+1}$ from \eqref{equ:phi},
in light of \eqref{equ:utilde_soln}, we 
define two field variables $\phi_1^{n+1}$ and $\phi_2^{n+2}$ by
\begin{align}
  &
  \phi_1^{n+1} = \nabla\cdot\tilde{\mbs u}_1^{n+1}; \label{equ:phi_1} \\
  &
  \phi_2^{n+1} = \nabla\cdot\tilde{\mbs u}_2^{n+1}. \label{equ:phi_2}
\end{align}
Then the solution to \eqref{equ:phi} is given by
\begin{equation}
  \phi^{n+1} = \phi_1^{n+1} + \xi\phi_2^{n+1},
  \label{equ:phi_soln}
\end{equation}
where $\xi$ is still to be determined.
The weak forms for equations \eqref{equ:phi_1} and \eqref{equ:phi_2}
are given by
\begin{align}
  &
  \int_{\Omega}\phi_1^{n+1}\varphi d\Omega =
  \int_{\Omega} \nabla\cdot\tilde{\mbs u}_1^{n+1}\varphi d\Omega,
  \quad \forall \varphi \in H^1(\Omega);
  \label{equ:phi1_weak} \\
  &
  \int_{\Omega}\phi_2^{n+1}\varphi d\Omega =
  \int_{\Omega} \nabla\cdot\tilde{\mbs u}_2^{n+1}\varphi d\Omega,
  \quad \forall \varphi \in H^1(\Omega).
  \label{equ:phi2_weak}
\end{align}
We define $\bar{\phi}^{n+1}$ in equation \eqref{equ:def_P} as
\begin{equation}\label{equ:phi_bar}
\bar\phi^{n+1} = \phi_1^{n+1} + \phi_2^{n+1}.
\end{equation}
Note that this is a second-order approximation of $\phi^{n+1}$.

To solve equations \eqref{equ:p_1}--\eqref{equ:p_4} for $p^{n+1}$
and $\mbs u^{n+1}$, we first derive the weak form of
the equations.
Let $q\in H^1(\Omega)$ denote a test function. 
Taking the $L^2$ inner product between equation \eqref{equ:p_1}
and $\nabla q$ leads to
\begin{equation}
  \int_{\Omega}\nabla p^{n+1}\cdot\nabla q d\Omega
  = \int_{\Omega}\left(
  \frac{\gamma_0}{\Delta t}\tilde{\mbs u}^{n+1} + \nabla p^n - \nu\nabla\phi^{n+1}
  \right)\cdot \nabla q d\Omega
  - \frac{\gamma_0}{\Delta t}\int_{\partial\Omega}\mbs n\cdot\mbs w^{n+1}q dA,
  \quad \forall q\in H^1(\Omega),
  \label{equ:p_weak}
\end{equation}
where we have used integration by part and equation \eqref{equ:p_3}.
In light of equations \eqref{equ:utilde_soln} and \eqref{equ:phi_soln},
we define two field variables $p_1^{n+1}$ and $p_2^{n+1}$ as
solutions to the following equations:
\begin{subequations}
  \begin{align}
    &
    \int_{\Omega}\nabla p_1^{n+1}\cdot\nabla q d\Omega
    = \int_{\Omega}\left(
    \frac{\gamma_0}{\Delta t}\tilde{\mbs u}_1^{n+1} + \nabla p^n - \nu\nabla\phi_1^{n+1}
    \right)\cdot \nabla q d\Omega
    - \frac{\gamma_0}{\Delta t}\int_{\partial\Omega}\mbs n\cdot\mbs w^{n+1}q dA,
    \quad \forall q\in H^1(\Omega);
    \label{equ:p1_1} \\
    &
    \int_{\Omega} p_1^{n+1}d\Omega = 0;
    \label{equ:p1_2}
  \end{align}
\end{subequations}
\begin{subequations}
  \begin{align}
    &
    \int_{\Omega}\nabla p_2^{n+1}\cdot\nabla q d\Omega
    = \int_{\Omega} \left(
    \frac{\gamma_0}{\Delta t}\tilde{\mbs u}_2^{n+1}
    - \nu\nabla\phi_2^{n+1}
    \right) \cdot \nabla q d\Omega,
    \quad \forall q\in H^1(\Omega);
    \label{equ:p2_1} \\
    &
    \int_{\Omega} p_2^{n+1}d\Omega = 0.
    \label{equ:p2_2}
  \end{align}
\end{subequations}
Then for given $\xi$ the solution to equations
\eqref{equ:p_weak} and \eqref{equ:p_4} is
\begin{equation} \label{equ:p_soln}
  p^{n+1} = p_1^{n+1} + \xi p_2^{n+1}.
\end{equation}
With $p_1^{n+1}$ and $p_2^{n+1}$ given by
equations \eqref{equ:p1_1}--\eqref{equ:p2_2},
we define $\bar p^{n+1}$ in \eqref{equ:def_P} as
\begin{equation}\label{equ:def_pbar}
  \bar p^{n+1} = p_1^{n+1} + p_2^{n+1}.
\end{equation}

Now we are ready to determine the scalar value $\xi$.
Note that the combination of equations
\eqref{equ:vel_1}, \eqref{equ:p_1} and \eqref{equ:R_disp}
leads to equation \eqref{equ:eng_1}.
In light of \eqref{equ:vel_2}, equation \eqref{equ:eng_1}
yields the following formula for computing $\xi$,
\begin{equation}
  \xi = \frac{
    \left(R^{n+1/2} \right)^2
    + \left(|A_1| + |A_2| \right)\Delta t
  }{
    E[\bar{\mbs u}^{n+3/2}]
    + \left[
      \nu\int_{\Omega}\|\nabla \bar{\mbs u}^{n+1} \|^2d\Omega
      + \left(|A_1| - A_1 \right)
      + \left(|A_2| - A_2 \right)
      \right] \Delta t
  },
  \label{equ:cal_xi}
\end{equation}
where $R^{n+1/2}$ is given by \eqref{equ:def_Rn32},
$\bar{\mbs u}^{n+1}$ and
$\bar{\mbs u}^{n+3/2}$ are given by \eqref{equ:def_ubar},
and $A_1$ and $A_2$ are given by \eqref{equ:def_A12}.
Then $R^{n+1}$ is computed as follows,
\begin{equation}\label{equ:Rnp1}
  \left\{
  \begin{split}
    &
    R^{n+3/2} = \sqrt{\xi E[\bar{\mbs u}^{n+3/2}]}, \\
    &
    R^{n+1} = \frac23 R^{n+3/2} + \frac13 R^n.
  \end{split}
  \right.
\end{equation}
It can be noted that $\xi> 0$ and $R^{n+3/2}> 0$
for all time steps, and also $R^{n+1}> 0$ for all
time steps, if $\left.R^{n+1/2}\right|_{n=0}> 0$,
irrespective of the $\Delta t$ value or
the external force $\mbs f$ and the boundary velocity $\mbs w$.
The Appendix A outlines a method for approximating
the variables for the first time step, which ensures
that $R^1> 0$ and $\left.R^{n+1/2}\right|_{n=0}> 0$.

Combining the above discussions, we end up with
the solution algorithm listed in Algorithm \ref{alg:alg_1}.
This algorithm has the following properties:
(i) The computations for the velocity and pressure
are de-coupled.
(ii) Only linear equations need to be solved within a time step.
(iii) 
The resultant linear algebraic systems upon discretization
involve quasi-constant coefficient matrices, which
can be updated every $k_0$ time step sizes ($k_0$ denoting
an integer parameter).
(iv) The computed values for the auxiliary variable
are guaranteed to be positive.
(v) Two copies of the field variables
(velocity, pressure and $\nabla\cdot\tilde{\mbs u}^{n+1}$)
are computed within a time step.
(vi) The algorithm satisfies a discrete energy stability property.

\begin{algorithm}[htb]
  \SetKwInOut{Input}{input}\SetKwInOut{Output}{output}

  \Input{($\mbs u^{n}$, $\tilde{\mbs u}^{n}$, $p^n$, $R^n$), and these variables of
  previous time steps}
  \Output{($\mbs u^{n+1}$, $\tilde{\mbs u}^{n+1}$, $p^{n+1}$, $R^{n+1}$, $\phi^{n+1}$) }
  \BlankLine\BlankLine
  \Begin{
      Solve equations \eqref{equ:u1_weak} for $\tilde{\mbs u}_1^{n+1}$\;
      Solve equations \eqref{equ:u2_weak} for $\tilde{\mbs u}_2^{n+1}$\;
      \BlankLine
      Solve equation \eqref{equ:phi1_weak} for $\phi_1^{n+1}$\;
      Solve equation \eqref{equ:phi2_weak} for $\phi_2^{n+1}$\;
      \BlankLine
      Solve equations \eqref{equ:p1_1}--\eqref{equ:p1_2} for $p_1^{n+1}$\;
      Solve equations \eqref{equ:p2_1}--\eqref{equ:p2_2} for $p_2^{n+1}$\;
      \BlankLine
      Compute $\bar{\mbs u}^{n+1}$, $\bar{\mbs u}^{n+3/2}$, $\bar\phi^{n+1}$,
      $\bar p^{n+1}$ and $\bar{P}^{n+1}$ based on equations \eqref{equ:def_ubar},
      \eqref{equ:phi_bar}, \eqref{equ:def_pbar}, and \eqref{equ:def_P}\;
      Compute $A_1$ and $A_2$ based on equation \eqref{equ:def_A12}\;
      Compute $\xi$ based on equation \eqref{equ:cal_xi}\;
      \BlankLine
      Compute $\tilde{\mbs u}^{n+1}$ based on equation \eqref{equ:utilde_soln}\;
      Compute $\phi^{n+1}$ based on equation \eqref{equ:phi_soln}\;
      Compute $p^{n+1}$ based on equation \eqref{equ:p_soln}\;
      Compute $R^{n+1}$ based on equation \eqref{equ:Rnp1}\;
      Compute $\mbs u^{n+1}$ by equation \eqref{equ:p_1} as follows,
      \begin{equation}\label{equ:unp1}
        \mbs u^{n+1} = \tilde{\mbs u}^{n+1} - \frac{\Delta t}{\gamma_0}\nabla\left(
        p^{n+1} - p^n + \nu\phi^{n+1}
        \right);
      \end{equation}
    }
    \caption{Solution algorithm within a time step.}
  \label{alg:alg_1}
\end{algorithm}

Equations \eqref{equ:u1_weak}--\eqref{equ:u2_weak},
\eqref{equ:phi1_weak}--\eqref{equ:phi2_weak},
and \eqref{equ:p1_1}--\eqref{equ:p2_2} for the field
functions $\tilde{\mbs u}_i^{n+1}$, $\phi_i^{n+1}$
and $p_i^{n+1}$ ($i=1,2$) are already in weak forms,
and they can be implemented using $C^0$ type finite elements
or spectral elements in a straightforward fashion.
In the current work, these equations are discretized in
space using $C^0$ type spectral
elements~\cite{SherwinK1995,KarniadakisS2005}.
Upon discretization, the pressure linear algebraic systems
have a symmetric coefficient matrix and are solved using
the conjugate gradient (CG) linear solver.
The coefficient matrix in the velocity linear algebraic systems
is non-symmetric (but positive definite) and is solved
using the bi-conjugate gradient stabilized (BiCGStab) linear solver.


\begin{remark}
  \label{rem:rem_1}
  In equation \eqref{equ:nse_1}  we can also choose
  \begin{equation}
    \mbs M(\mbs u) = 0,
    \label{equ:def_M2}
  \end{equation}
  and use the same algorithm represented by
  equations \eqref{equ:vel_1}--\eqref{equ:R_disp}.
  The energy stability property, Theorem \ref{thm:thm_1},
  still holds for this modified algorithm.
  The advantage of this modification lies in that
  the resultant linear algebraic systems upon discretization now
  involve only constant and time-independent
  coefficient matrices, which can be pre-computed.
  However, we observe that  this modified algorithm
  is less accurate than the current algorithm when the time step
  size increases to moderate or fairly large values. 
  This point will be demonstrated by numerical experiments
  in Section \ref{sec:tests}.
  
\end{remark}

\section{Representative Numerical Tests}
\label{sec:tests}

We next use several flow problems in two
dimensions to test the performance of the method
developed in the previous section.
The spatial/temporal convergence rates of the method
are first investigated using a manufactured analytic solution.
Then the Kovasznay flow and the flow
past a hemisphere in a narrow periodic channel are simulated to study 
the accuracy and stability of the method at large
(or fairly large) time step sizes.

\subsection{Convergence Rates}

\begin{figure}
  \centerline{
    \includegraphics[width=2.8in]{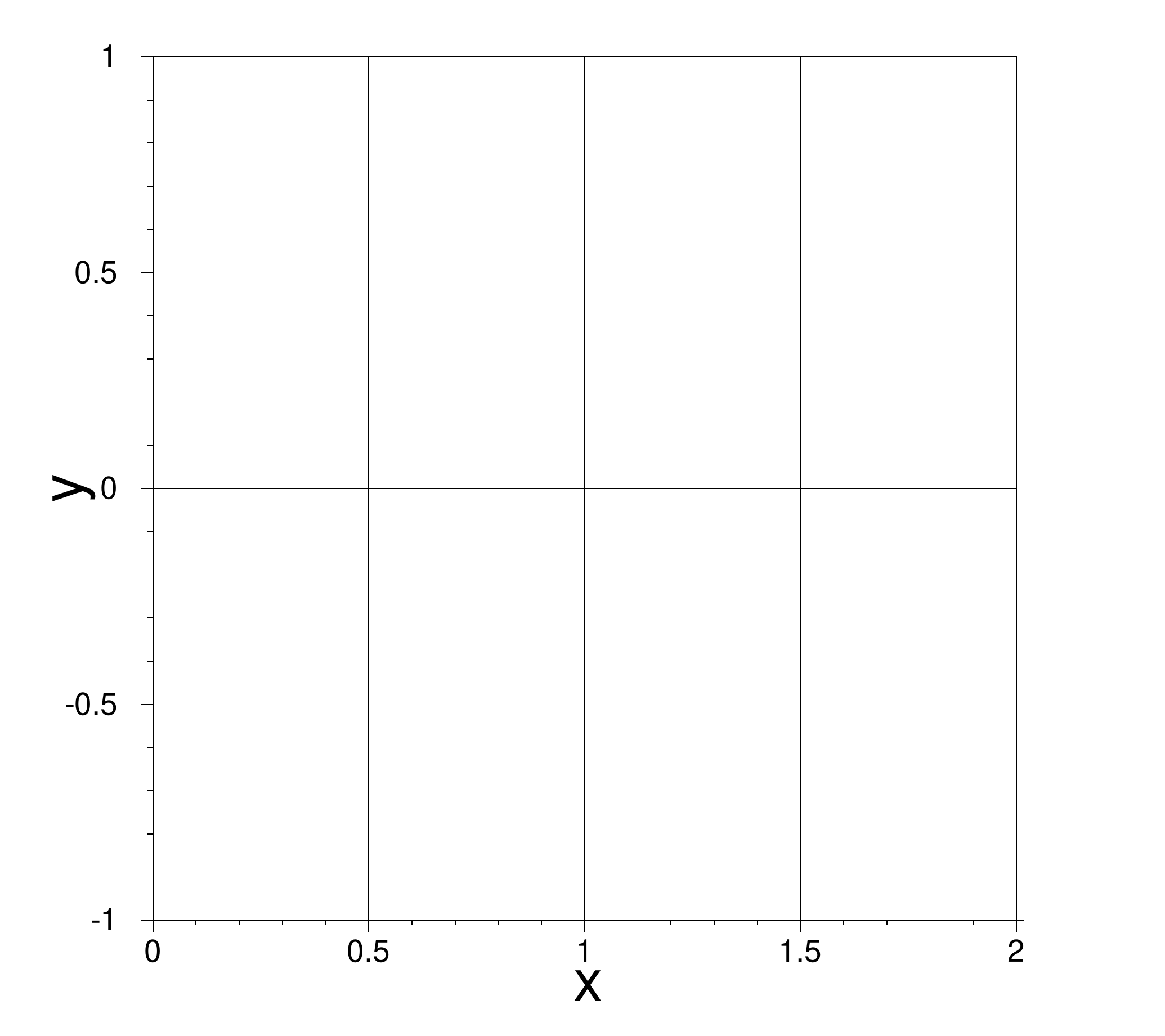}(a)
  }
  \centerline{
    \includegraphics[width=3in]{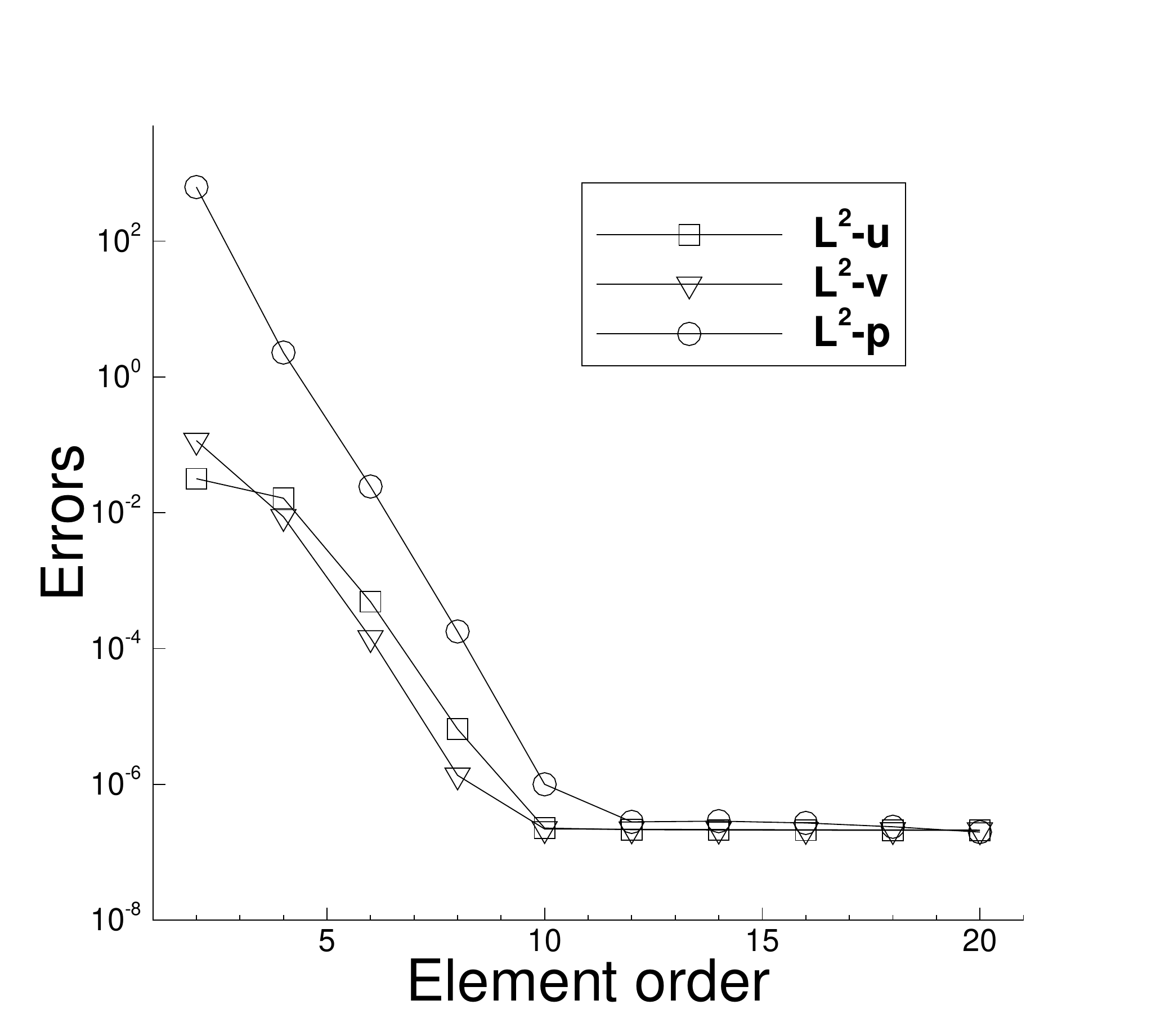}(b)
    \includegraphics[width=3in]{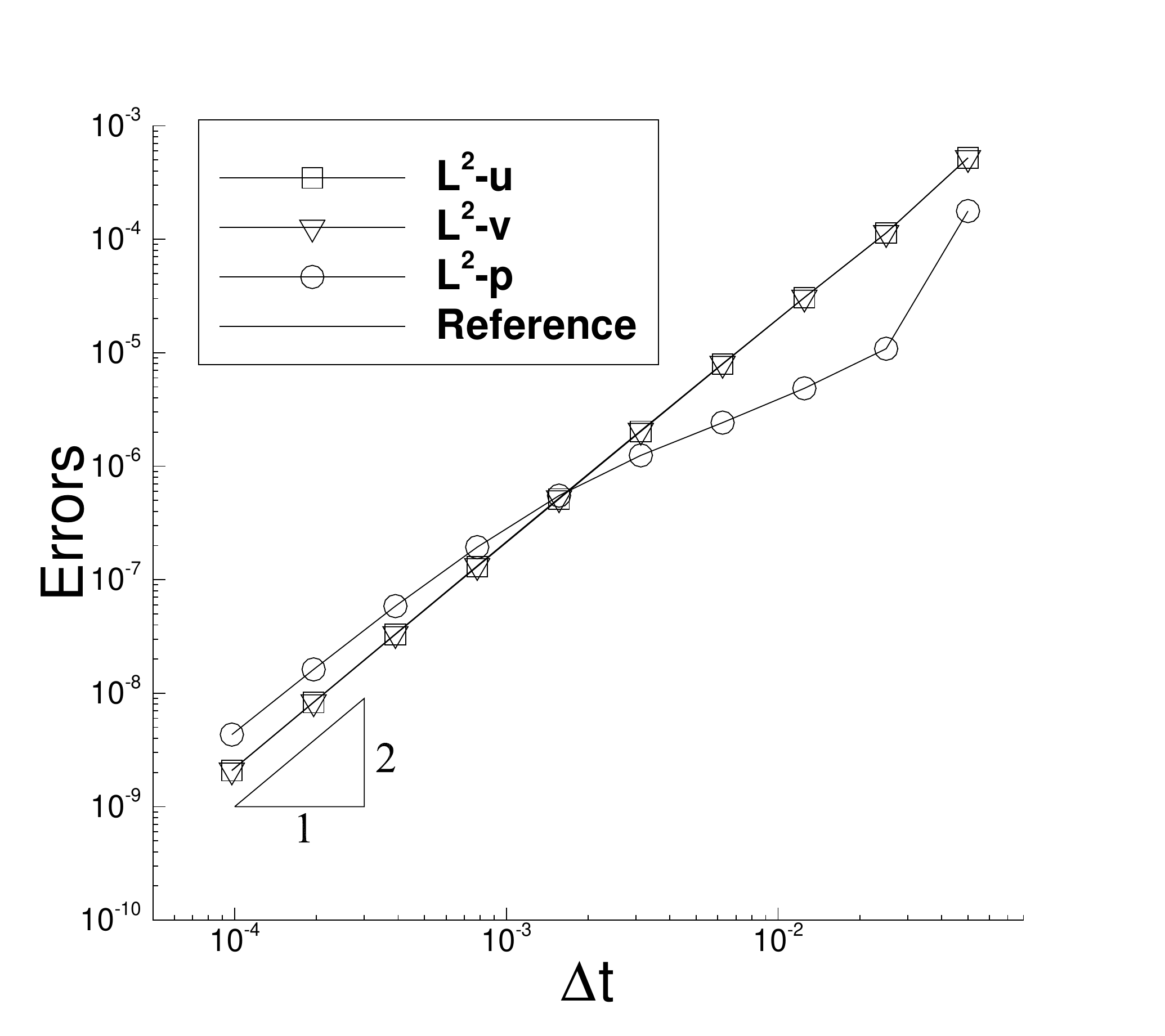}(c)
  }
  \caption{
    Convergence rates: (a) Computational domain and mesh.
    $L^2$ errors of the flow variables
    as a function of (b) the element order (with fixed $t_f=0.1$ and
    $\Delta t=0.001$), and (c) the time step size $\Delta t$
    (with fixed $t_f=0.1$ and element order $14$).
  }
  \label{fig:conv}
\end{figure}

We first demonstrate the spatial and temporal convergence rates of
the method developed herein using a manufactured analytic solution
to the incompressible Navier-Stokes equations.
Consider the rectangular domain  shown in Figure \ref{fig:conv}(a),
$0\leqslant x\leqslant 2$ and $-1\leqslant y\leqslant 1$,
and the following analytic expressions for the flow variables
on this domain,
\begin{equation}
  \left\{
  \begin{split}
    &
    u = 2\sin(\pi x)\cos(\pi y)\sin t, \\
    &
    v = -2\cos(\pi x)\sin(\pi y)\sin t, \\
    &
    p = 2\sin(\pi x)\sin(\pi y)\cos t,
  \end{split}
  \right.
  \label{equ:anal_soln}
\end{equation}
where $(u,v)$ are the $x$ and $y$ components of the velocity $\mbs u$,
respectively.
In equation \eqref{equ:nse}
the external body force $\mbs f$  is chosen such that this equation is
satisfied by the analytic
expressions given in \eqref{equ:anal_soln}. It can be verified
that these expressions also satisfy the equation \eqref{equ:cont}.


We discretize the domain using a mesh of $8$ quadrilateral elements
as shown in Figure \ref{fig:conv}(a), with $4$ elements along the
$x$ direction and $2$ along the $y$ direction.
The scheme from Section \ref{sec:method} is employed to solve
the incompressible Navier-Stokes equations
\eqref{equ:nse}--\eqref{equ:cont}. Dirichlet boundary
condition \eqref{equ:bc} is imposed on all boundaries,
in which the boundary velocity $\mbs w$ is chosen according to
the analytical expressions given in \eqref{equ:anal_soln}.
The initial velocity ${\mbs u}_{in}$ is obtained by setting
$t=0$ in the expressions of \eqref{equ:anal_soln}.
We employ a fixed $C_0=1000$ in the tests of this subsection.
The field $\mbs u_0$ in $\mbs M(\mbs u)$ (see equation \eqref{equ:def_M1})
is updated every $20$ time steps ($k_0=20$).

We integrate the Navier-Stokes equations from $t=0$ to
$t=t_f$ ($t_f$ to be specified below), and compare the numerical
solution at $t=t_f$ against the analytical solution given
by \eqref{equ:anal_soln}. The $L^2$ norms of the errors for
different flow variables have been computed.
The element order and the time step size $\Delta t$ are
varied in the spatial and temporal convergence tests,
in order to study their effects on the errors of
the numerical solutions.

Figure \ref{fig:conv}(b) illustrates the spatial convergence
behavior of the method. Here we use a fixed $t_f=0.1$ and
time step size $\Delta t=0.001$, and then
vary the element order systematically between
$2$ and $20$. This figure shows the $L^2$ errors of
different variables corresponding to these element orders.
A clear exponential convergence rate can be observed
for element orders below $10$. The error curves are observed to level off
for element orders above $10$, due to the saturation of
the temporal truncation errors.

Figure \ref{fig:conv}(c) is an illustration of the temporal
convergence behavior of the method.
Here the integration time is fixed at $t_f=0.1$
and the element order is fixed at $14$.
We vary the time step size systematically between
$\Delta t=0.05$ and $\Delta t=9.765625e-5$, and plot
the $L^2$ errors of the flow variables
as a function of $\Delta t$. The temporal convergence rate for
the velocity is clearly second-order.
It is also observed to be second-order for the pressure
when $\Delta t$ is small. But the pressure convergence
behavior is not as uniform as the velocity.

\subsection{Kovasznay Flow}

\begin{figure}
  \centerline{
    \includegraphics[width=3in]{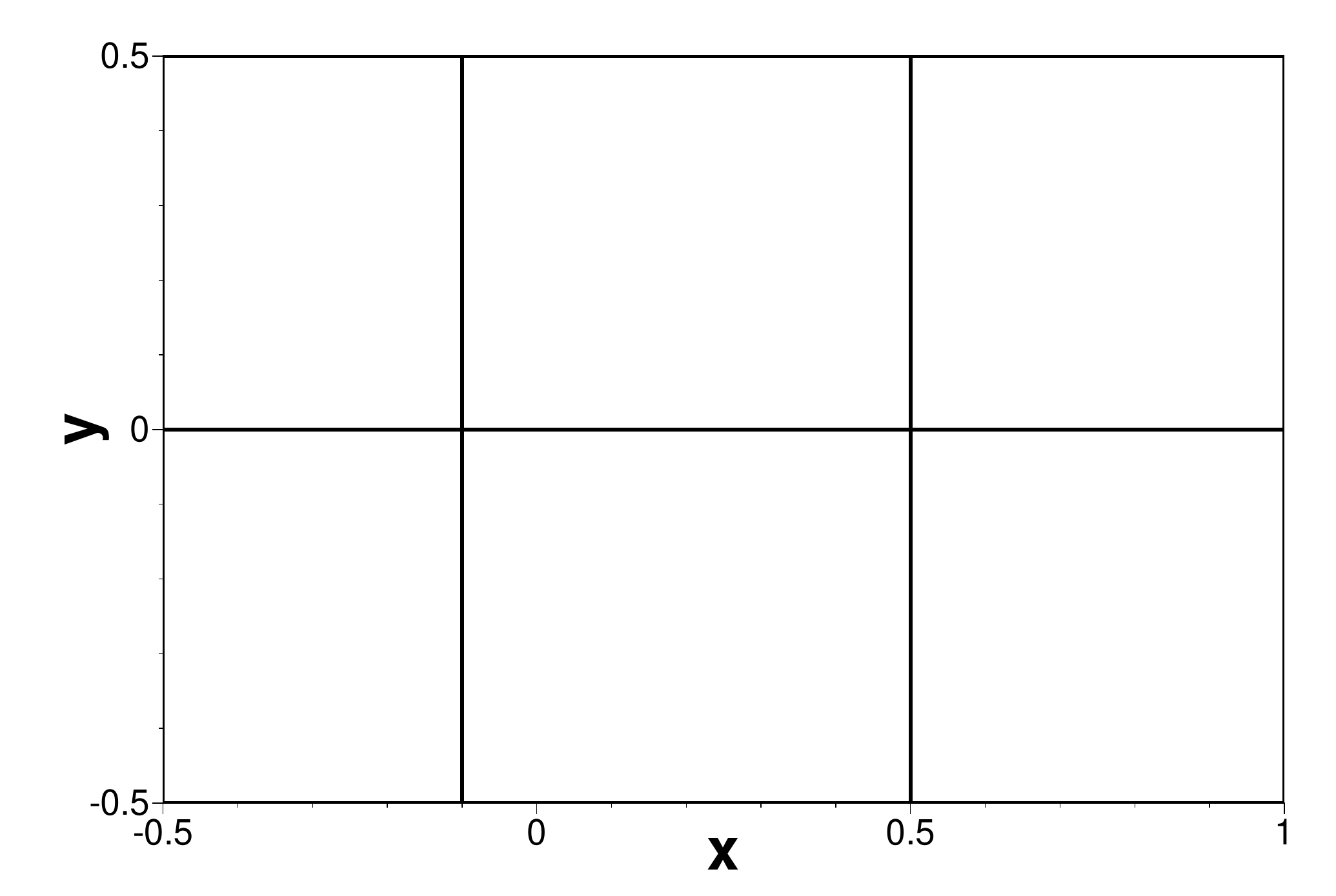}(a)
    \includegraphics[width=3in]{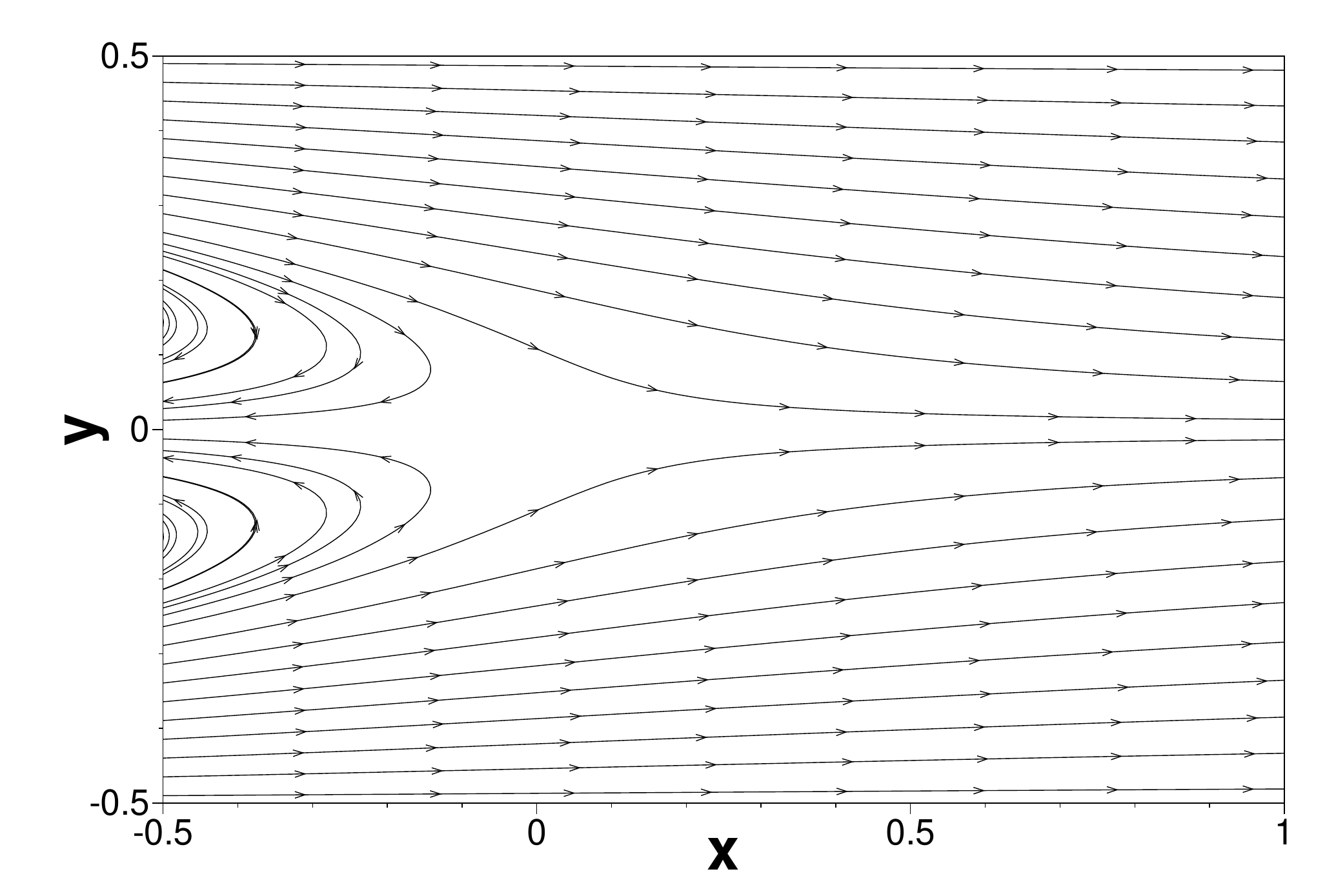}(b)
  }
  \caption{
    Kovasznay flow: (a) mesh of $6$ quadrilateral elements,
    and (b) flow patterns visualized by streamlines.
  }
  \label{fig:kovas}
\end{figure}


In this subsection we employ the Kovasznay flow,
 a steady-state problem with a known analytic solution,
to test the accuracy and stability of the current method.
This problem has been studied in a
number of previous works (see e.g.~\cite{BlackburnS2004,KarniadakisS2005,DongS2010},
among others).

Consider the domain, $0.5\leqslant x\leqslant 1$
and $-0.5\leqslant y\leqslant 0.5$, as shown in
Figure \ref{fig:kovas}(a).
The Kovasznay flow is given by the following
expressions for the flow variables~\cite{Kovasznay1948},
\begin{equation}
  \left\{
  \begin{split}
    &
    u = 1 - \exp(\lambda x)\cos(2\pi y) \\
    &
    v = \frac{\lambda}{2\pi}\exp(\lambda x)\sin(2\pi y) \\
    &
    p = \frac{1}{2}(1-\exp(2\lambda x))
  \end{split}
  \right.
  \label{equ:kovas_soln}
\end{equation}
with the constant
$\lambda = \frac{1}{2\nu}\left(1-\sqrt{1+16\pi^2\nu^2} \right)$.
These expressions satisfy the Navier-Stokes equations
\eqref{equ:nse}--\eqref{equ:cont} with $\mbs f=0$.
Figure \ref{fig:kovas}(b) is a visualization of
the flow patterns based on the streamlines.
We employ a fixed $\nu=\frac{1}{40}$ in this test.


We employ the method presented in Section \ref{sec:method}
to simulate the Kovasznay flow. The flow domain is first
discretized using a mesh of $6$ quadrilateral spectral elements,
as given in Figure \ref{fig:kovas}(a).
The element order is varied in the tests, which will be
specified below. The external body force in the Navier-Stokes
equation \eqref{equ:nse} is set to $\mbs f=0$.
Dirichlet boundary condition \eqref{equ:bc}
is imposed on all domain boundaries, with the
boundary velocity $\mbs w$ chosen according to the
analytical expressions from \eqref{equ:kovas_soln}.
Zero initial velocity ($\mbs u_{in}=0$ in \eqref{equ:ic}) has been
employed in all the tests below. The governing equations
are integrated to a sufficiently long time so that
the flow has reached the steady state. The steady-state solutions
are then compared with
the analytical expressions from \eqref{equ:kovas_soln}
to compute their errors in different norms.
The simulation parameter values are varied to investigate
their effects on the results, which will be specified
in the discussions below.

\begin{figure}
  \centerline{
    \includegraphics[width=3in]{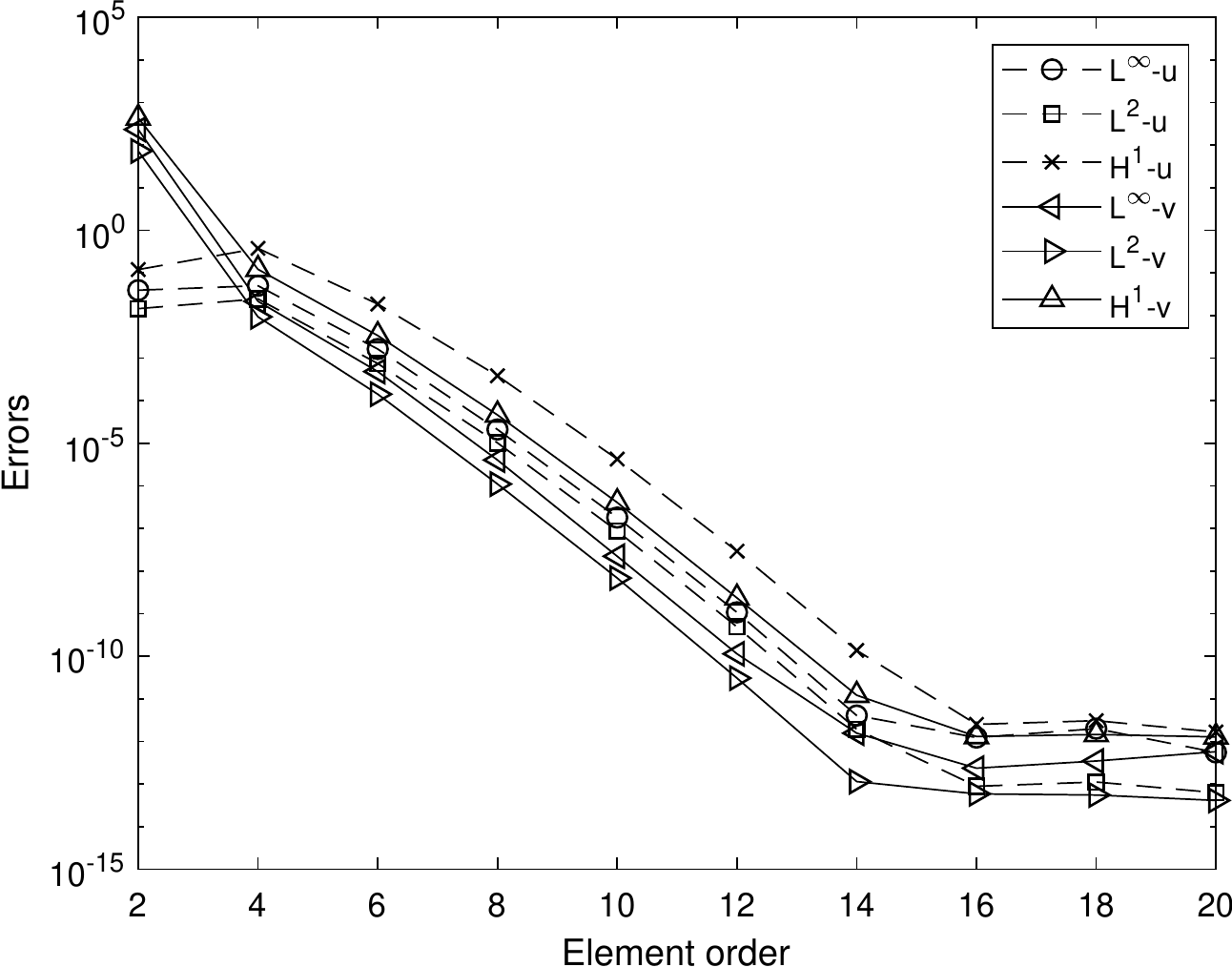}
  }
  \caption{
    Kovasznay flow: Numerical errors of the steady-state velocity
    versus the element order.
  }
  \label{fig:kovas_order}
\end{figure}


We vary the element order systematically and have computed the
errors of the steady-state solution against the analytic
solution in \eqref{equ:kovas_soln} corresponding to different
element orders. Figure \ref{fig:kovas_order} shows
the numerical errors of the steady-state velocity
in $L^{\infty}$, $L^2$ and $H^1$ norms as a function of
the element order. These results are computed with
$C_0=1000$ and $\Delta t=0.001$, and the field $\mbs u_0$ (and hence
the coefficient matrix) is updated
every $20$ time steps ($k_0=20$).
The numerical errors decrease exponentially
with increasing element order for orders below $14$,
and the errors saturate at a level around $10^{-13}$ with element orders
beyond $14$.


\begin{figure}
  \centering
  \includegraphics[width=3in]{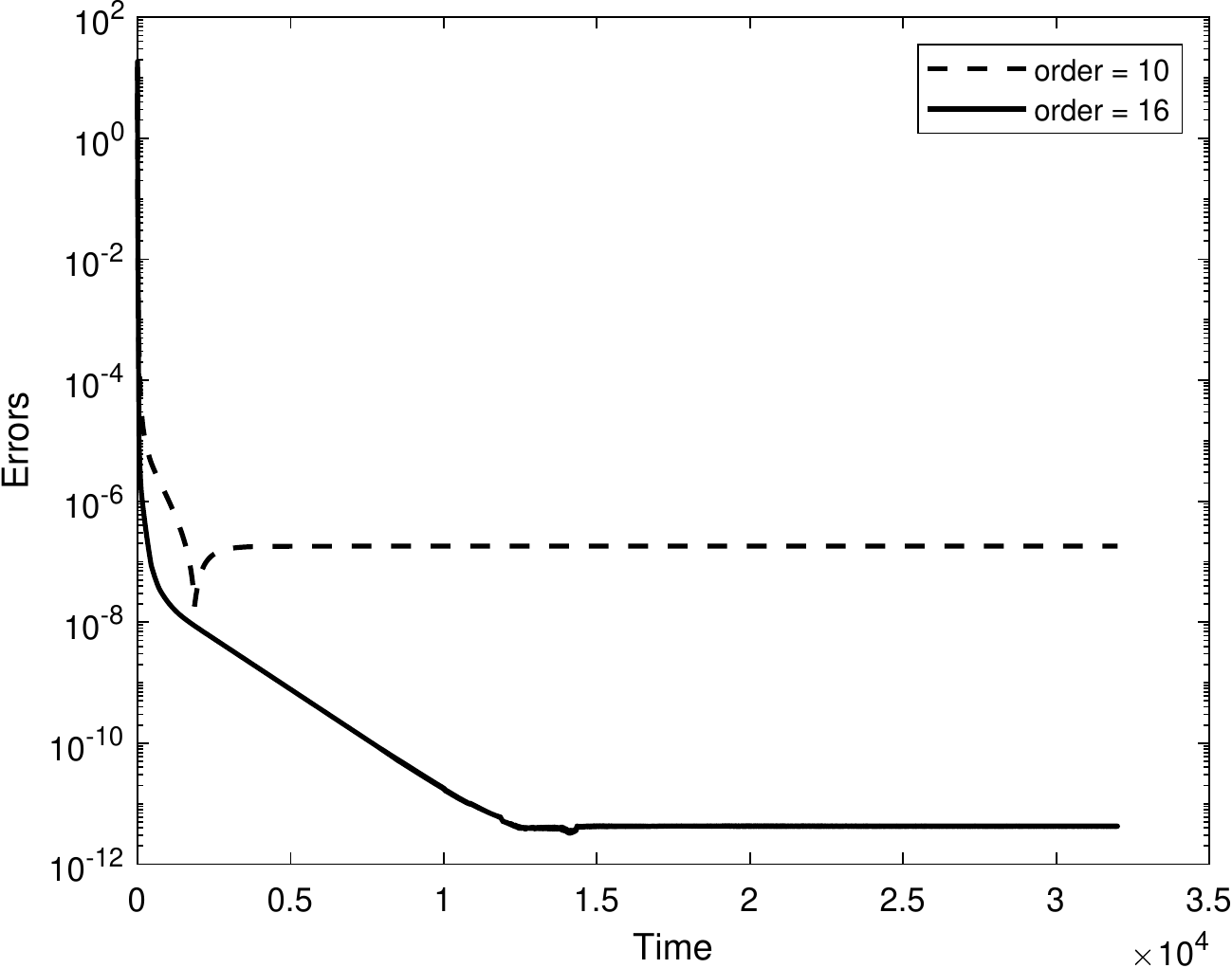}
  \caption{
    Kovasznay flow: time histories of the $L^{\infty}$ error of
    the $x$ velocity obtained with two element orders ($10$ and $16$) and
    a time step size $\Delta t=0.4$.
  }
  \label{fig:kovas_hist}
\end{figure}

Figure \ref{fig:kovas_hist} illustrates the typical convergence 
behavior of the method for the Kovasznay flow. It shows the time histories
of the $L^{\infty}$ error of the $x$ velocity component
computed with a time step size $\Delta t=0.4$
with two element orders $10$ and $16$. The field $\mbs u_0$ is updated
every $20$ time steps ($k_0=20$), and $C_0=1000$ in the simulations.
The error decreases over time and eventually reaches a steady-state
level, around $10^{-7}$ with element order $10$ and
around $10^{-12}$ with  order $16$.
It can take a quite long time for
the simulation to reach the steady state,
for instance about $t=1.5\times 10^4$ with
the element order $16$.


\begin{table}
  \centering
  \begin{tabular}{ll|llll}
    \hline
    Element order & $\Delta t$ & $L^{\infty}-u$ & $L^2-u$ & $L^{\infty}-v$ & $L^2-v$ \\
    \hline
    10 & 0.004 & 1.807e-7 & 8.712e-8 & 1.807e-7 & 8.712e-8   \\
    & 0.005 & 1.806e-7 & 8.707e-8 & 2.284e-8 & 6.884e-9  \\
    & 0.006 & 1.799e-7 & 8.683e-8  & 2.277e-8 & 6.878e-9 \\
    & 0.007 & 1.800e-7 & 8.685e-8  & 2.289e-8 & 6.891e-9 \\
    & 0.008 & 1.799e-7 & 8.681e-8 & 2.295e-8 & 6.900e-9 \\
    & 0.01 & 1.800e-7 & 8.683e-8 & 2.313e-8 & 6.926e-9 \\
    & 0.02 & 1.800e-7 & 8.683e-8 & 2.371e-8 & 7.034e-9 \\
    & 0.03 & 1.801e-7 & 8.687e-8 & 2.409e-8 & 7.119e-9 \\
    & 0.04 & 1.802e-7 & 8.689e-8 & 1.802e-7 & 8.689e-8 \\
    & 0.05 & 1.803e-7 & 8.690e-8 & 2.455e-8 & 7.230e-9 \\
    & 0.06 & 1.803e-7 & 8.691e-8 & 2.470e-8 & 7.269e-9 \\
    & 0.07 & 1.803e-7 & 8.691e-8 & 2.482e-8 & 7.300e-9 \\
    & 0.08 & 1.804e-7 & 8.692e-8 & 2.492e-8 & 7.326e-9 \\
    & 0.1 & 1.804e-7 & 8.692e-8 & 2.508e-8 & 7.366e-9 \\
    & 0.2 & 1.804e-7 & 8.694e-8 & 2.548e-8 & 7.465e-9 \\
    & 0.3 & 1.804e-7 & 8.694e-8 & 2.564e-8 & 7.506e-9 \\
    & 0.4 & 1.804e-7 & 8.694e-8 & 2.573e-8 & 7.529e-9 \\
    & 0.5 & 2.233e-2 & 7.558e-3 & 2.233e-2 & 7.558e-3 \\
    & 0.75 & 2.254e-2 & 4.468e-3 & 8.565e-3 & 1.561e-3 \\
    & 1.0 & 4.117e-2 & 7.967e-3 & 9.873e-3 & 2.179e-3 \\
    & 2.0 & 7.901e-2 & 1.687e-2 & 2.680e-2 & 6.113e-3 \\
    & 10.0 & 1.939e-1 & 4.906e-2 & 4.124e-2 & 1.064e-2 \\
    & 100.0 & 4.895e-1 & 1.165e-1 & 9.866e-2 & 2.442e-2 \\
    \hline
    16 & 0.004 & 7.154e-13 & 3.241e-14 & 7.397e-14 & 1.620e-14  \\
    & 0.005 & 4.681e-13 & 1.820e-14 & 5.831e-14 & 9.803e-15 \\
    & 0.006 & 2.809e-13 & 1.245e-14 & 1.103e-13 & 8.323e-15 \\
    & 0.007 & 3.720e-13 & 2.829e-14 & 1.982e-13 & 8.572e-15 \\
    & 0.008 & 6.074e-13 & 5.108e-14 & 1.256e-13 & 8.638e-15 \\
    & 0.01 & 1.198e-12 & 9.969e-14 & 4.689e-13 & 1.281e-14 \\
    & 0.02 & 1.249e-12 & 1.054e-13 & 1.649e-12 & 9.735e-14 \\
    & 0.03 & 1.879e-12 & 1.625e-13 & 5.806e-13 & 1.501e-14 \\
    & 0.04 & 7.096e-13 & 4.212e-14 & 2.645e-13 & 1.725e-14 \\
    & 0.05 & 2.174e-12 & 2.078e-13 & 5.763e-13 & 1.710e-14 \\
    & 0.06 & 1.084e-12 & 1.296e-13 & 2.345e-12 & 1.666e-13 \\
    & 0.07 & 2.303e-12 & 2.222e-13 & 6.019e-13 & 1.819e-14 \\
    & 0.08 & 2.301e-12 & 2.069e-13 & 6.044e-13 & 1.776e-14 \\
    & 0.1 & 2.151e-12 & 1.768e-13 & 6.692e-13 & 1.772e-14 \\
    & 0.2 & 2.425e-12 & 1.671e-13 & 5.800e-13 & 1.853e-14 \\
    & 0.3 & 1.985e-12 & 1.733e-13 & 4.923e-13 & 1.827e-14 \\
    & 0.4 & 4.238e-12 & 1.451e-12 & 8.329e-13 & 2.231e-13 \\
    & 0.5 & 3.287e-2 & 1.002e-2 & 6.049e-3 & 1.504e-3 \\
    & 0.75 & 4.682e-2 & 6.360e-3 & 1.788e-2 & 2.714e-3 \\
    & 1.0 & 4.653-2 & 6.507e-3 & 1.301e-2 & 2.322e-3 \\
    & 2.0 & 7.620e-2 & 1.329e-2 & 1.993e-2 & 3.747e-3 \\
    & 10.0 & 2.214e-1 & 5.381e-2 & 4.510e-2 & 1.112e-2 \\
    & 100.0 & 4.788e-1 & 1.100e-1 & 1.033e-1 & 2.434e-2 \\
    \hline
  \end{tabular}
  \caption{
    Kovasznay flow: effect of $\Delta t$ on the accuracy of simulation results.
    $C_0=1000$ and $\mbs u_0$ is updated every $20$ time steps ($k_0=20$) in simulations.
  }
  \label{tab:kovas_dt}
\end{table}

Thanks to the energy stability property 
(Theorem \ref{thm:thm_1}), stable simulation results can be obtained
using the current method
with various time step sizes, ranging from small to very large
values. This point is demonstrated by the results in Table \ref{tab:kovas_dt}.
This table lists the $L^{\infty}$ and $L^2$ errors of the steady-state velocity
($x$ component $u$ and $y$ component v) from current simulations
corresponding to various $\Delta t$ values ranging from
$\Delta t=0.004$ to $\Delta t=100$. The results for two element orders,
order $10$ and order $16$, are provided.
In these tests $C_0=1000$, and the coefficient matrix is updated every
$20$ time steps ($k_0=20$).
With $\Delta t=0.5$ and larger, we observe that the numerical errors
fluctuate over time about some level in the long-time simulations.
So the errors provided in this table corresponding to such $\Delta t$
are the time-averaged values. 
We can make several observations from these results.
First, they verify that the current method is indeed stable
in long-term simulations, even with large time step sizes.
Second, the computation using the current method starts to lose accuracy
with time step sizes beyond a certain value, which corresponds to
$\Delta t=0.5$ and larger for the current problem.
Table \ref{tab:kovas_dt} shows that, for element order $10$
the errors for the computed steady-state velocity are
at levels $10^{-9}\sim 10^{-7}$ with time step sizes
$\Delta t\leqslant 0.4$, and they increase to a level
$\sim10^{-2}$ with $\Delta t\geqslant 0.5$.
For element order $16$ the numerical errors are at a level
$10^{-14}\sim 10^{-12}$ with $\Delta t\leqslant 0.4$ and
they increase to a level $\sim 10^{-2}$ with $\Delta t\geqslant 0.5$.
It is evident that the current method can produce accurate
 results at quite large time step sizes.
The borderline time step size, beyond which the simulation accuracy
starts to deteriorate, is around $\Delta t=0.4$ for the
Kovasznay flow. This is a very large $\Delta t$ value
for all practical purposes.


\begin{table}
  \centering
  \begin{tabular}{lllll}
    \hline
    $C_0$ & $L^{\infty}-u$ & $L^2-u$ & $L^{\infty}-v$ & $L^2-v$ \\
    1e-4 & 1.806e-7 & 8.709e-8 & 2.285e-8 & 6.885e-9 \\
    1e-3 & 1.806e-7 & 8.709e-8 & 2.285e-8 & 6.885e-9 \\
    1e-2 & 1.806e-7 & 8.709e-8 & 2.285e-8 & 6.885e-9 \\
    1e-1 & 1.806e-7 & 8.709e-8 & 2.285e-9 & 6.885e-9 \\
    1.0 & 1.806e-7 & 8.709e-8 & 2.285e-8 & 6.885e-9 \\
    10 & 1.806e-7 & 8.709e-8 & 2.285e-8 & 6.885e-9 \\
    100 & 1.806e-7 & 8.709e-8 & 2.285e-8 & 6.885e-9 \\
    1e3 & 1.806e-7 & 8.707e-8 & 2.284e-8 & 6.884e-9 \\
    1e5 & 1.806e-7 & 8.706e-8 & 2.283e-8 & 6.883e-9 \\
    1e7 & 1.806e-7 & 8.706e-8 & 2.283e-8 & 6.883e-9 \\
    1e9 & 1.806e-7 & 8.706e-8 & 2.283e-8 & 6.883e-9 \\
    1e11 & 1.798e-7 & 8.678e-8 & 2.262e-8 & 6.865e-9 \\
    1e12 & 1.806e-7 & 8.706e-8 & 2.283e-8 & 6.883e-9 \\
    \hline
  \end{tabular}
  \caption{
    Kovasznay flow: effect of $C_0$ on the errors of results.
  }
  \label{tab:kovas_C0}
\end{table}

When defining the biased energy $E(t)$ in \eqref{equ:def_E}, we
need a chosen energy constant $C_0$ to ensure that $E(t)>0$ is
satisfied for all time, so that the expression $\frac{R^2}{E(t)}$
is well-defined in the algorithm. We observe that with the current method
the simulation result is not sensitive to the  value of $C_0$.
This is demonstrated by the data in Table \ref{tab:kovas_C0}.
Here we have varied $C_0$ systematically in a range of values between
$10^{-4}$ and $10^{12}$, and listed the numerical errors of the
steady-state velocity corresponding to these  values.
These results are computed with an element order $10$ and a time step size
$\Delta t=0.005$, and the $\mbs u_0$ field in $\mbs M(\mbs u)$ is
updated every $20$ time steps ($k_0=20$).
It is evident that the different $C_0$ values have  little
or basically no influence on the numerical errors of the results.


\begin{table}
  \centering
  \begin{tabular}{llllll}
    \hline
    $\Delta t$ & $k_0$ & $L^{\infty}-u$ & $L^2-u$ & $L^{\infty}-v$ & $L^2-v$ \\
    \hline
    0.005 & 10 & 3.942e-13 & 1.728e-14 & 7.102e-14 & 8.136e-15 \\
    & 20 & 4.681e-13 & 1.820e-14 & 5.831e-14 & 9.803e-15 \\
    & 50 & 1.662e-13 & 1.918e-14 & 5.122e-14 & 8.171e-15 \\
    & 100 & 4.318e-13 & 1.860e-14 & 6.476e-14 & 9.647e-15 \\
    & 200 & 7.341e-13 & 1.724e-14 & 6.066e-14 & 8.516e-15 \\
    & 500 & 3.113e-13 & 1.779e-14 & 5.350e-14 & 1.367e-14 \\
    & 1000 & 2.643e-13 & 1.420e-14 & 5.748e-14 & 7.974e-15 \\ \hline
    0.1 & 10 & 1.725e-12 & 1.884e-13 & 5.914e-13 & 1.643e-14 \\
    & 20 & 2.151e-12 & 1.768e-13 & 6.692e-13 & 1.772e-14 \\
    & 50 & 2.185e-12 & 1.780e-13 & 6.876e-13 & 1.735e-14 \\
    & 100 & 2.177e-12 & 1.775e-13 & 6.913e-13 & 1.730e-14 \\
    & 200 & 2.186e-12 & 1.773e-13 & 6.874e-13 & 1.743e-14 \\
    & 500 & 2.191e-12 & 1.744e-13 & 6.804e-13 & 1.730e-14 \\
    & 1000 & 2.187e-12 & 1.732e-13 & 6.856e-13 & 1.755e-14 \\
    \hline
  \end{tabular}
  \caption{
    Kovasznay flow: effect of the frequency for $\mbs u_0$
    update 
    on simulation errors.
    Note that $\mbs u_0$ is updated every $k_0$ time steps.
    Element order is $16$.
  }
  \label{tab:kovas_k0}
\end{table}

In the current method the field function $\mbs u_0$ in
$\mbs M(\mbs u)$ is updated every $k_0$
time steps, using a historical velocity field
at the time step that is the largest multiple of $k_0$,
as discussed in Section \ref{sec:method}.
We observe that for the Kovasznay flow
the accuracy of simulation results is not sensitive to
the frequency for $\mbs u_0$ update ($k_0$ value) in the algorithm.
This is demonstrated by Table \ref{tab:kovas_k0},
which lists the errors of the steady-state velocity obtained with
various $k_0$ values ranging from $k_0=10$ to $k_0=1000$
under two time step sizes ($\Delta t=0.005$ and $0.1$).
In this group of tests $C_0=1000$ and the element order is $16$. 
The results are evidently not sensitive to
how frequently $\mbs u_0$ 
is updated for this problem. The errors are comparable when $\mbs u_0$
is updated every $1000$ time steps or every $10$ time steps.
The observed insensitivity is due to the fact that
the Kovasznay flow eventually reaches a steady state. If the problem
is unsteady, very large $k_0$ values can lead to the deterioration
in accuracy of the simulation results, which will be
shown in later numerical tests.


\begin{table}
  \centering
  \begin{tabular}{ll|ll|ll|ll}
    \hline
    & & $C_0=10^3$ & & $C_0=10^6$ & & $C_0=10^9$ & \\ \cline{3-8}
    Element order & $\Delta t$ & $L^{\infty}-u$ & $L^2-u$ & $L^{\infty}-u$ & $L^2-u$ &
    $L^{\infty}-u$ & $L^2-u$  \\ \hline
    10 & 0.001 & 3.201e-3 & 1.440e-3 & 3.367e-6 & 1.517e-6 & 1.830e-7 & 8.746e-8 \\
    & 0.002 & 4.662e-3 & 2.097e-3 & 5.125e-6 & 2.306e-6 & 1.837e-7 & 8.733e-8  \\
    & 0.003 & 5.366e-3 & 2.413e-3 & 6.151e-6 & 2.767e-6 & 1.840e-7 & 8.724e-8 \\
    & 0.004 & 2.580e-1 & 1.136e-1 & 4.358e-1 & 1.836e-1 & 4.351e-1 & 1.834e-1 \\
    & 0.005 & 2.232e-1 & 9.872e-2 & 4.221e-1 & 1.786e-1 & 4.332e-1 & 1.827e-1 \\
    & 0.006 & 2.058e-1 & 9.133e-1 & 4.268e-1 & 1.803e-1 & 4.346e-1 & 1.832e-1 \\
    & 0.007 & 1.967e-1 & 8.744e-2 & 4.240e-1 & 1.793e-1 & 4.348e-1 & 1.833e-1 \\
    & 0.008 & 6.646e-1 & 8.564e-2 & 4.231e-1 & 1.790e-1 & 4.331e-1 & 1.827e-1 \\
    & 0.01 & 3.251e-1 & 1.410e-1 & 4.356e-1 & 1.836e-1 & 4.344e-1 & 1.831e-1 \\
    & 0.02 & 3.938e-1 & 1.560e-1 & 4.267e-1 & 1.803e-1 & 4.347e-1 & 1.832e-1 \\
    & 0.03 & 3.676e-1 & 1.488e-1 & 4.280e-1 & 1.808e-1 & 4.355e-1 & 1.835e-1 \\
    & 0.04 & 3.756e-1 & 1.529e-1 & 4.347e-1 & 1.832e-1 & 4.353e-1 & 1.835e-1 \\
    & 0.05 & 3.827e-1 & 1.532e-1 & 4.345e-1 & 1.832e-1 & 4.348e-1 & 1.833e-1 \\
    & 0.06 & 4.031e-1 & 1.548e-1 & 4.267e-1 & 1.80e-1 & 4.352e-1 & 1.834e-1 \\
    & 0.07 & 4.141e-1 & 1.570e-1 & 4.339e-1 & 1.830e-1 & 4.349e-1 & 1.833e-1 \\
    & 0.08 & 4.218e-1 & 1.584e-1 & 4.292e-1 & 1.812e-1 & 4.351e-1 & 1.834e-1 \\
    & 0.1 & 4.314e-1 & 1.604e-1 & 4.281e-1 & 1.808e-1 & 4.352e-1 & 1.834e-1 \\
    & 0.2 & 5.938e-1 & 1.703e-1 & 4.309e-1 & 1.819e-1 & 4.351e-1 & 1.834e-1 \\
    & 0.3 & 5.937e-1 & 1.728e-1 & 4.288e-1 & 1.811e-1 & 4.351e-1 & 1.834e-1 \\
    & 0.4 & 4.602e-1 & 1.690e-1 & 4.202e-1 & 1.779e-1 & 4.348e-1 & 1.833e-1 \\
    & 0.5 & 4.752e-1 & 1.709e-1 & 4.250e-1 & 1.797e-1 & 4.351e-1 & 1.834e-1 \\
    & 0.75 & 5.450e-1 & 1.777e-1 & 4.181e-1 & 1.772e-1 & 4.356e-1 & 1.835e-1 \\
    & 1.0 & 5.409e-1 & 1.797e-1 & 4.166e-1 & 1.766e-1 & 4.353e-1 & 1.835e-1 \\
    & 2.0 & 5.487e-1 & 2.147e-1 & 4.023e-1 & 1.712e-1 & 4.354e-1 & 1.835e-1 \\
    & 10.0 & 5.540e-1 & 2.113e-1 & 4.315e-1 & 1.817e-1 & 4.357e-1 & 1.836e-1 \\
    & 100.0 & 5.477e-1 & 2.070e-1 & 4.326e-1 & 1.814e-1 & 4.347e-1 & 1.833e-1 \\
    \hline
    16 & 0.001 & 3.362e-3 & 1.513e-3 & 3.522e-6 & 1.585e-6 & 3.513e-9 & 1.581e-9 \\
    & 0.002 & 2.699e-1 & 1.524e-2 & 4.293e-1 & 1.813e-1 & 4.345e-1 & 1.832e-1 \\
    & 0.003 & 2.853e-1 & 1.249e-1 & 4.308e-1 & 1.818e-1 & 4.351e-1 & 1.834e-1 \\
    & 0.004 & 2.755e-1 & 1.207e-1 & 4.277e-1 & 1.807e-1 & 4.349e-1 & 1.833e-1 \\
    & 0.005 & 2.512e-1 & 1.106e-1 & 4.304e-1 & 1.817e-1 & 4.354e-1 & 1.835e-1 \\
    & 0.006 & 2.305e-1 & 1.018e-1 & 4.308e-1 & 1.818e-1 & 4.338e-1 & 1.829e-1 \\
    & 0.007 & 2.232e-1 & 9.873e-2 & 4.300e-1 & 1.815e-1 & 4.350e-1 & 1.834e-1 \\
    & 0.008 & 6.037e-1 & 1.093e-1 & 4.264e-1 & 1.802e-1 & 4.351e-1 & 1.834e-1 \\
    & 0.01 & 6.084e-1 & 1.180e-1 & 4.301e-1 & 1.816e-1 & 4.355e-1 & 1.835e-1 \\
    & 0.02 & 3.926e-1 & 1.358e-1 & 4.299e-1 & 1.815e-1 & 4.354e-1 & 1.835e-1 \\
    & 0.03 & 6.019e-1 & 1.473e-1 & 4.261e-1 & 1.801e-1 & 4.357e-1 & 1.836e-1 \\
    & 0.04 & 6.038e-1 & 1.523e-1 & 4.312e-1 & 1.820e-1 & 4.352e-1 & 1.834e-1 \\
    & 0.05 & 6.156e-1 & 1.558e-1 & 4.306e-1 & 1.817e-1 & 4.350e-1 & 1.833e-1 \\
    & 0.06 & 6.268e-1 & 1.583e-1 & 4.297e-1 & 1.814e-1 & 4.353e-1 & 1.834e-1 \\
    & 0.07 & 6.185e-1 & 1.603e-1 & 4.284e-1 & 1.809e-1 & 4.353e-1 & 1.835e-1 \\
    & 0.08 & 6.493e-1 & 1.620e-1 & 4.281e-1 & 1.808e-1 & 4.354e-1 & 1.835e-1 \\
    & 0.1 & 4.291e-1 & 1.623e-1 & 4.296e-1 & 1.814e-1 & 4.356e-1 & 1.836e-1 \\
    & 0.2 & 4.392e-1 & 1.658e-1 & 4.257e-1 & 1.800e-1 & 4.353e-1 & 1.835e-1 \\
    & 0.3 & 4.484e-1 & 1.674e-1 & 4.243e-1 & 1.794e-1 & 4.352e-1 & 1.834e-1 \\
    & 0.4 & 4.539e-1 & 1.683e-1 & 4.238e-1 & 1.792e-1 & 4.353e-1 & 1.835e-1 \\
    & 0.5 & 4.535e-1 & 1.689e-1 & 4.228e-1 & 1.789e-1 & 4.352e-1 & 1.834e-1 \\
    & 0.75 & 5.630e-1 & 1.779e-1 & 4.191e-1 & 1.774e-1 & 4.353e-1 & 1.835e-1 \\
    & 1.0 & 5.527e-1 & 1.798e-1 & 4.156e-1 & 1.761e-1 & 4.351e-1 & 1.834e-1 \\
    & 2.0 & 5.501e-1 & 2.147e-1 & 4.027e-1 & 1.712e-1 & 4.354e-1 & 1.835e-1 \\
    & 10.0 & 5.549e-1 & 2.113e-1 & 4.174e-1 & 1.768e-1 & 4.356e-1 & 1.836e-1 \\
    & 100.0 & 5.480e-1 & 2.070e-1 & 4.313e-1 & 1.801e-1 & 4.347e-1 & 1.833e-1 \\
    \hline
  \end{tabular}
  \caption{
    Kovasznay flow: Errors of steady-state velocity obtained
    using the modified algorithm (${\mbs M}(\mbs u)=0$).
  }
  \label{tab:kovas_Mu0}
\end{table}

In Remark~\ref{rem:rem_1} we have suggested a modified algorithm,
corresponding to $\mbs M(\mbs u)=0$ in the formulation.
This modified scheme has a lower computational cost, because
the associated velocity coefficient matrix is constant and can be
pre-computed. However, it is  inferior in accuracy to
the current method at moderate and large $\Delta t$ values,
and its accuracy has a strong dependence on the  energy constant $C_0$.
This point is demonstrated by Table~\ref{tab:kovas_Mu0},
which lists the $L^{\infty}$ and $L^2$ errors of the $x$-velocity component
at steady state
obtained using this modified algorithm with
various $\Delta t$ ranging from small to
large values. The errors 
corresponding to two element orders ($10$ and $16$) and several $C_0$
values are provided.
This table can be compared with Table~\ref{tab:kovas_dt},
which is obtained using the current method.
One can observe that, with the modified algorithm
($\mbs M(\mbs u)=0$), the simulation loses accuracy
with time step sizes $\Delta t=0.004$ and larger
for element order $10$ (and $\Delta t=0.002$ and larger for
element order $16$).
The errors reach a level around $10^{-1}$.
In contrast, with the current method, accurate results can be
obtained with time step sizes up to $\Delta t=0.4$;
see Table \ref{tab:kovas_dt}.
Table \ref{tab:kovas_Mu0} further indicates
that with small $\Delta t$ the accuracy
of the modified algorithm ($\mbs M(\mbs u)=0$)
strongly depends on the energy constant $C_0$.
A larger $C_0$ leads to considerably more accurate results.
This behavior is  different from that of the
current method, 
whose accuracy is not sensitive to $C_0$
as shown by Table \ref{tab:kovas_C0}.
%
%
A comparison between Table \ref{tab:kovas_Mu0} and
the data from~\cite{LinYD2019} indicates that this
modified algorithm seems also inferior in accuracy to
the method of~\cite{LinYD2019}. We note that
the modified algorithm here from Remark \ref{rem:rem_1}
is based on the pressure-correction type strategy,
while the method from~\cite{LinYD2019} is
more aligned with a velocity-correction type
scheme, which is likely the cause for the observed
difference in accuracy.


\subsection{Flow past a Hemisphere in a Narrow Periodic Channel}
\label{sec:hemis}


\begin{figure}
  \centering
  \includegraphics[width=6in]{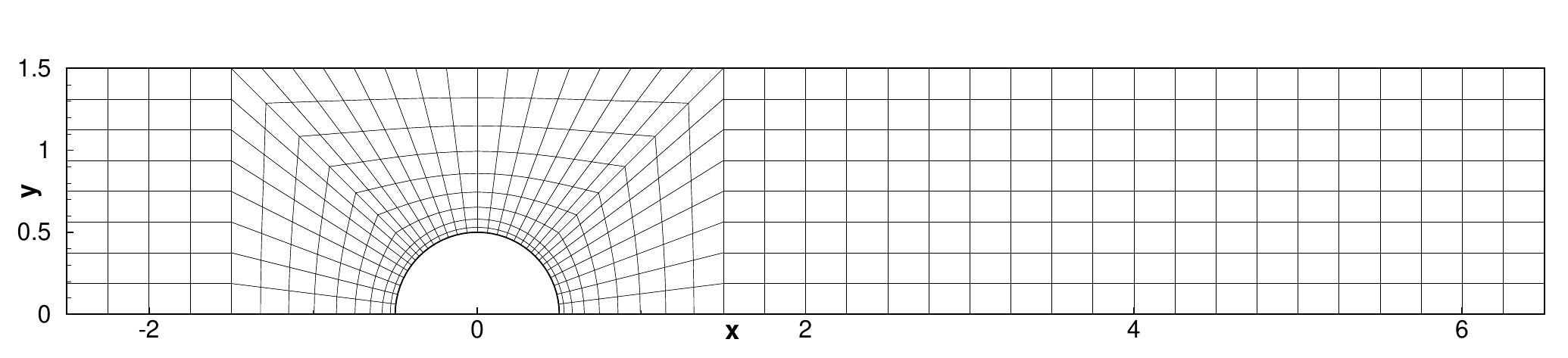}(a)
  \includegraphics[width=6in]{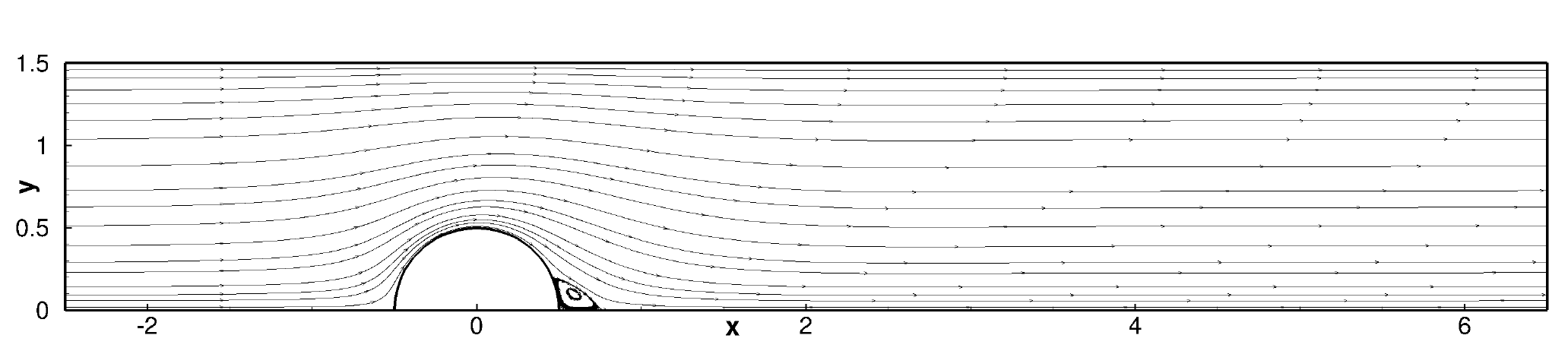}(b)
  \includegraphics[width=6in]{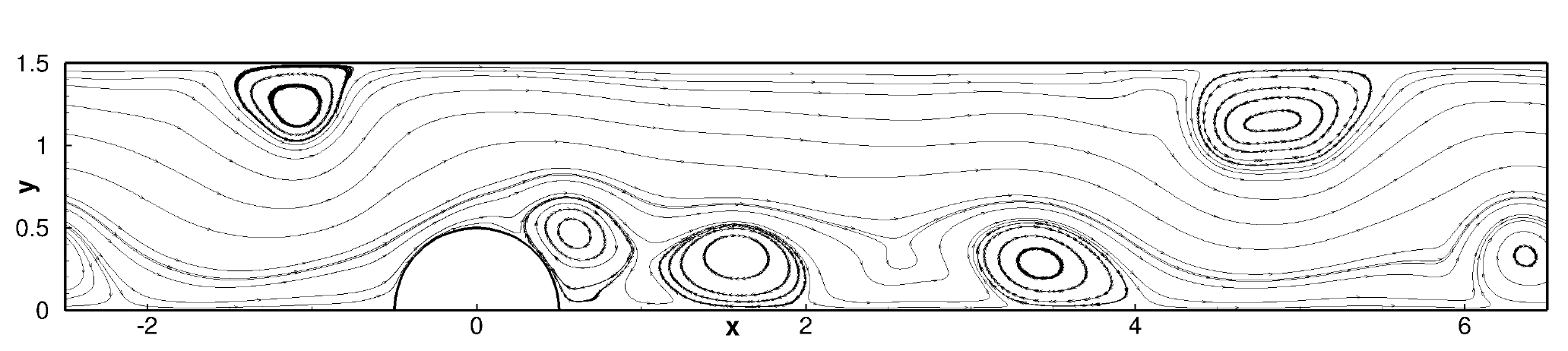}(c)
  \caption{Flow past a hemisphere:
    (a) flow configuration and a mesh of $480$ quadrilateral
    elements.
    Flow patterns visualized by streamlines corresponding to
    $\nu=0.02$ (b) and $\nu=0.001$ (c).
  }
  \label{fig:hemis_config}
\end{figure}

In this subsection we test the current method 
with the flow past a hemisphere in a narrow periodic channel in two dimensions.
Specifically, we consider the domain shown in Figure \ref{fig:hemis_config}(a).
A hemisphere (or half-disk) with diameter $d$ is mounted on the bottom
of a narrow channel, which occupies the domain
$-2.5d\leqslant x\leqslant 6.5d$ and $0\leqslant y\leqslant 1.5d$.
The hemispheric center coincides with the origin of
the coordinate system.
The top and bottom of the channel ($y=0, 1.5d$) are walls, and in the
horizontal direction ($x=-2.5d, 6.5d$) the channel is assumed to be
periodic. The flow is driven by a horizontal pressure gradient.
This configuration mimics the flow
past an infinite array of hemispheres in an infinitely long channel.
We choose the hemisphere diameter $d$ as the length scale
and a unit velocity scale $U_0=1$. All the other physical
variables and parameters are then normalized accordingly.


We discretize the domain using a mesh of $480$ quadrilateral elements;
see Figure \ref{fig:hemis_config}(a).
On the top and bottom channel walls and on the surface of
the hemisphere we impose the no-slip condition,
i.e.~boundary condition~\eqref{equ:bc} with $\mbs w=0$.
In the horizontal direction periodic condition is imposed
for all the flow variables. The Navier-Stokes equations
\eqref{equ:nse}--\eqref{equ:cont}, with a horizontal body
force (pressure gradient) of normalized magnitude  $|\mbs f|=0.03$,
are solved
using the algorithm from Section \ref{sec:method}.
The element order, the time step size $\Delta t$,
the energy constant $C_0$, the Reynolds number,
and other algorithmic parameters 
are varied to study their effects on the simulation
results.


An overview of the characteristics of this flow is provided by
Figures \ref{fig:hemis_config}(b,c),
which visualize the flow patterns  at two Reynolds numbers
corresponding to $\nu=0.02$ and $\nu=0.001$
using streamlines.
At low Reynolds numbers one observes a steady flow
(Figure \ref{fig:hemis_config}(b)).
As the Reynolds number increases, vortex shedding
can be seen in the hemisphere wake. Due to periodicity,
these vortices re-enter the domain from the left,
and can interact with the hemisphere and generate complicated dynamics.
For instance, vortices can at times be observed near the
top channel wall (Figure \ref{fig:hemis_config}(c)).


\begin{figure}
  \centering
  \includegraphics[width=3in]{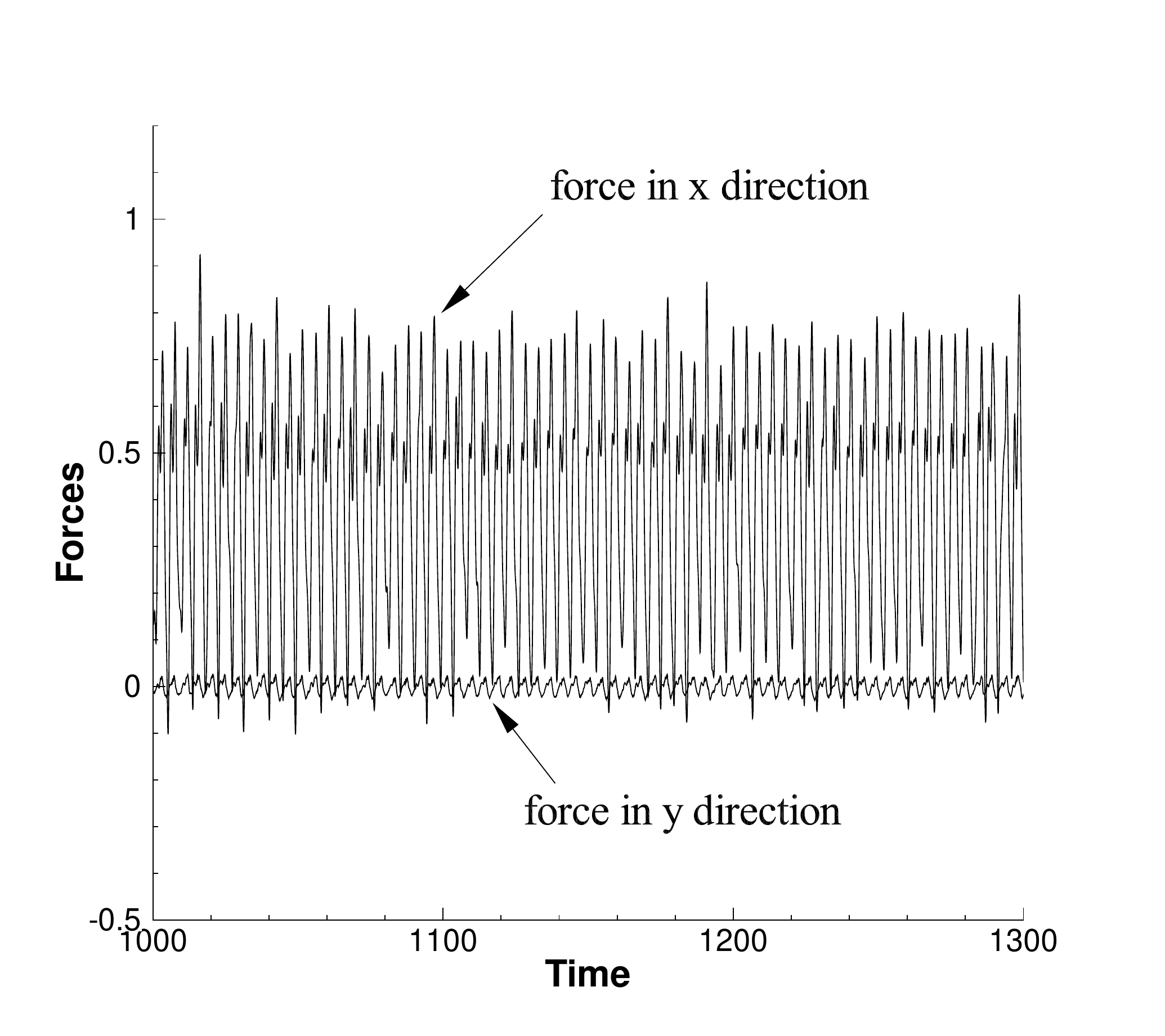}
  \caption{
    Flow past a hemisphere: Time histories of the forces on the walls
    with $\nu=0.001$.
  }
  \label{fig:hemis_force}
\end{figure}

We have monitored the total force exerting on the wall surfaces (channel walls
and the hemispheric surface). 
Figure \ref{fig:hemis_force} shows a typical signal of
 the  force ($x$ and $y$ components)
at the Reynolds number corresponding to $\nu=0.001$.
These are attained using an element order $7$,
$\Delta t=0.001$ and $C_0=1000$ in the simulations.
The force signals are fluctuational due to
the vortex shedding. The horizontal force (drag)
exhibits large fluctuations in magnitude,
while the vertical force is much weaker in comparison.
The long history and the signal characteristics
indicate that the flow has reached a statistically stationary state.


\begin{table}[htb]
  \centering
  \begin{tabular}{lllllll}
    \hline
    $\nu$ & Element order & ${\bar f}_x$ & $f'_x$ & ${\bar f}_y$ & $f'_y$ & Driving force \\
    \hline
    0.02 & 4 & 0.396 & 0 & 7.6e-4 & 0 & 0.393 \\
    & 5 & 0.394 & 0 & -5.6e-4 & 0 & 0.393 \\
    & 6 & 0.393 & 0 & 8.7e-5 & 0 & 0.393 \\
    & 7 & 0.393 & 0 & 1.4e-5 & 0 & 0.393 \\
    & 8 & 0.393 & 0 & 1.1e-5 & 0 & 0.393 \\
    \hline
    0.005 & 4 & 0.395 & 0 & -0.192 & 0 & 0.393 \\
    & 5 & 0.393 & 0 & -4.6e-3 & 0 & 0.393 \\
    & 6 & 0.393 & 0 & 1.6e-2 & 0 & 0.393 \\
    & 7 & 0.393 & 0 & 7.0e-4 & 0 & 0.393 \\
    & 8 & 0.393 & 0 & -1.3e-4 & 0 & 0.393 \\
    \hline
    0.001 & 4 & 0.388 & 0.278 & 0.129 & 0.467 & 0.393 \\
    & 5 & 0.405 & 0.252 & -0.0153 & 0.0443 & 0.393 \\
    & 6 & 0.394 & 0.238 & -0.0636 & 0.0559 & 0.393 \\
    & 7 & 0.386 & 0.237 & -9.29e-4 & 0.0134 & 0.393 \\
    & 8 & 0.391 & 0.241 & 0.0245 & 0.0135 & 0.393 \\
    \hline
    0.0002 & 4 & 0.351 & 0.796 & 1.353 & 1.638 & 0.393 \\
    & 5 & 0.437 & 0.875 & 0.103 & 0.276 & 0.393 \\
    & 6 & 0.388 & 0.812 & 0.102 & 0.401 & 0.393 \\
    & 7 & 0.405 & 0.760 & -1.43e-3 & 0.0829 & 0.393 \\
    & 8 & 0.399 & 0.725 & -0.0865 & 0.167 & 0.393 \\
    \hline
  \end{tabular}
  \caption{ Flow past a hemisphere:
    Effect of spatial resolution on the forces on walls.
    $\bar{f}_x$ and $\bar{f}_y$ are the time-averaged mean
    forces in $x$ and $y$ directions, and $f'_x$ and
    $f'_y$ are the rms forces in the two directions.
  }
  \label{tab:hemis_order}
\end{table}

From the force histories we can compute the statistical quantities
such as the time-averaged mean and root-mean-square (rms) of the forces
on the walls. In Table \ref{tab:hemis_order} we have listed
the mean and rms forces at several Reynolds numbers
(for $\nu$ ranging from $\nu=0.02$ to $\nu=0.0002$),
which are computed using element orders ranging from
$4$ to $8$. In these simulations fixed values of
$\Delta t=0.001$ and $C_0=1000$ are employed, and
the field $\mbs u_0$ is updated every $20$ time steps ($k_0=20$).
The total driving force on the domain,
i.e.~(driving pressure gradient)$\times$(domain area)
= $0.03\times (1.5\times 9-\pi/8)\approx 0.393$,
has also been listed in the table.
At a steady state or a statistically stationary state,
the time-averaged total horizontal force on the wall
should physically match the total driving force in the domain.
Therefore, these  values can serve as a basic check on
the simulation results.
At $\nu=0.02$ and $\nu=0.005$, it is a steady flow.
So given in the table are the steady-state forces, and no
time-averaging is performed for these cases.
It can be observed that with element orders beyond about $5$
the computed values of the horizontal force are quite close to
(or for the lower Reynolds numbers the same as)
the total driving force on the domain.
The rms horizontal force $f'_x$ also appears to exhibit a sense
of convergence with increasing element order.
The mean and rms vertical forces ($\overline{f}_y, f'_y$) are
quite small when compared with the horizontal counterpart.


\begin{figure}
  \centerline{
    \includegraphics[width=1.1in]{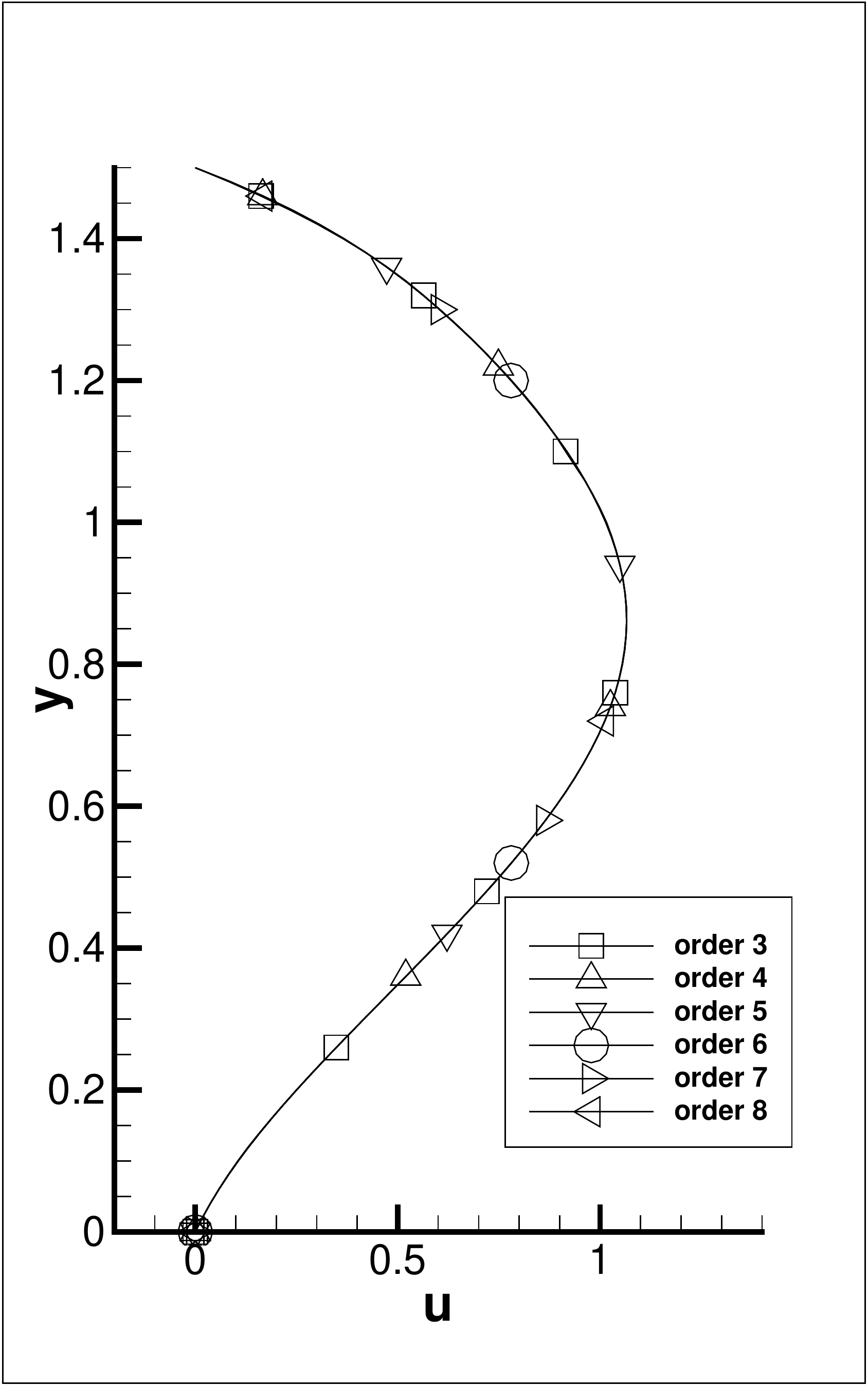}(a)
    \includegraphics[width=1.1in]{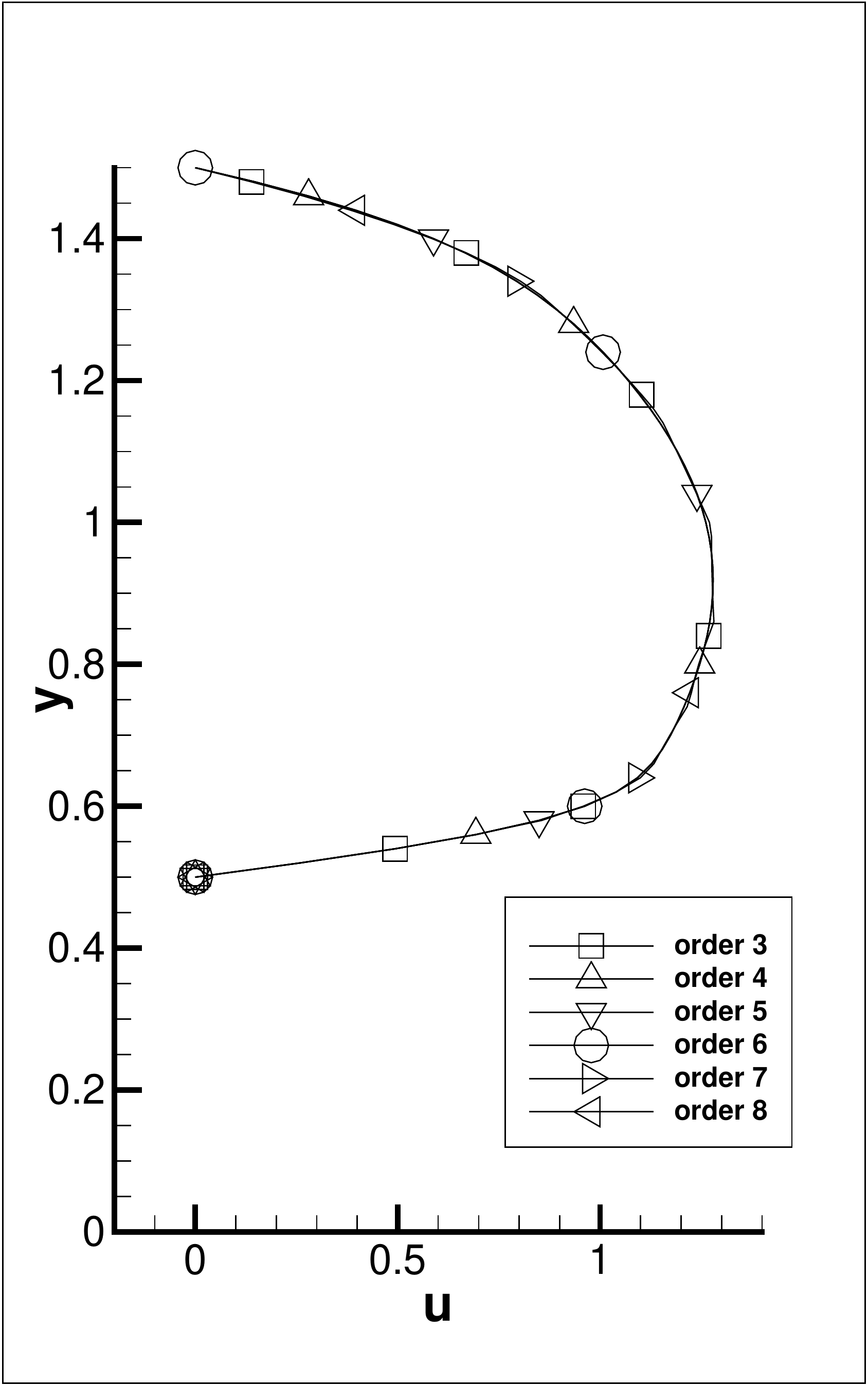}(b)
    \includegraphics[width=1.1in]{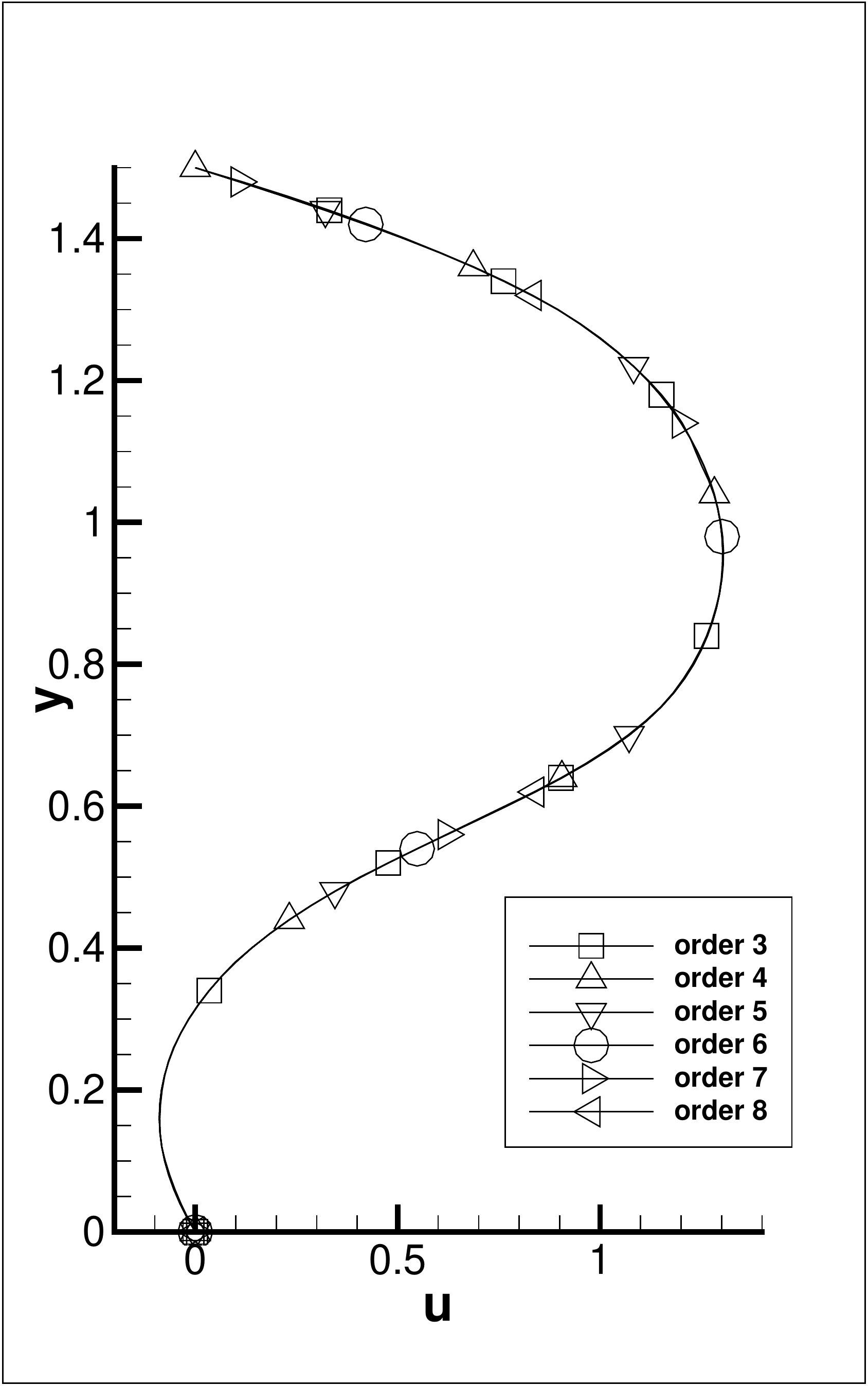}(c)
    \includegraphics[width=1.1in]{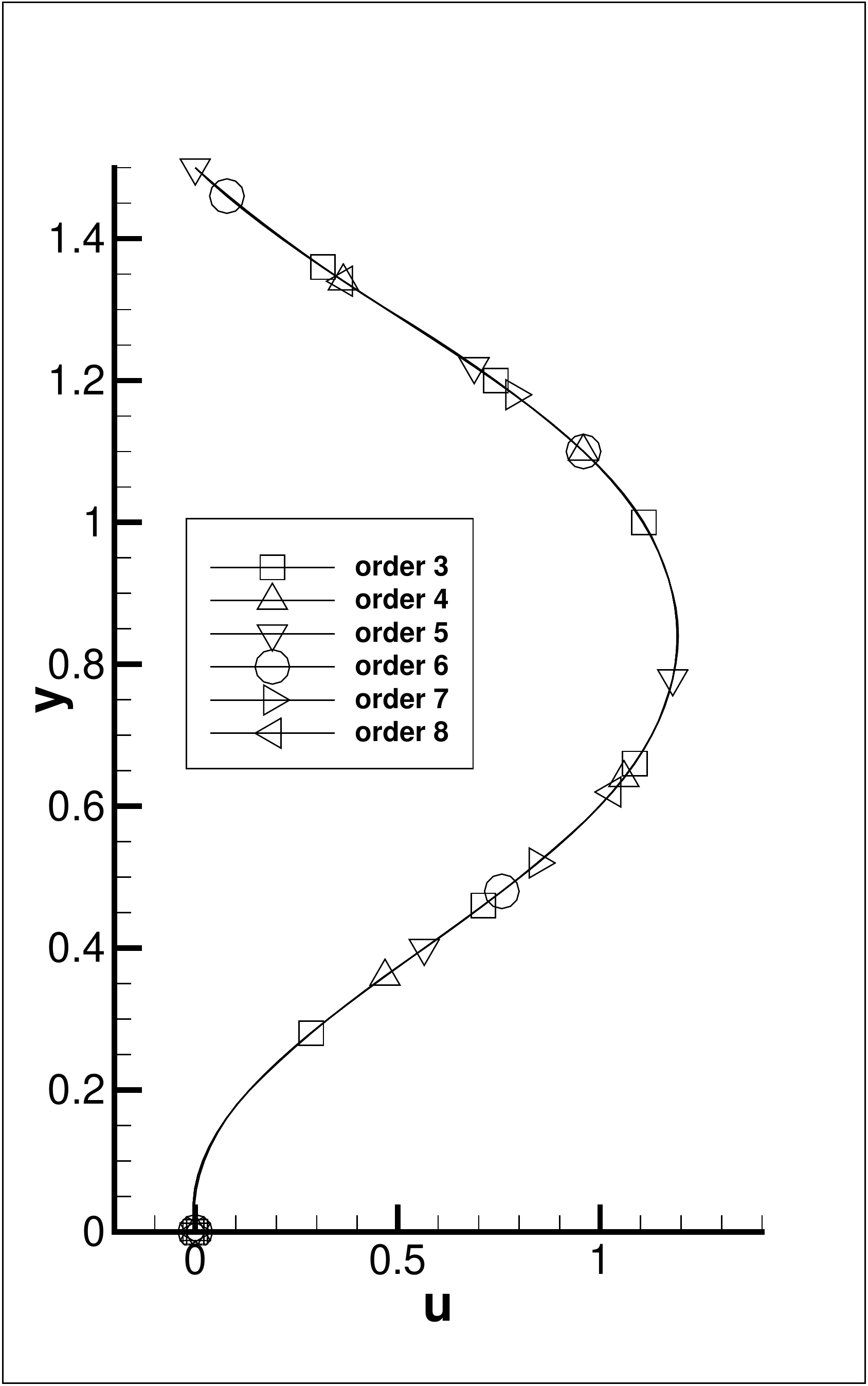}(d)
    \includegraphics[width=1.1in]{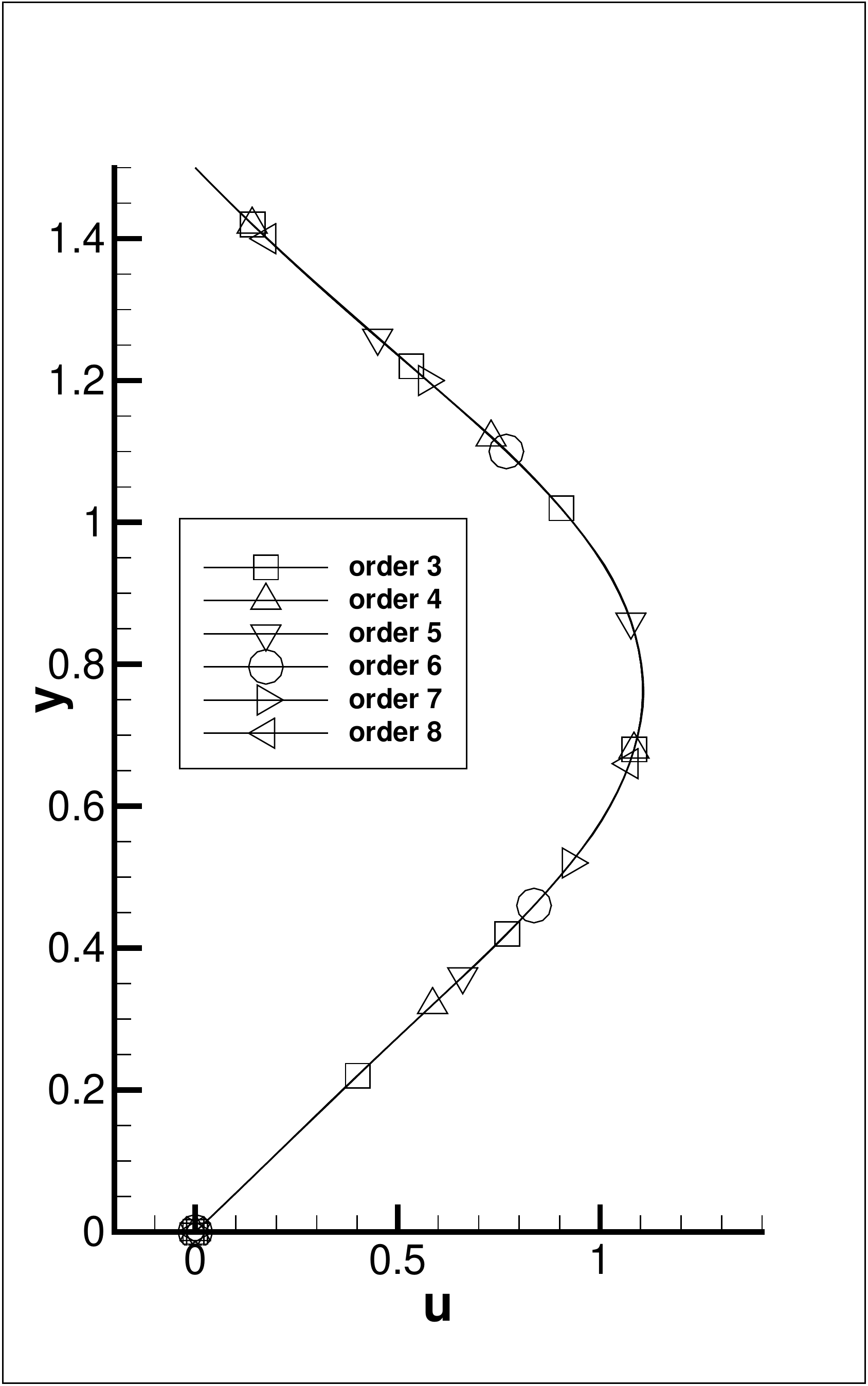}(e)
  }
  \centerline{
    \includegraphics[width=1.1in]{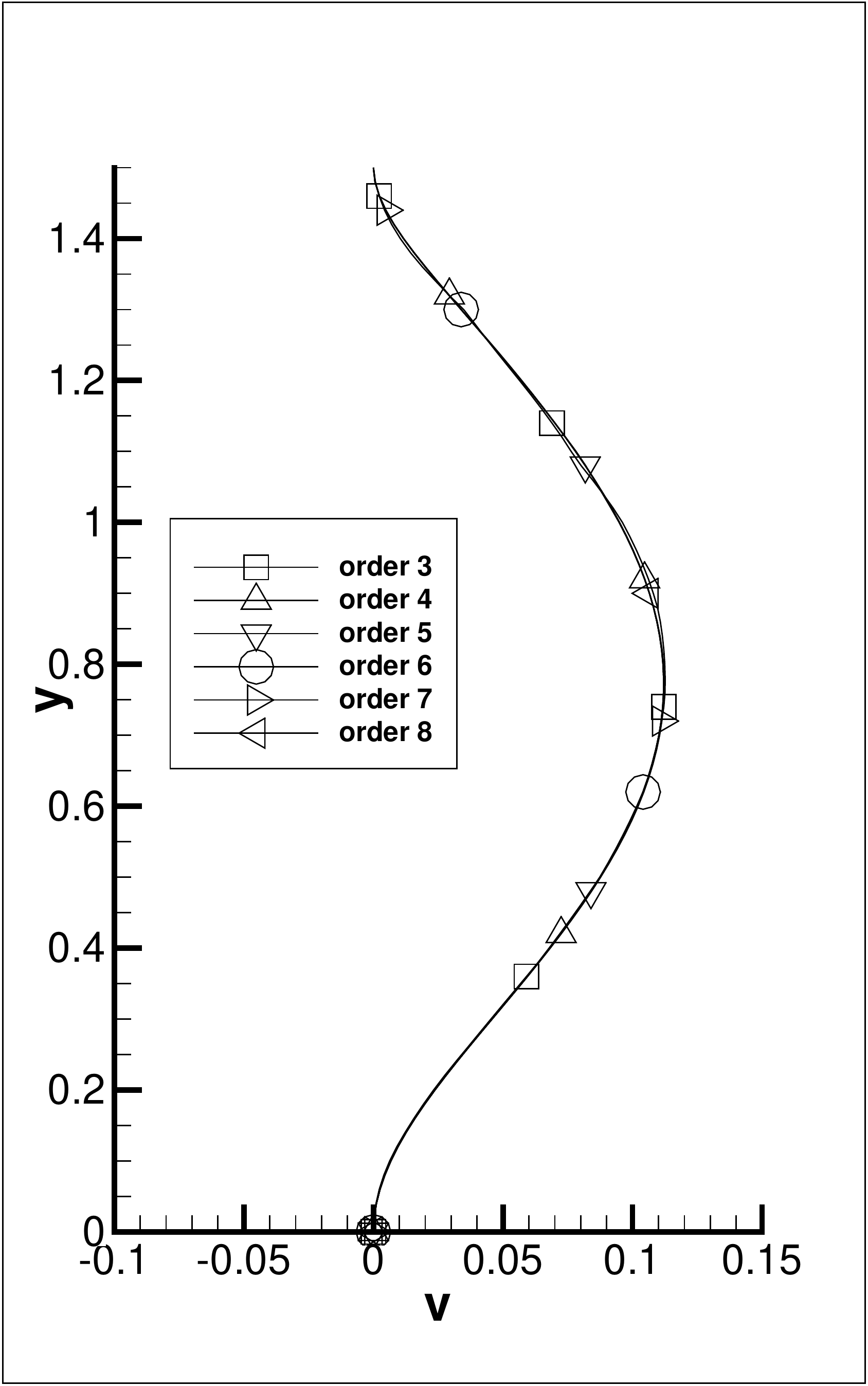}(f)
    \includegraphics[width=1.1in]{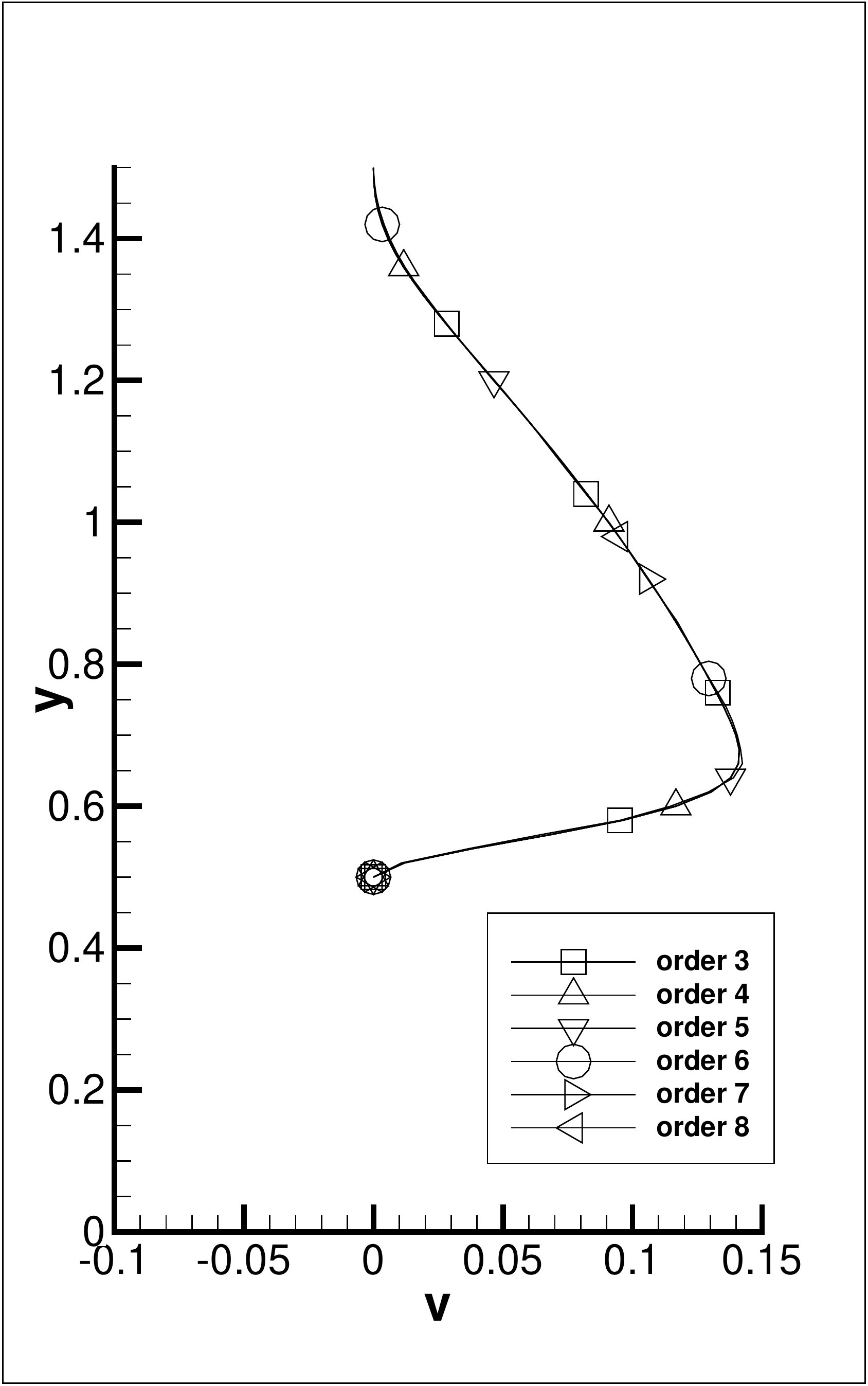}(g)
    \includegraphics[width=1.1in]{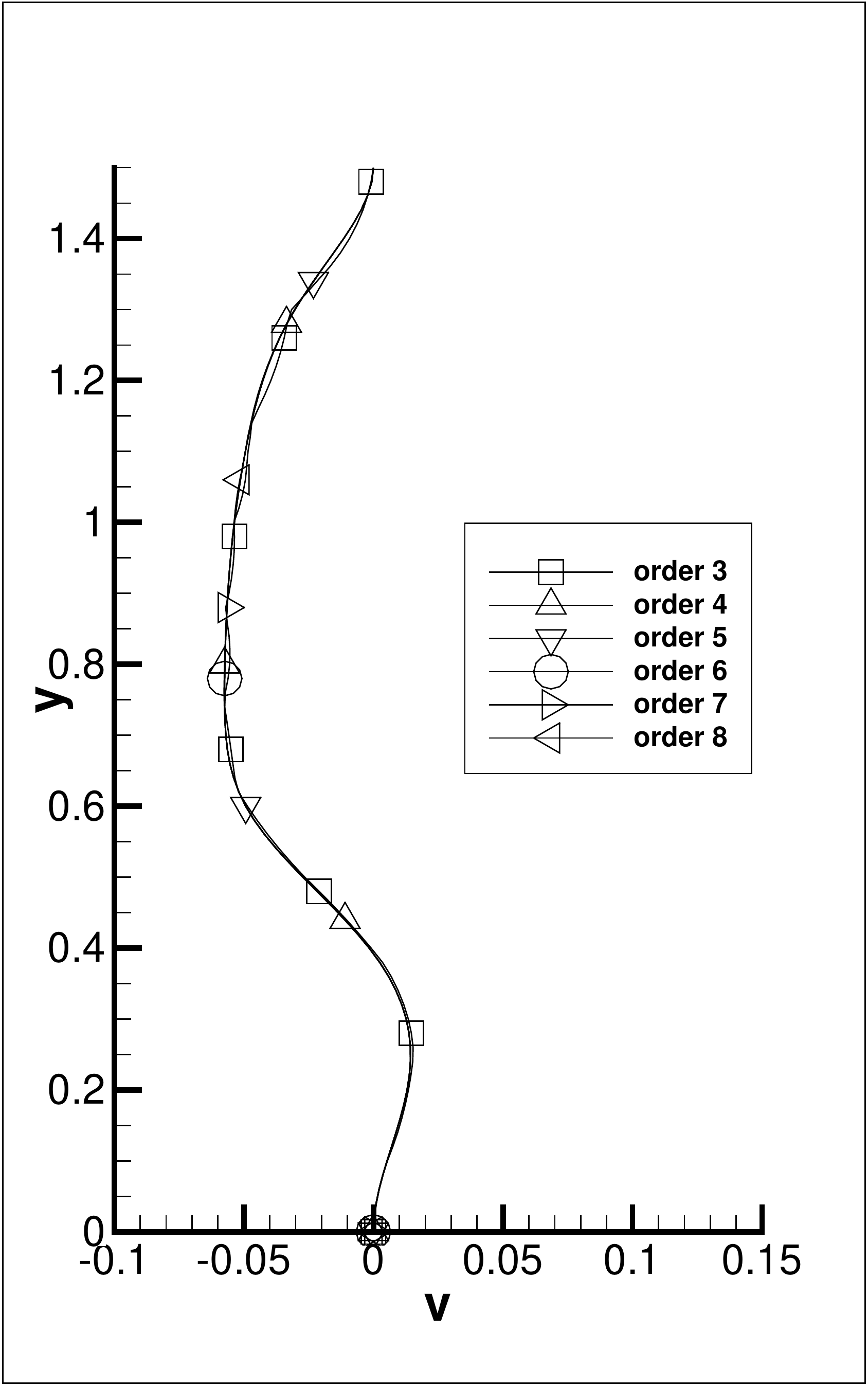}(h)
    \includegraphics[width=1.1in]{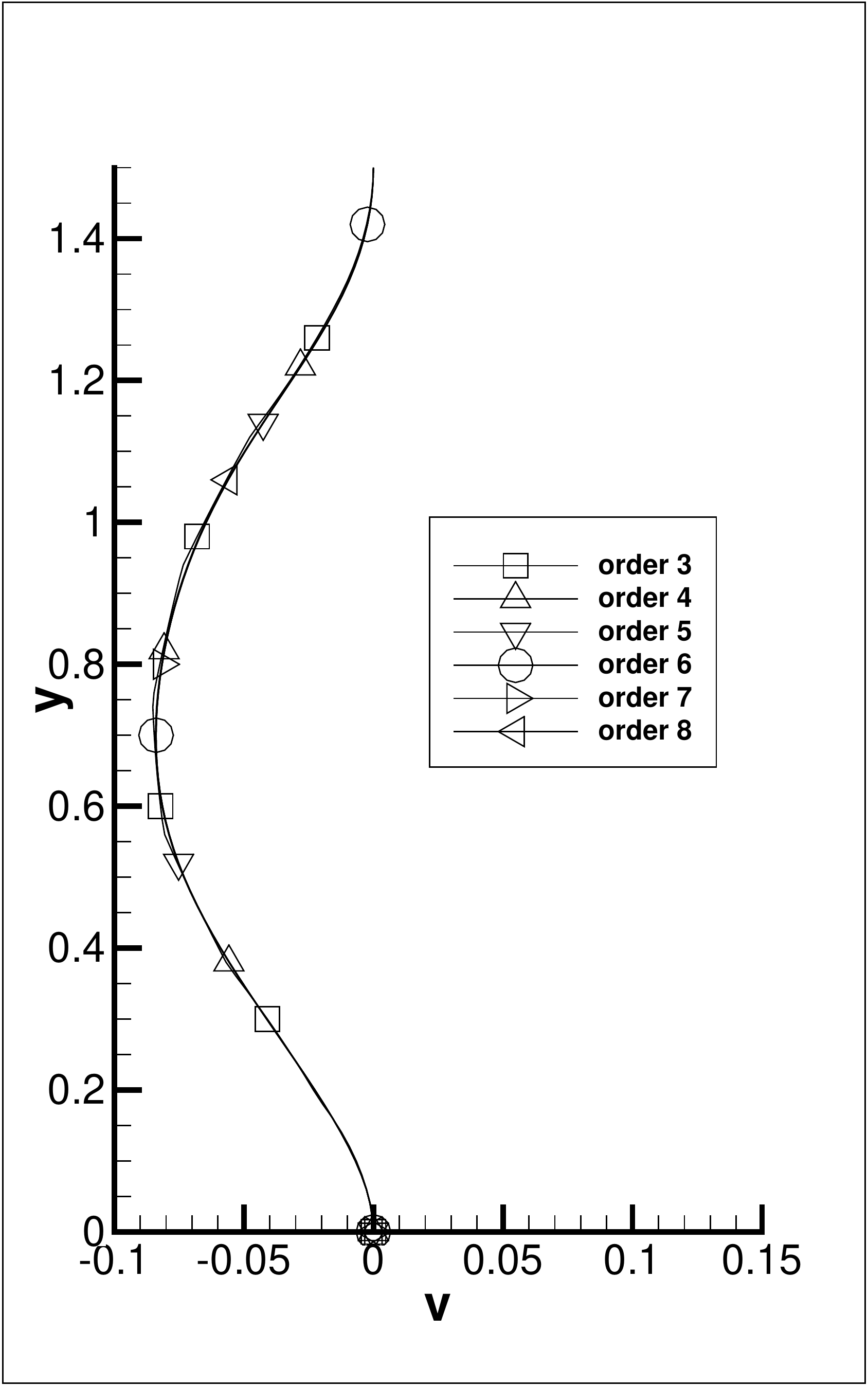}(i)
    \includegraphics[width=1.1in]{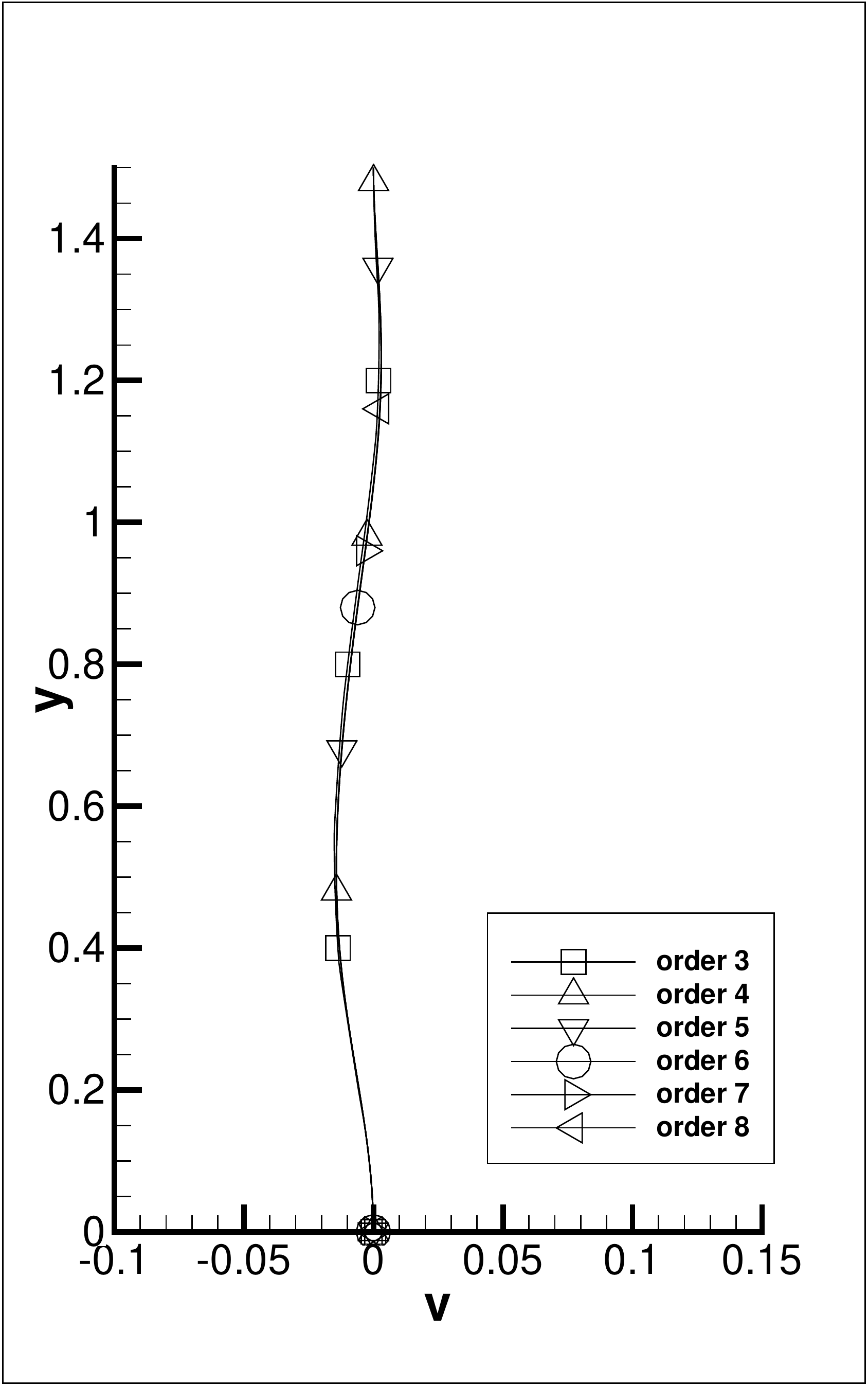}(j)
  }
  \caption{
    Flow past a hemisphere ($\nu=0.005$):
    comparison of profiles of the streamwise velocity $u$ (top row)
    and the vertical velocity $v$ (bottom row)
    at several downstream locations,
    $x/d=-1$ (a, f), $0$ (b, g), $1$ (c, h), $3$ (d, i), $5$ (e, j).
  }
  \label{fig:hemis_prof}
\end{figure}

Figure \ref{fig:hemis_prof} shows a comparison of the steady-state streamwise
and vertical velocity profiles across the channel at several downstream
locations at  $\nu=0.005$, computed with various element orders.
The velocity profiles corresponding to different element orders
essentially overlap with one another. This suggests that these
simulations produce essentially the same velocity distribution, and that they
have numerically converged with respect to the spatial resolution.
The majority of simulation results reported below are computed 
with an element order $6$ or $7$.


\begin{figure}
  \centerline{
    \includegraphics[width=3in]{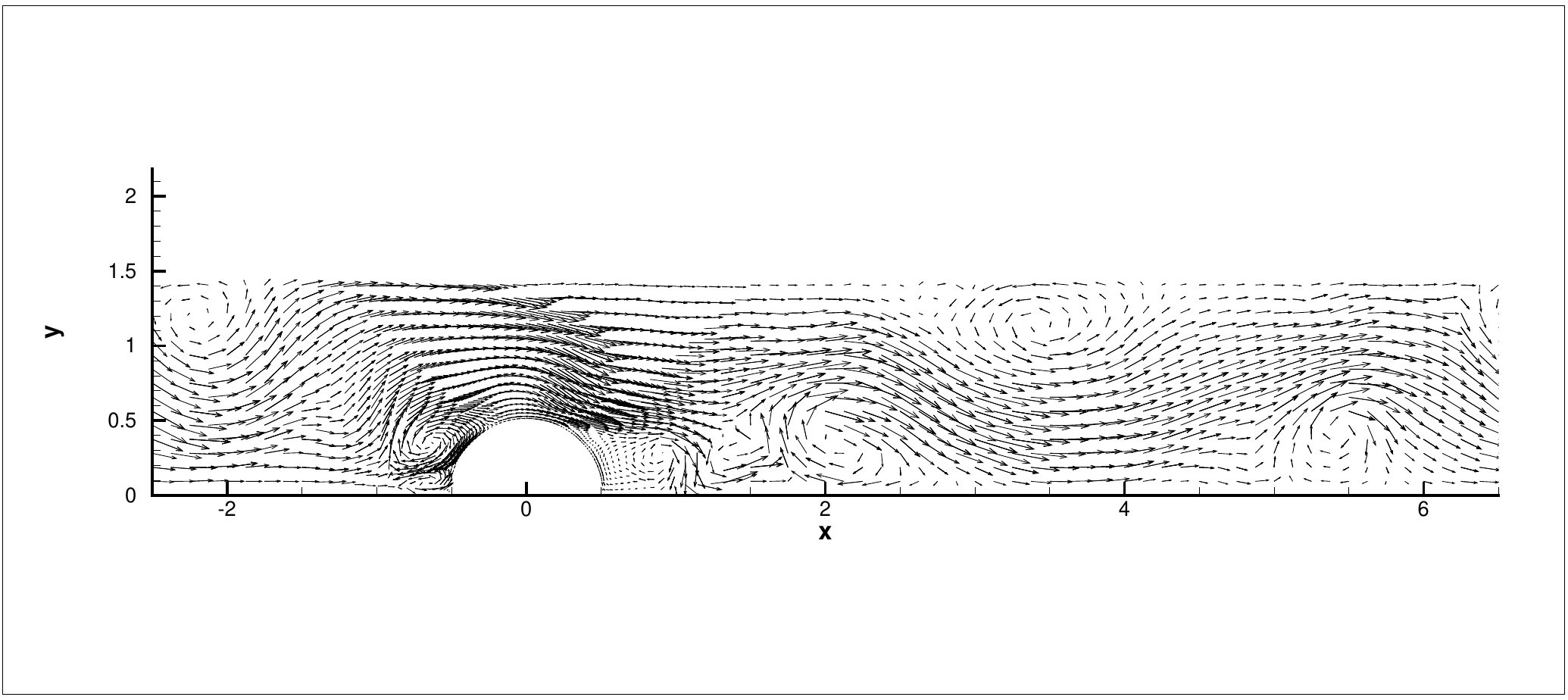}(a)
    \includegraphics[width=3in]{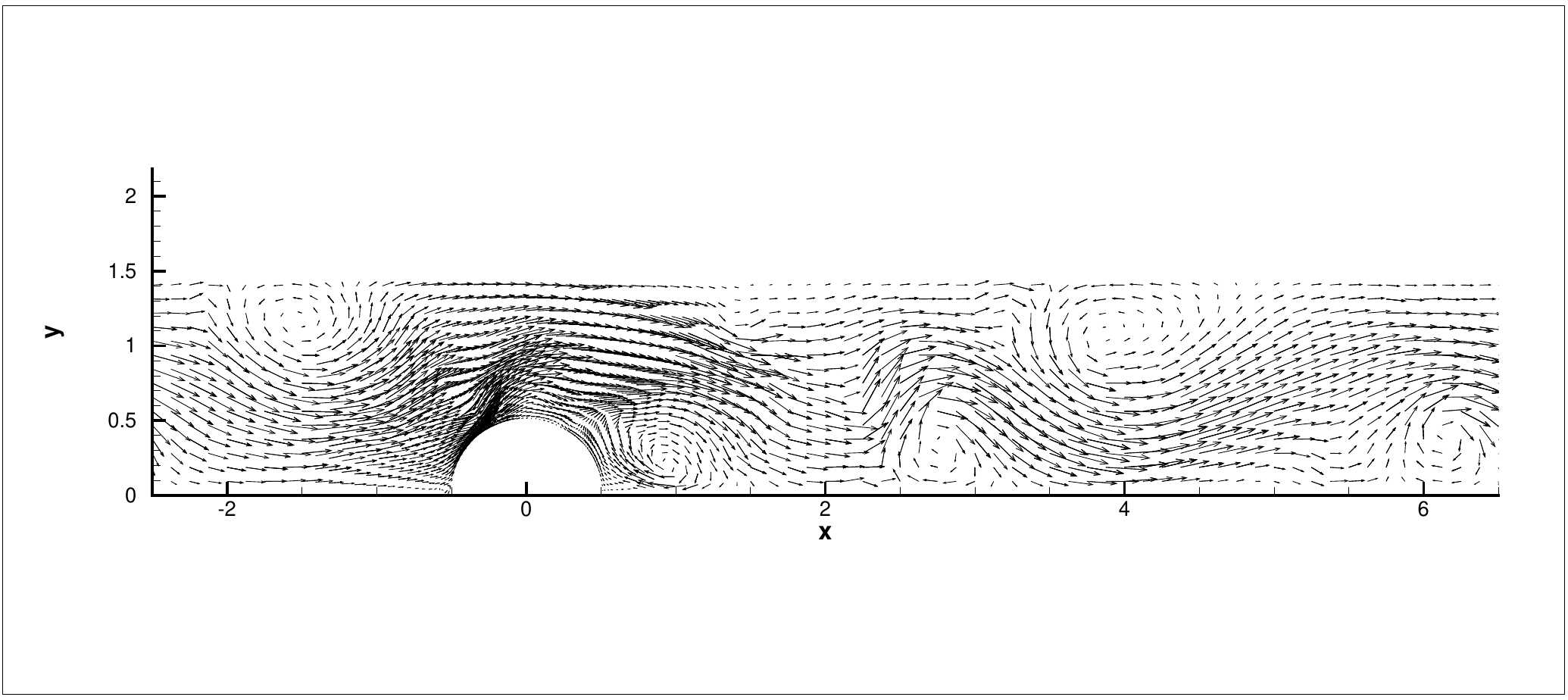}(b)
  }
  \centerline{
    \includegraphics[width=3in]{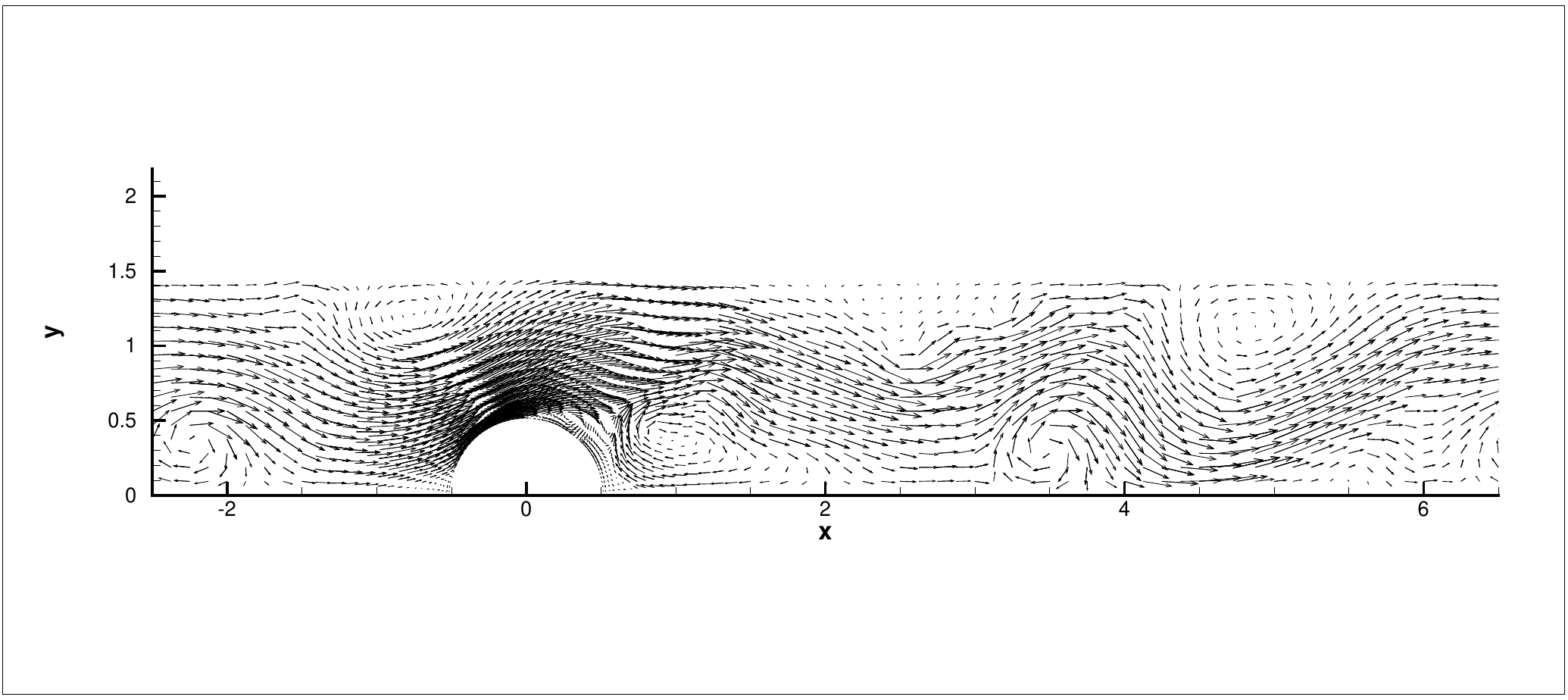}(c)
    \includegraphics[width=3in]{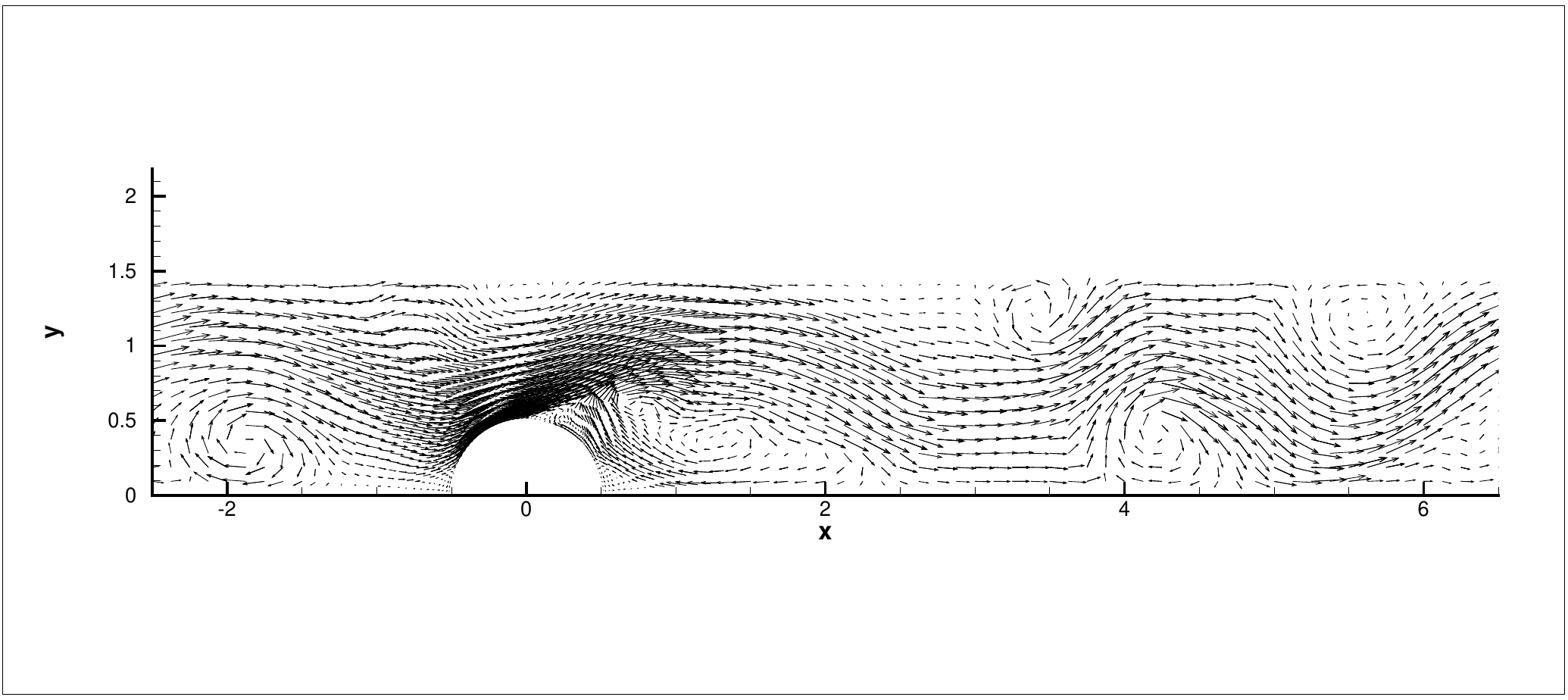}(d)
  }
  \centerline{
    \includegraphics[width=3in]{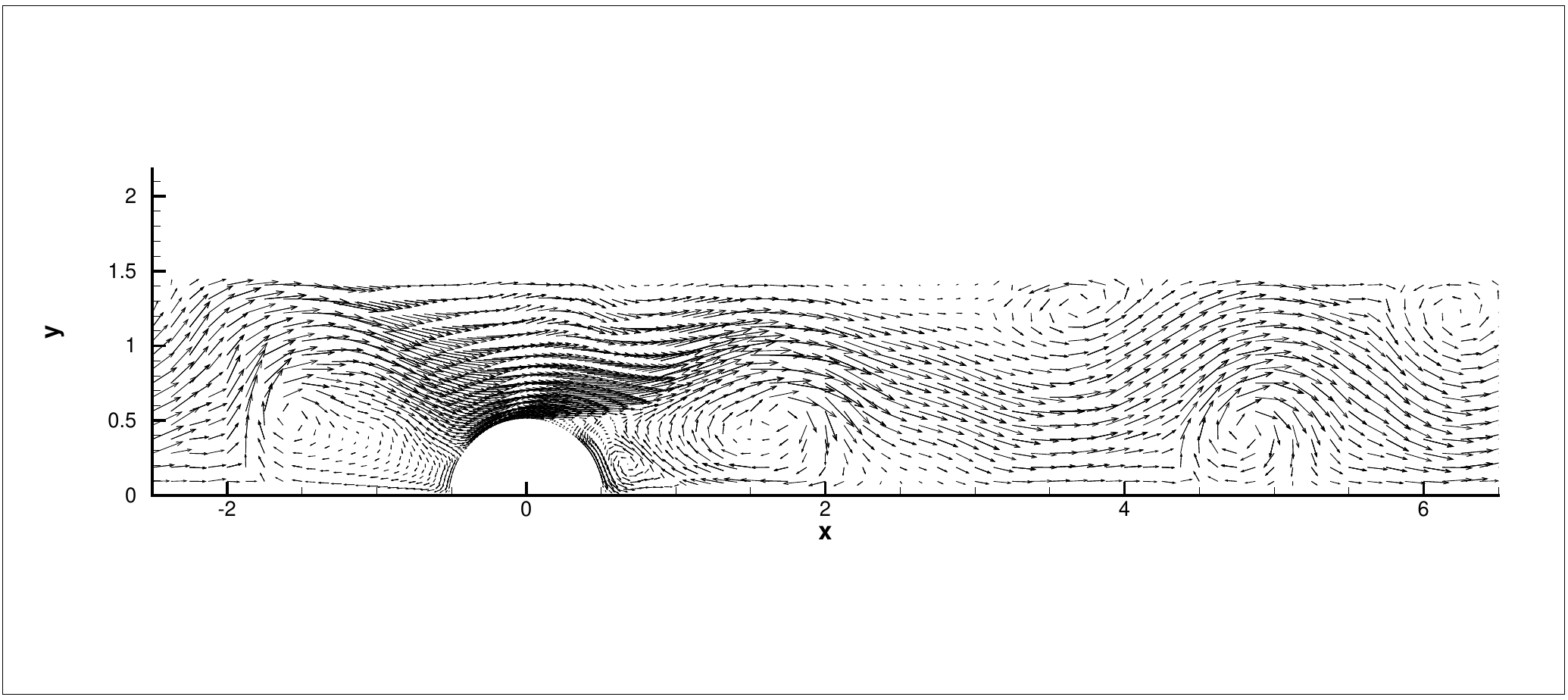}(e)
    \includegraphics[width=3in]{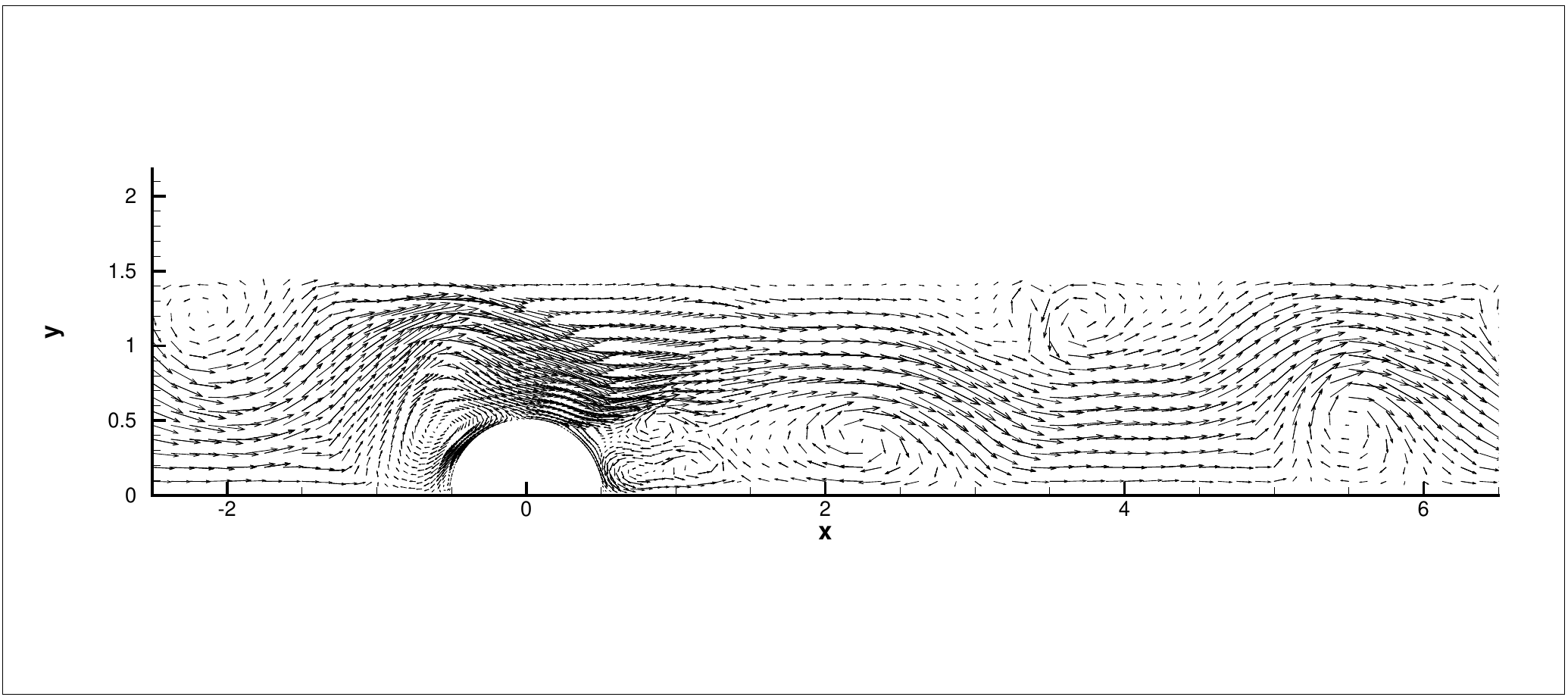}(f)
  }
  \centerline{
    \includegraphics[width=3in]{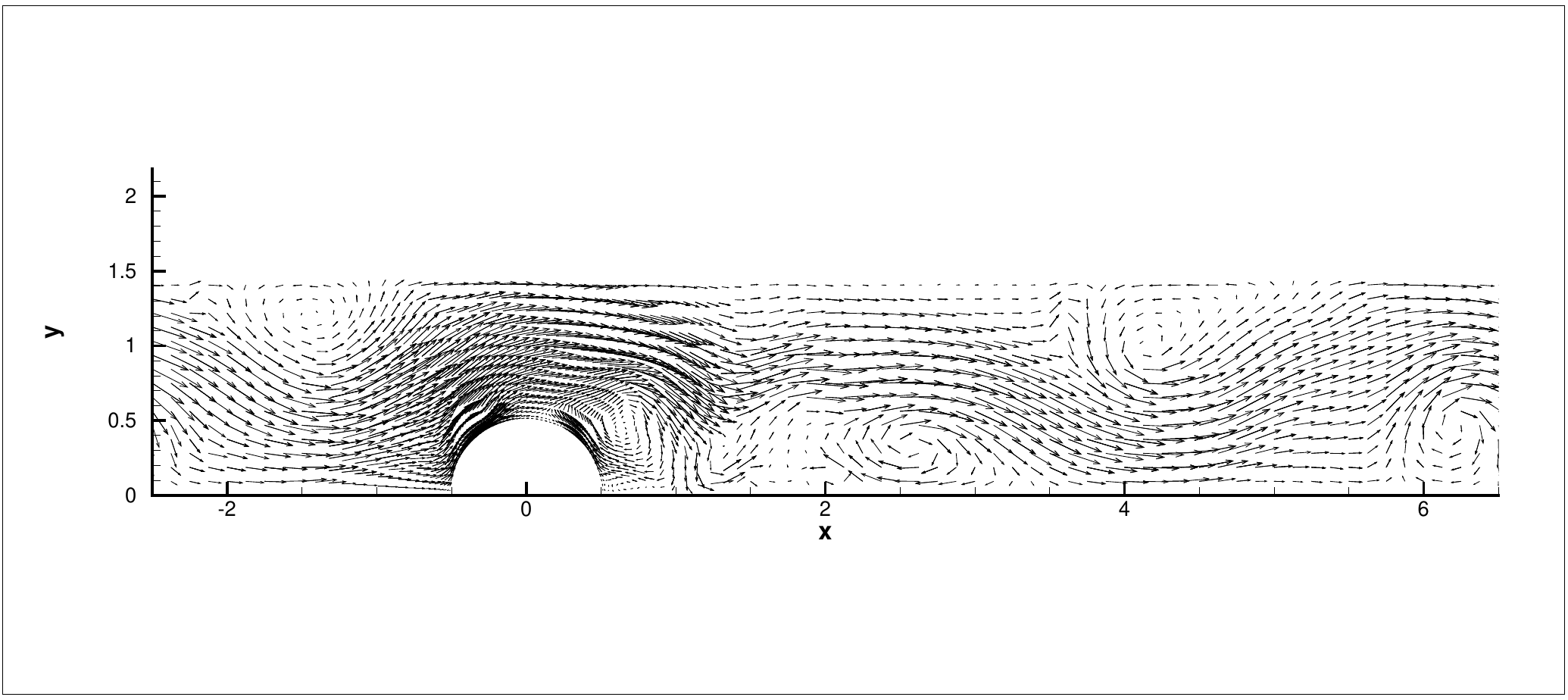}(g)
    \includegraphics[width=3in]{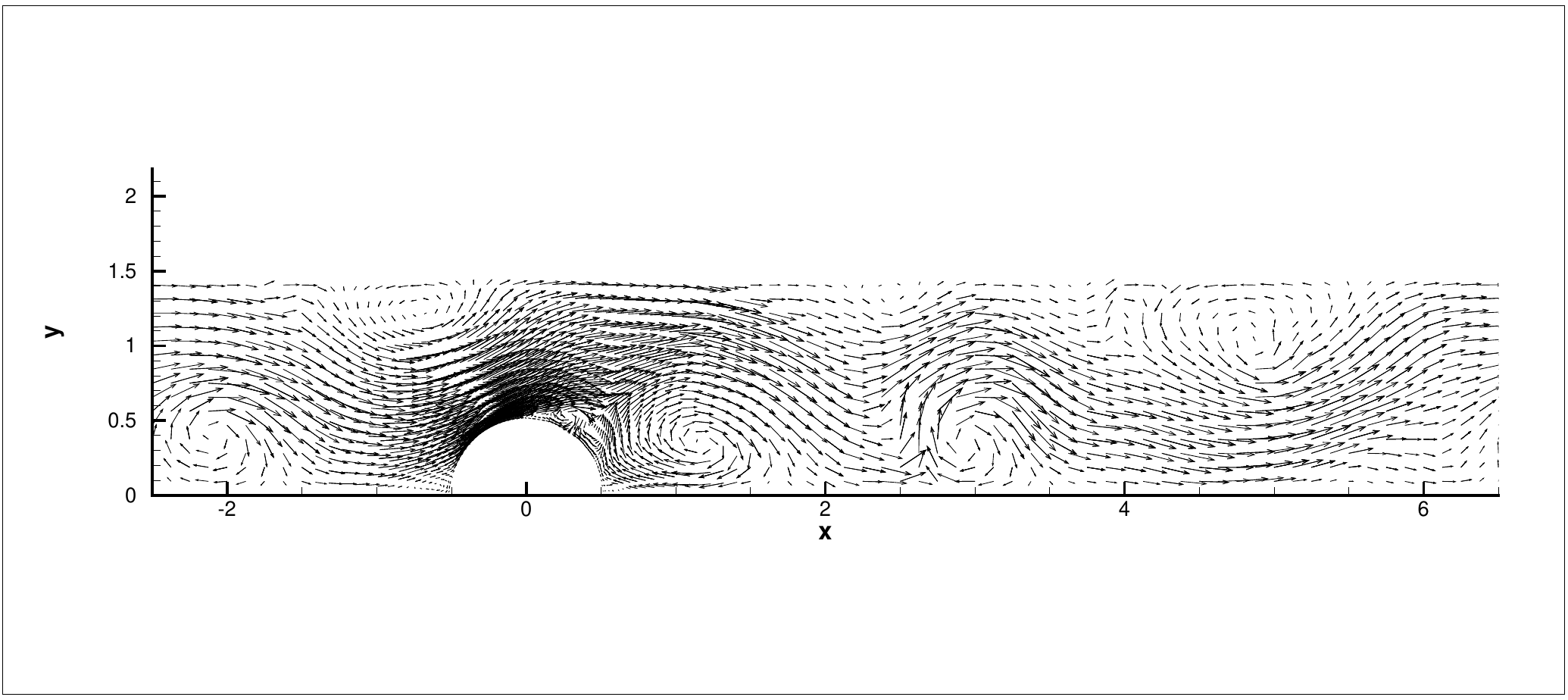}(h)
  }
  \centerline{
    \includegraphics[width=3in]{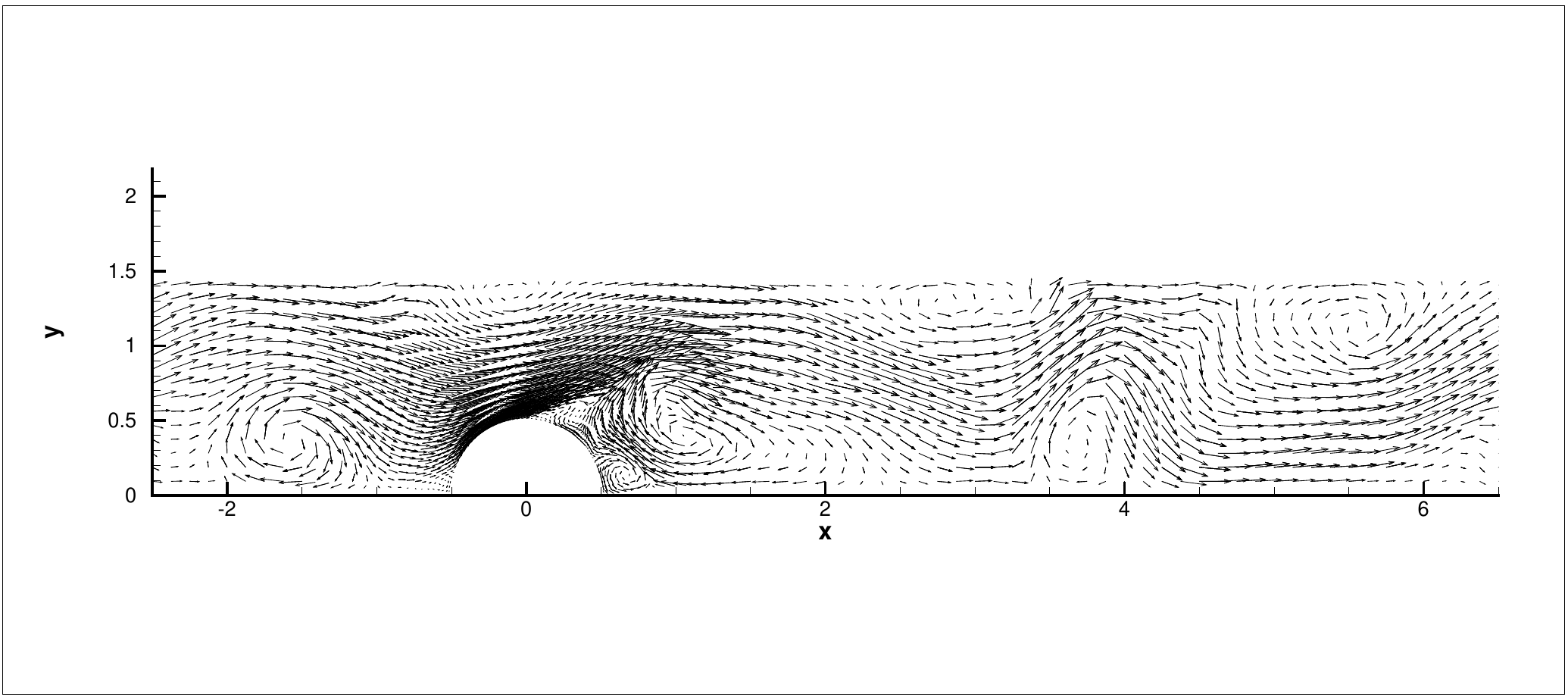}(i)
    \includegraphics[width=3in]{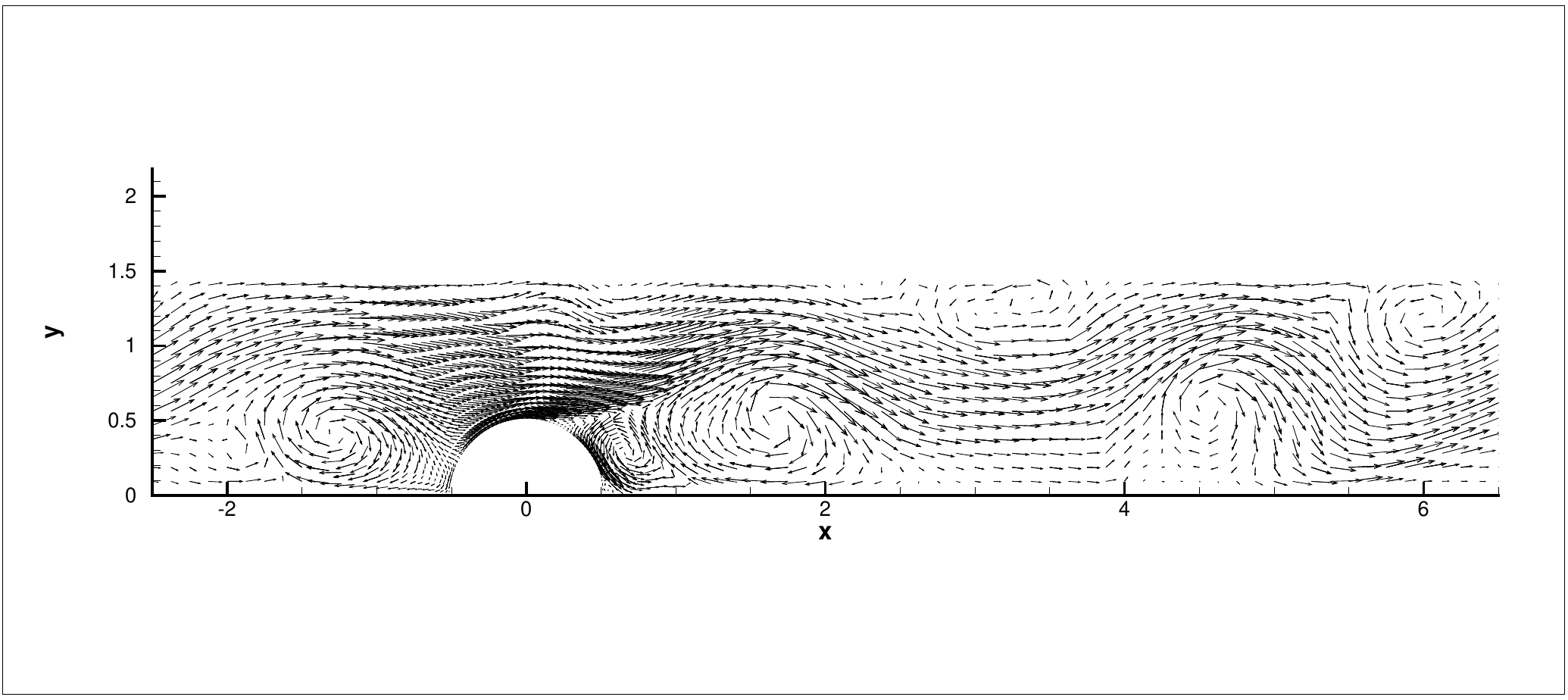}(j)
  }
  \centerline{
    \includegraphics[width=3in]{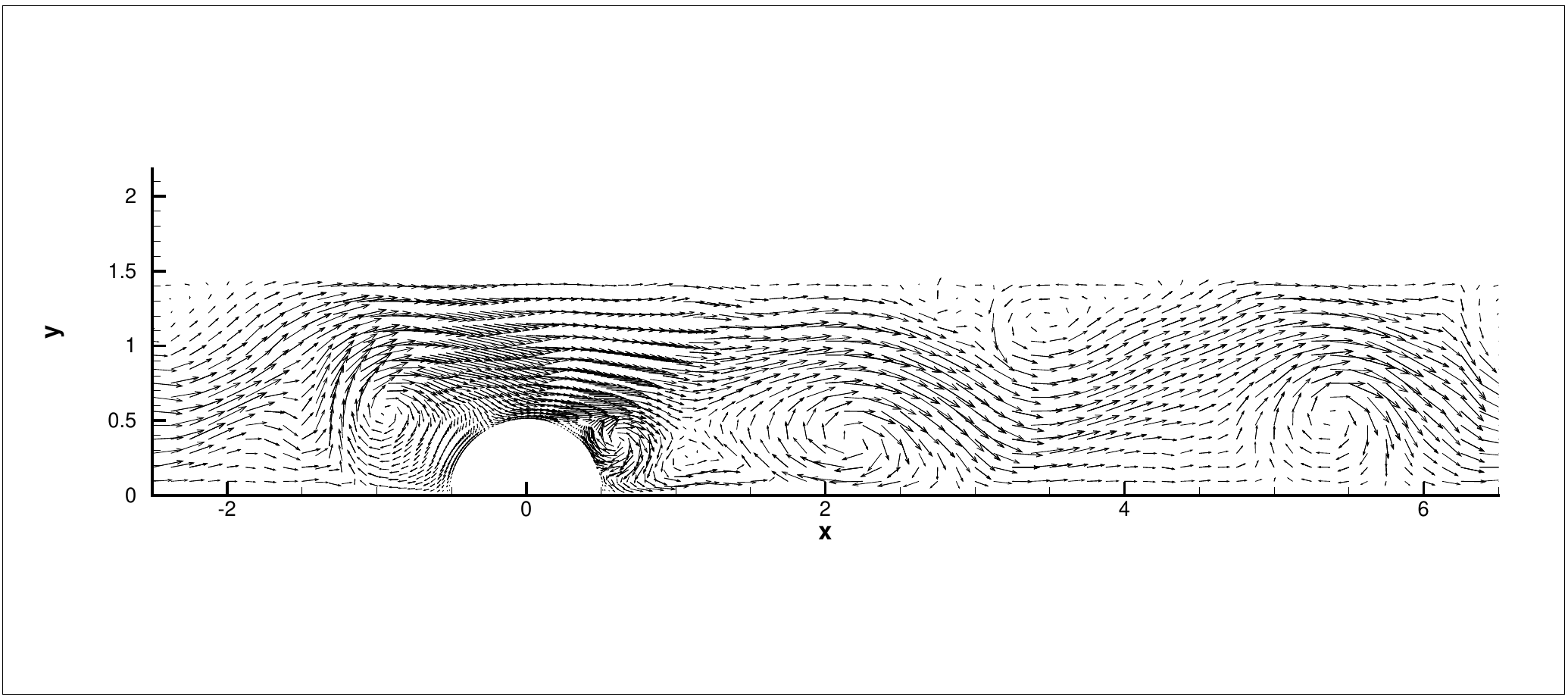}(k)
    \includegraphics[width=3in]{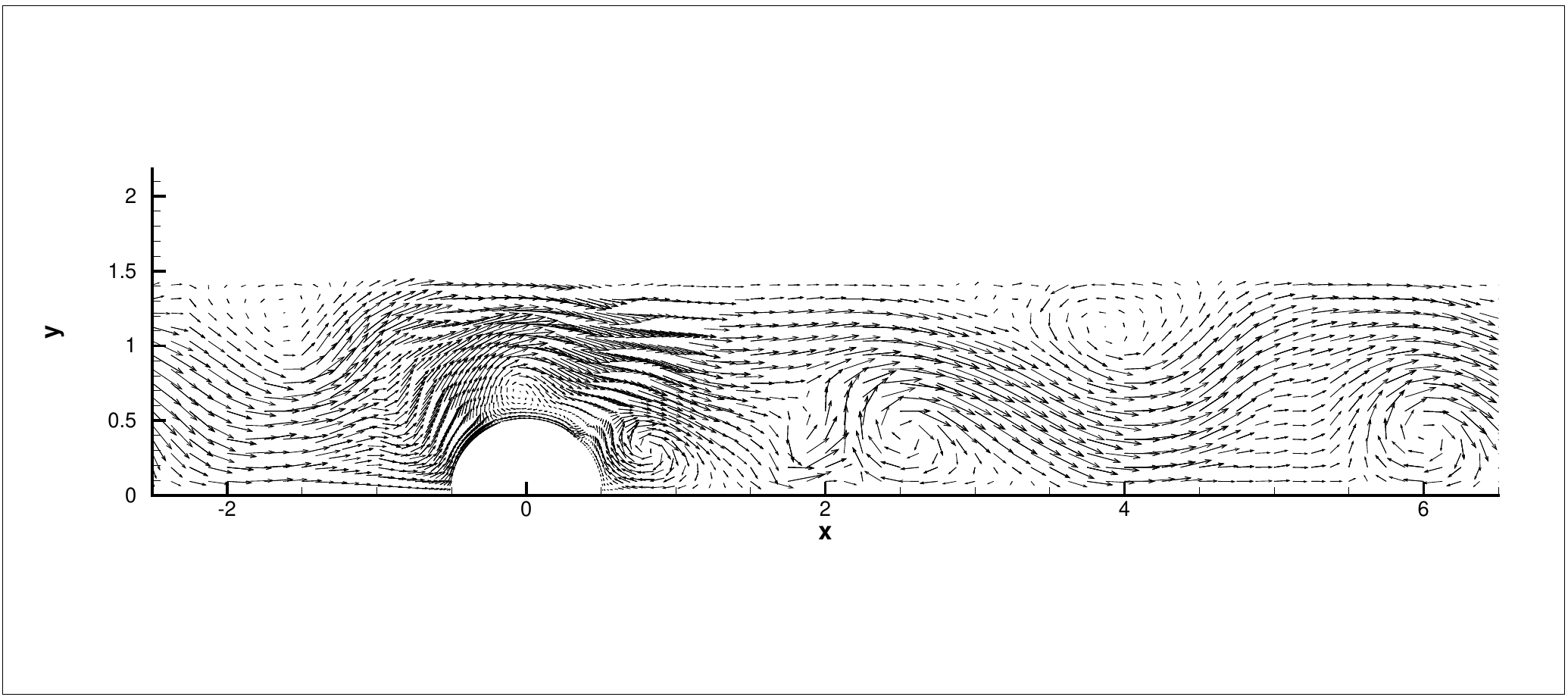}(l)
  }
  \caption{
    Flow past a hemisphere ($\nu=0.0005$): temporal sequence of snapshots of
    the velocity fields at time instants:
    (a) $t=t_0$,
    (b) $t=t_0+0.8$,
    (c) $t=t_0+1.6$,
    (d) $t=t_0+2.4$,
    (e) $t=t_0+3.2$,
    (f) $t=t_0+4.0$,
    (g) $t=t_0+4.8$,
    (h) $t=t_0+5.6$,
    (i) $t=t_0+6.4$,
    (j) $t=t_0+7.2$,
    (k) $t=t_0+8.0$,
    (l) $t=t_0+8.8$.
    $t_0$ denotes the initial time instant.
    Velocity vectors are plotted on a sparser grid on clarity.
  }
  \label{fig:hemis_dyn}
\end{figure}

Figure \ref{fig:hemis_dyn} illustrates the dynamics of the hemisphere flow
with a temporal sequence of snapshots of the velocity fields
at the Reynolds number corresponding to $\nu=0.0005$.
These results are obtained using a time step size $\Delta t=0.001$ and
$C_0=1000$, and the field $\mbs u_0$ is updated every $20$ time steps
in the simulations.
%
Several effects seem to play a role in
the dynamics of this flow:
(i) vortex shedding behind the hemisphere; 
(ii) periodicity of the channel, which introduces vortices into
the domain upstream of the hemisphere;
(iii) confinement of the narrow channel.
The vortices in the hemisphere wake
appear to be mostly confined to the regions near the
top and bottom walls.


\begin{table}[tb]
  \centering
  \begin{tabular}{lllllll}
    \hline
    $\nu$ & $C_0$ & ${\bar f}_x$ & $f'_x$ & ${\bar f}_y$ & $f'_y$ & Driving force \\
    \hline
    0.02 & 1e-2 & 0.393 & 0 & 1.09e-4 & 0 & 0.393 \\
    & 1e0 & 0.393 & 0 & 1.19e-4 & 0 & 0.393 \\
    & 1e+3 & 0.393 & 0 & 8.7e-5 & 0 & 0.393 \\
    & 1e+5 & 0.393 & 0 & 4.3e-5 & 0 & 0.393 \\
    & 1e+7 & 0.393 & 0 & 8.9e-5 & 0 & 0.393 \\
    \hline
    0.001 & 1e-2 & 0.394 & 0.242 & -0.0696 & 0.0564 & 0.393 \\
    & 1e0 & 0.394 & 0.241 & -0.0643 & 0.0563 & 0.393 \\
    & 1e+3 & 0.394 & 0.238 & -0.0636 & 0.0559 & 0.393 \\
    & 1e+5 & 0.394 & 0.240 & -0.0715 & 0.0565 & 0.393 \\
    & 1e+7 & 0.394 & 0.238 & -0.0689 & 0.0558 & 0.393 \\
    \hline
    0.0002 & 1e-2 & 0.386 & 0.805 & 0.117 & 0.408 & 0.393 \\
    & 1e0 & 0.389 & 0.800 & 0.0772 & 0.385 & 0.393 \\
    & 1e+3 & 0.388 & 0.812 & 0.102 & 0.401 & 0.393 \\
    & 1e+5 & 0.388 & 0.820 & 0.121 & 0.421 & 0.393 \\
    & 1e+7 & 0.387 & 0.825 & 0.139 & 0.444 & 0.393 \\
    \hline
  \end{tabular}
  \caption{ Flow past a hemisphere:
    Effect of $C_0$ on the computed forces on the walls.
  }
  \label{tab:hemis_C0}
\end{table}

The effect of the energy constant $C_0$ 
on the simulation
results is studied in Table \ref{tab:hemis_C0}
for the hemisphere flow.
This table lists the mean and rms forces on the wall
with respect to a range of $C_0$ values at three
Reynolds numbers corresponding to $\nu=0.02$, $0.001$ and $0.0002$.
A time step size $\Delta t=0.001$ and element order $6$ have
been employed in this
group of tests, and the field $\mbs u_0$ is updated every $20$ time steps.
We observe that the obtained forces are essentially the same
or quite close corresponding to different $C_0$,
suggesting that they have a low sensitivity to $C_0$ using
the current method. This is consistent with what has been observed
with the Kovasznay flow in the previous section.


\begin{table}[tbb]
  \centering
  \begin{tabular}{lllllll}
    \hline
    $\nu$ & $\Delta t$ & ${\bar f}_x$ & $f'_x$ & ${\bar f}_y$ & $f'_y$ & Driving force \\
    \hline
    0.02 & 0.001 & 0.393 & 0 & 8.7e-5 & 0 & 0.393 \\
    & 0.005 & 0.393 & 0 & 1.05e-4 & 0 & 0.393 \\
    & 0.01 & 0.393 & 0 & -2.1e-5 & 0 & 0.393 \\
    & 0.1 & 0.393 & 0 & 8.7e-5 & 0 & 0.393 \\
    & 1.0 & 0.393 & 0 & -2.1e-5 & 0 & 0.393 \\
    \hline
    0.001 & 5e-4 & 0.393 & 0.238 & -0.0625 & 0.122 & 0.393 \\
    & 0.001 & 0.394 & 0.238 & -0.0636 & 0.0559 & 0.393 \\
    & 0.005 & 0.395 & 0.238 & -0.0706 & 0.0115 & 0.393 \\
    & 0.01 & 0.395 & 0.216 & -0.0681 & 0.00668 & 0.393 \\
    & 0.1 & (BiCGStab & fails & to & converge) \\
    & 1.0 & (BiCGStab & fails & to & converge) \\
    \hline
    0.0002 & 2.5e-4 & 0.385 & 0.841 & 0.172 & 1.033 & 0.393 \\
    & 5e-4 & 0.387 & 0.840 & 0.103 & 0.784 & 0.393 \\
    & 0.001 & 0.388 & 0.812 & 0.102 & 0.401 & 0.393 \\
    & 0.005 & 0.399 & 0.671 & 0.177 & 0.175 & 0.393 \\
    & 0.01 & 0.393 & 0.459 & -0.00365 & 0.127 & 0.393 \\
    & 0.1 & (BiCGStab & fails & to & converge) \\
    & 1.0 & (BiCGStab & fails & to & converge) \\
    \hline
  \end{tabular}
  \caption{ Flow past a hemisphere:
    Computed forces on the walls corresponding to a range of $\Delta t$ values.
  }
  \label{tab:hemis_ldt}
\end{table}

We next investigate the effect of  $\Delta t$
on the stability and accuracy of the simulations.
Thanks to the discrete energy stability property (Theorem~\ref{thm:thm_1}),
fairly large time step sizes can be employed in
actual simulations with the current method.
Table \ref{tab:hemis_ldt} lists the mean and rms forces
on the walls obtained using time step sizes
ranging from $\Delta t=2.5e-4$ to $\Delta t=1.0$
in the simulations of the hemisphere flow.
A fixed $C_0=1000$ and an element order $6$ are employed,
and the $\mbs u_0$
field is updated every $20$ time steps.
We observe that the current method can produce stable simulation
results with various $\Delta t$, ranging from small
to very large values, at lower Reynolds numbers; see
the case $\nu=0.02$ in Table \ref{tab:hemis_ldt}.
At higher Reynolds numbers, we observe that
the method produces stable results with small to fairly large
$\Delta t$ values. However, when $\Delta t$ becomes
very large the method seems less robust, in that
the BiCGStab linear solver may fail to converge for solving
the linear algebraic system of equations.
For example, for Reynolds numbers corresponding to
$\nu=0.001$ and $\nu=0.0002$, with $\Delta t=0.1$ and larger
we observe that the BiCGStab linear solver fails to converge
after some time into the computation using the current method.
Because the current method involves a non-symmetric velocity coefficient
matrix due to the $\mbs M(\mbs u)$ term,
with large $\Delta t$ the conditioning
of the velocity linear algebraic system can possibly become poor,
which can cause difficulty to the BiCGStab solver.
It should be noted that these large $\Delta t$ values, with which BiCGStab
solver encounters a difficulty here, are
considerably larger than those maximum $\Delta t$ values
a typical semi-implicit scheme can use in order to maintain
stability. For instance, for the hemisphere flow with $\nu=0.0002$,
using the semi-implicit scheme from~\cite{DongS2015} (which also employs a
pressure correction-type strategy),
the simulation is only stable with $\Delta t=2.5e-4$ or smaller
under the same mesh resolution.


\begin{figure}
  \centerline{
    \includegraphics[height=2.5in]{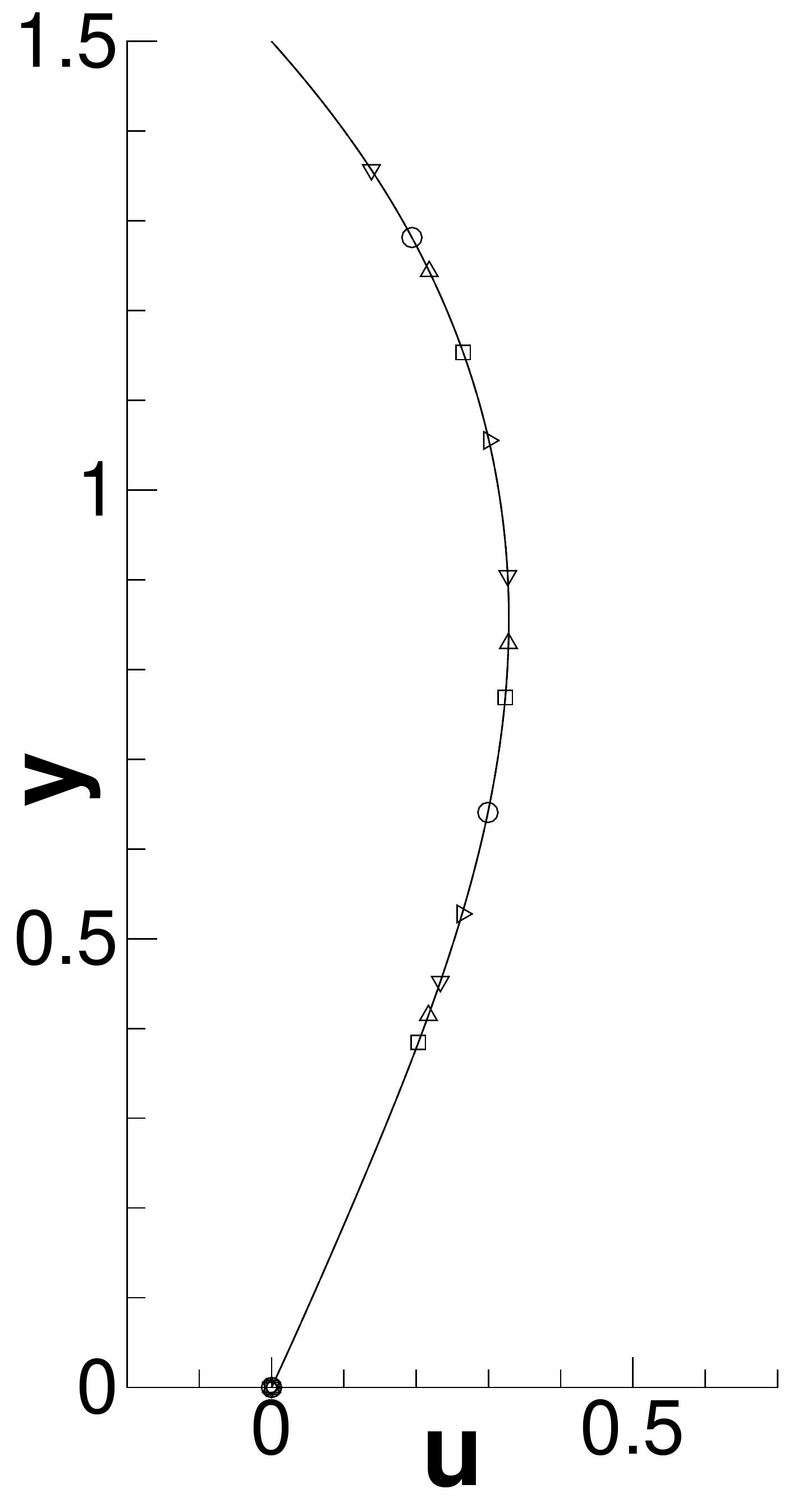}(a)
    \includegraphics[height=2.5in]{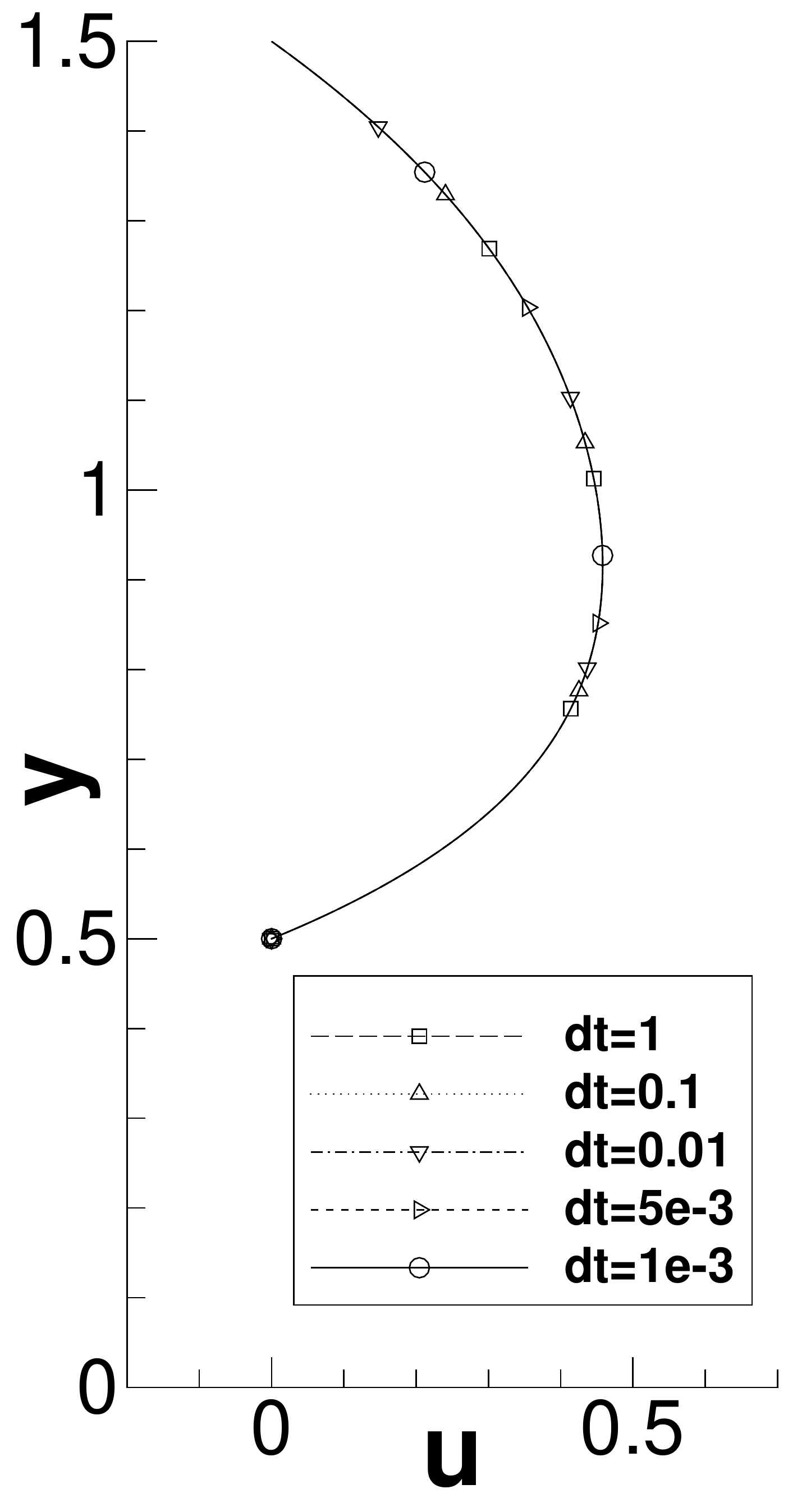}(b)
    \includegraphics[height=2.5in]{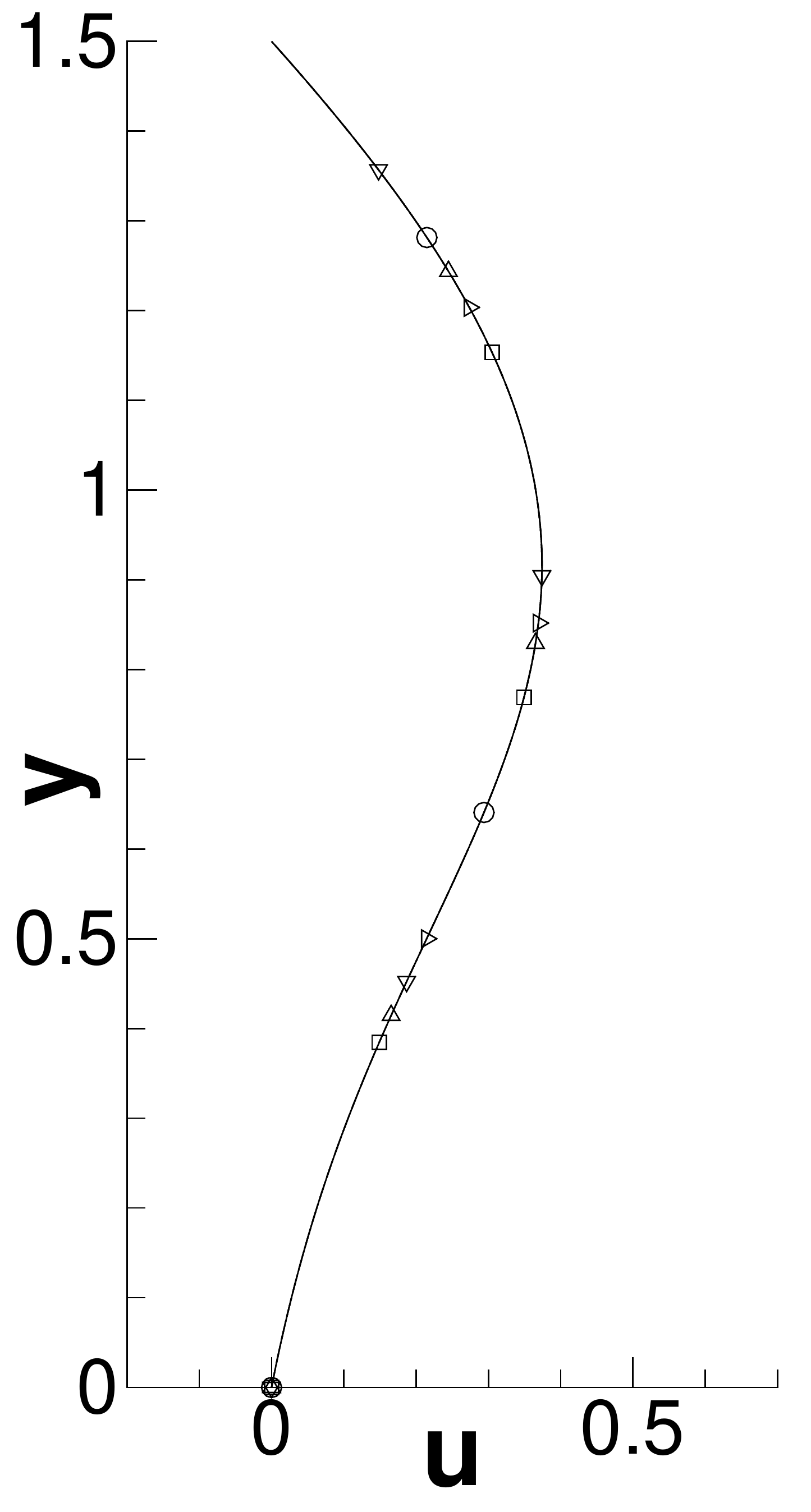}(c)
  }
  \centerline{
    \includegraphics[height=2.5in]{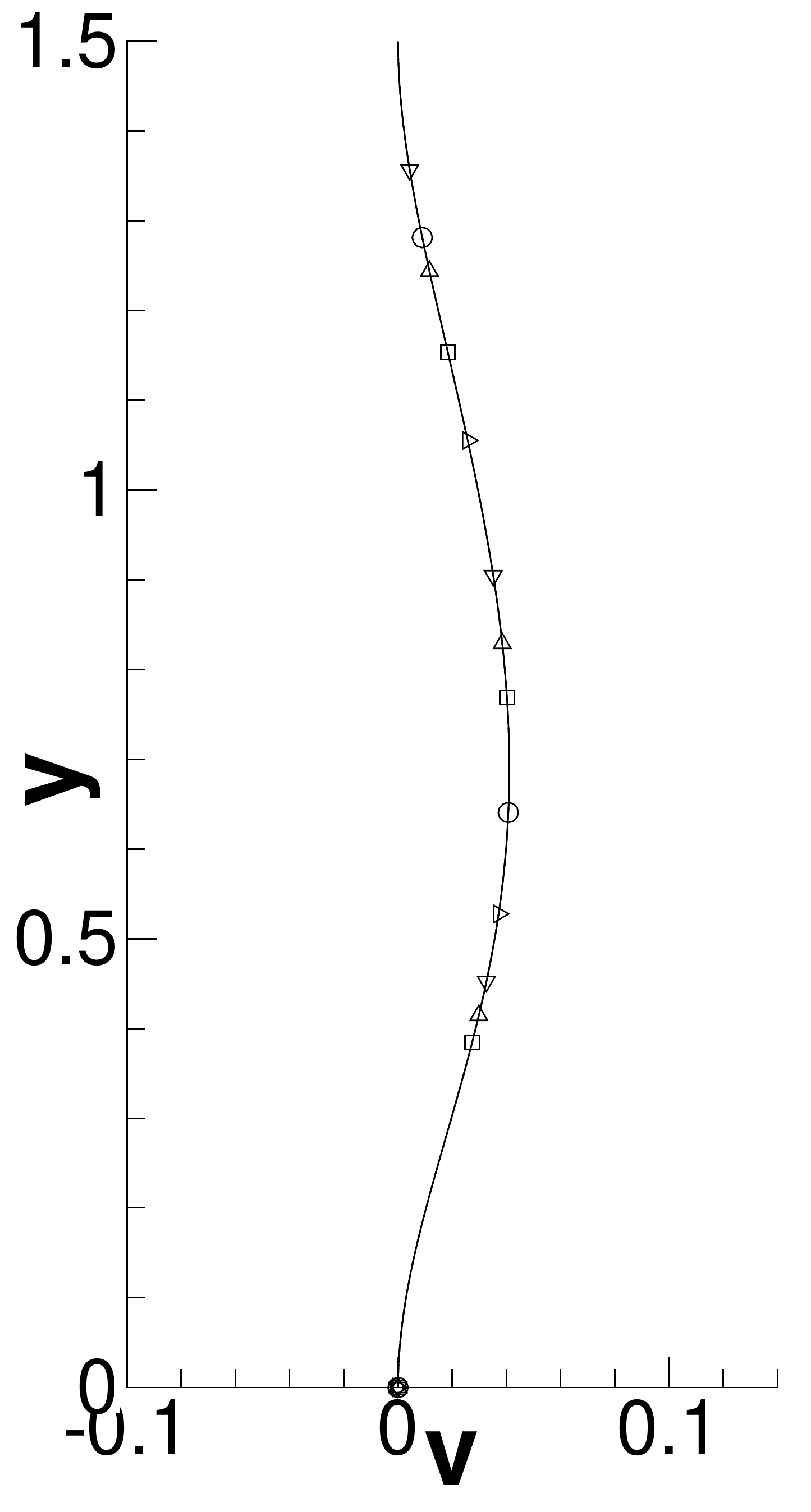}(d)
    \includegraphics[height=2.5in]{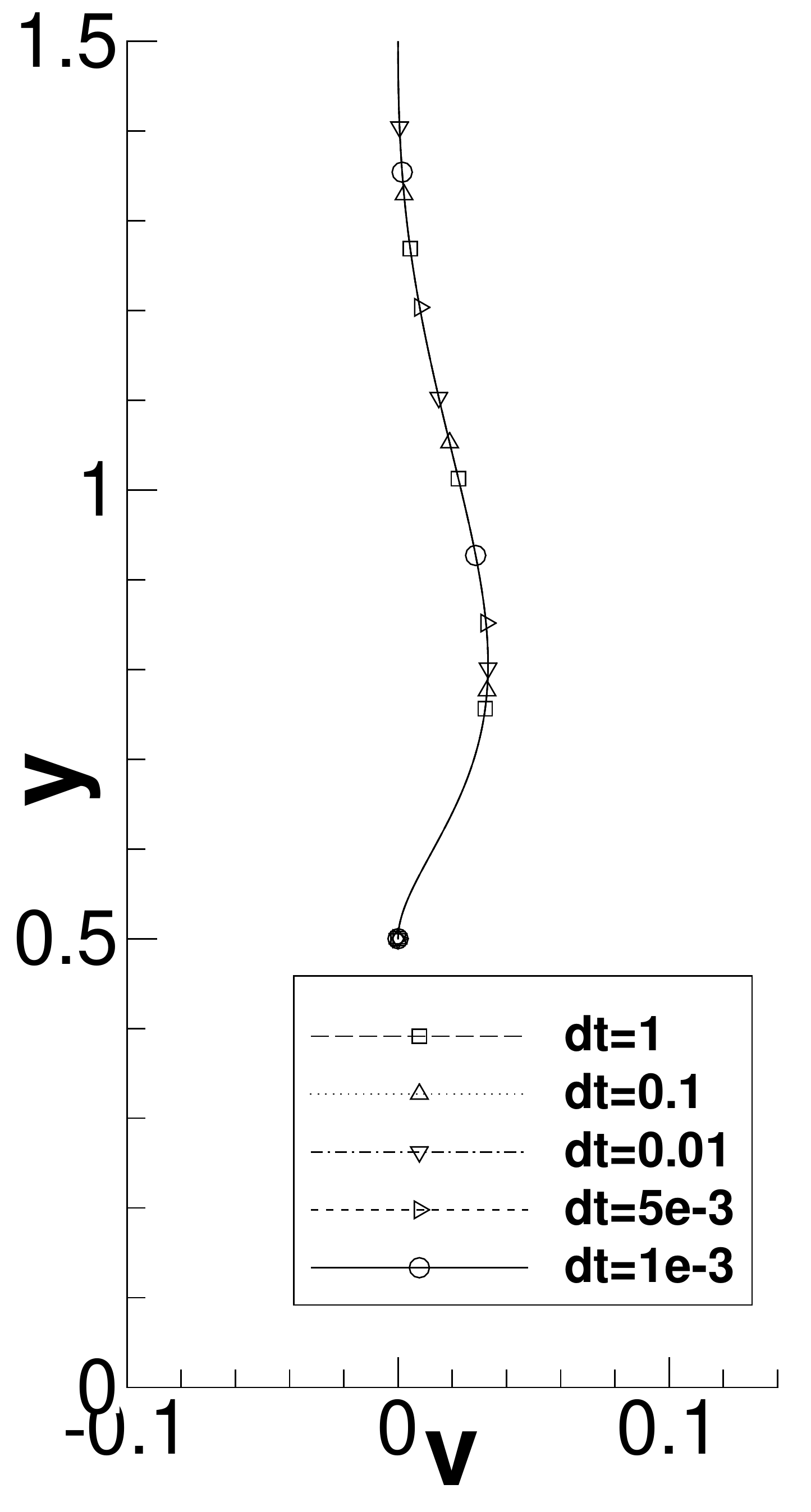}(e)
    \includegraphics[height=2.5in]{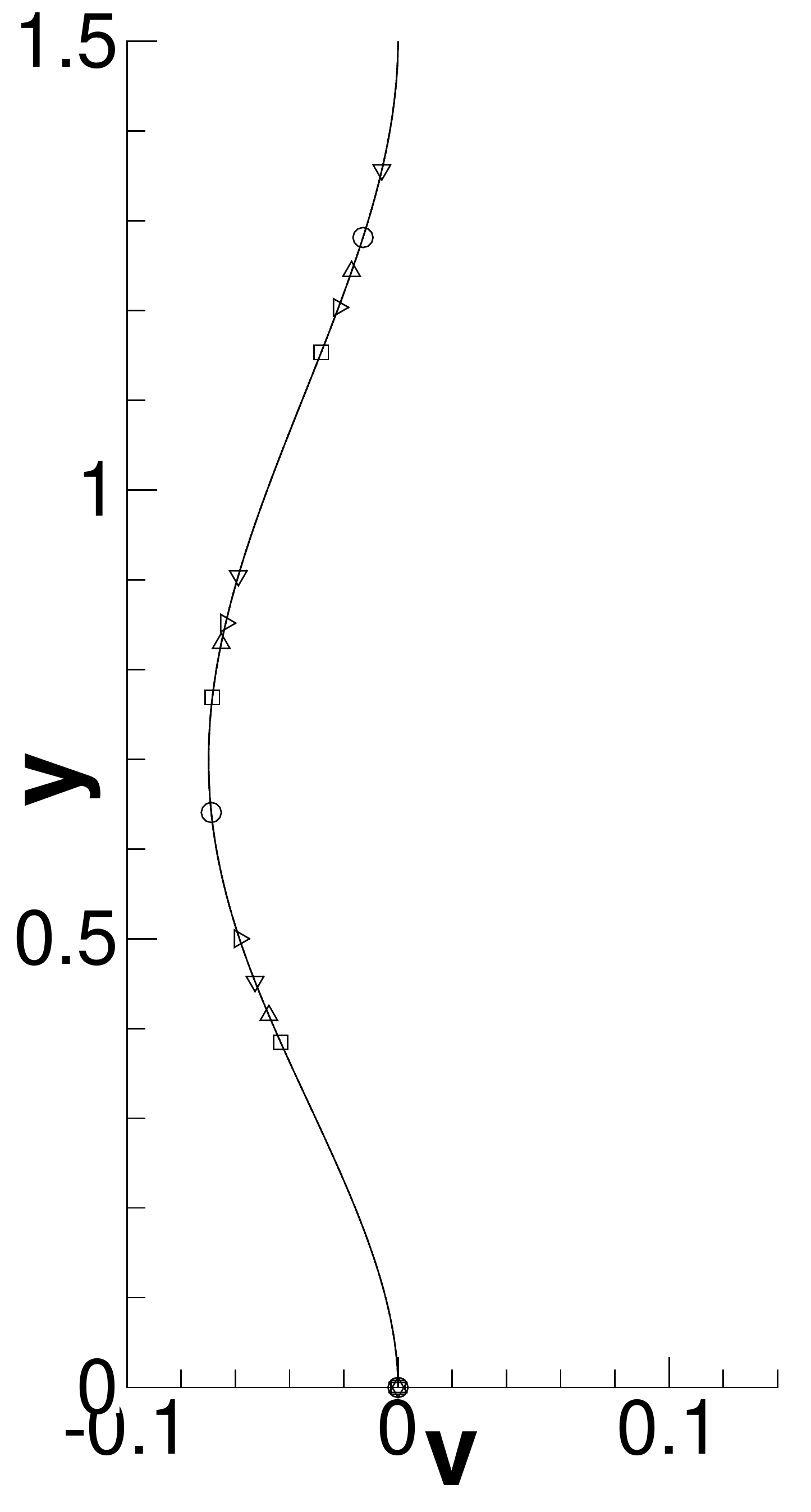}(f)
  }
  \caption{
    Flow past a hemisphere ($\nu=0.02$): Comparison of profiles of
    the streamwise velocity (top row) and vertical velocity (bottom row) at locations
    (a,d) $x/d=-1$, (b,e) $x/d=0$, and (c,f) $x/d=1$ computed using the current method
    with different time step sizes.
  }
  \label{fig:hemis_vcomp}
\end{figure}


Figure \ref{fig:hemis_vcomp} shows a comparison of profiles of
the steady-state streamwise and vertical velocities across
the channel at three downstream
locations ($x/d=0, \pm 1.0$) for  $\nu=0.02$.
These  profiles are computed using the current method with several
time step sizes ranging from $\Delta t=0.001$ to $\Delta t=1.0$.
In these simulations $C_0=1000$, the element order is $6$,
and the field $\mbs u_0$ is updated
every $20$ time steps.
The velocity profiles obtained with different $\Delta t$,
ranging from small to large values, exactly overlap
with one another. This suggests that the current method can produce
accurate results
with large $\Delta t$ for this problem.


\begin{figure}
  \centerline{
    \includegraphics[width=3in]{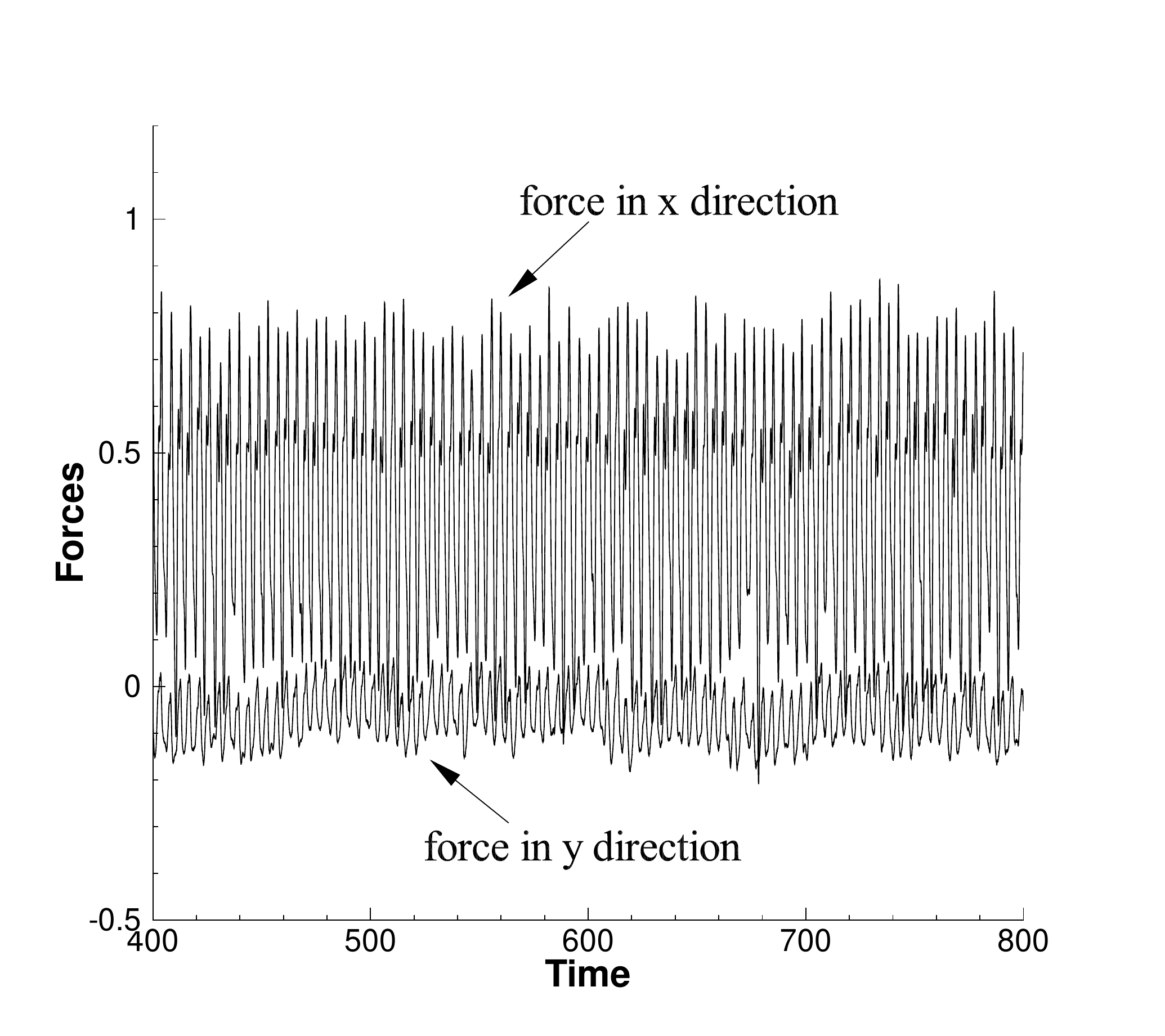}(a)
    \includegraphics[width=3in]{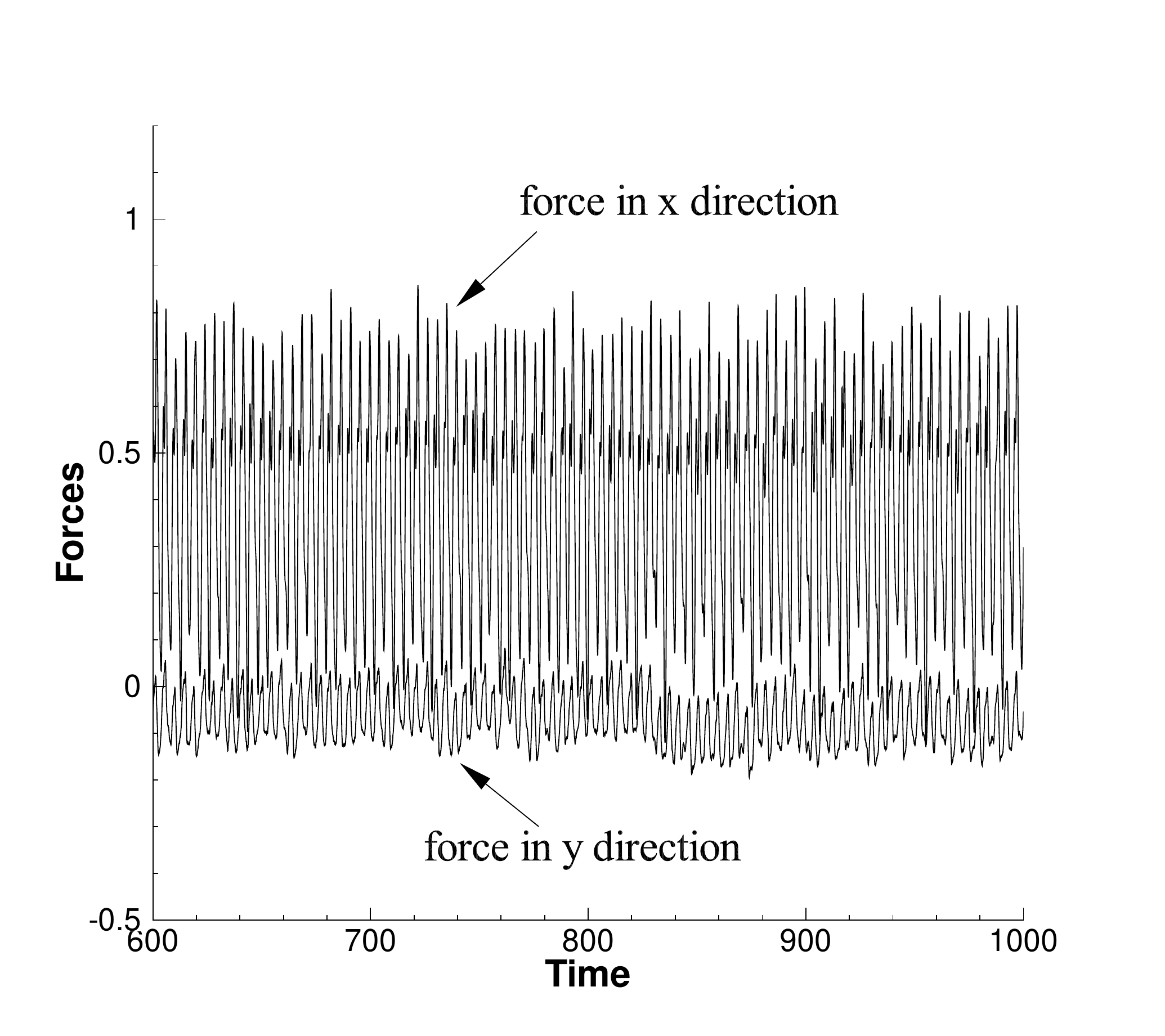}(b)
  }
  \centerline{
    \includegraphics[width=3in]{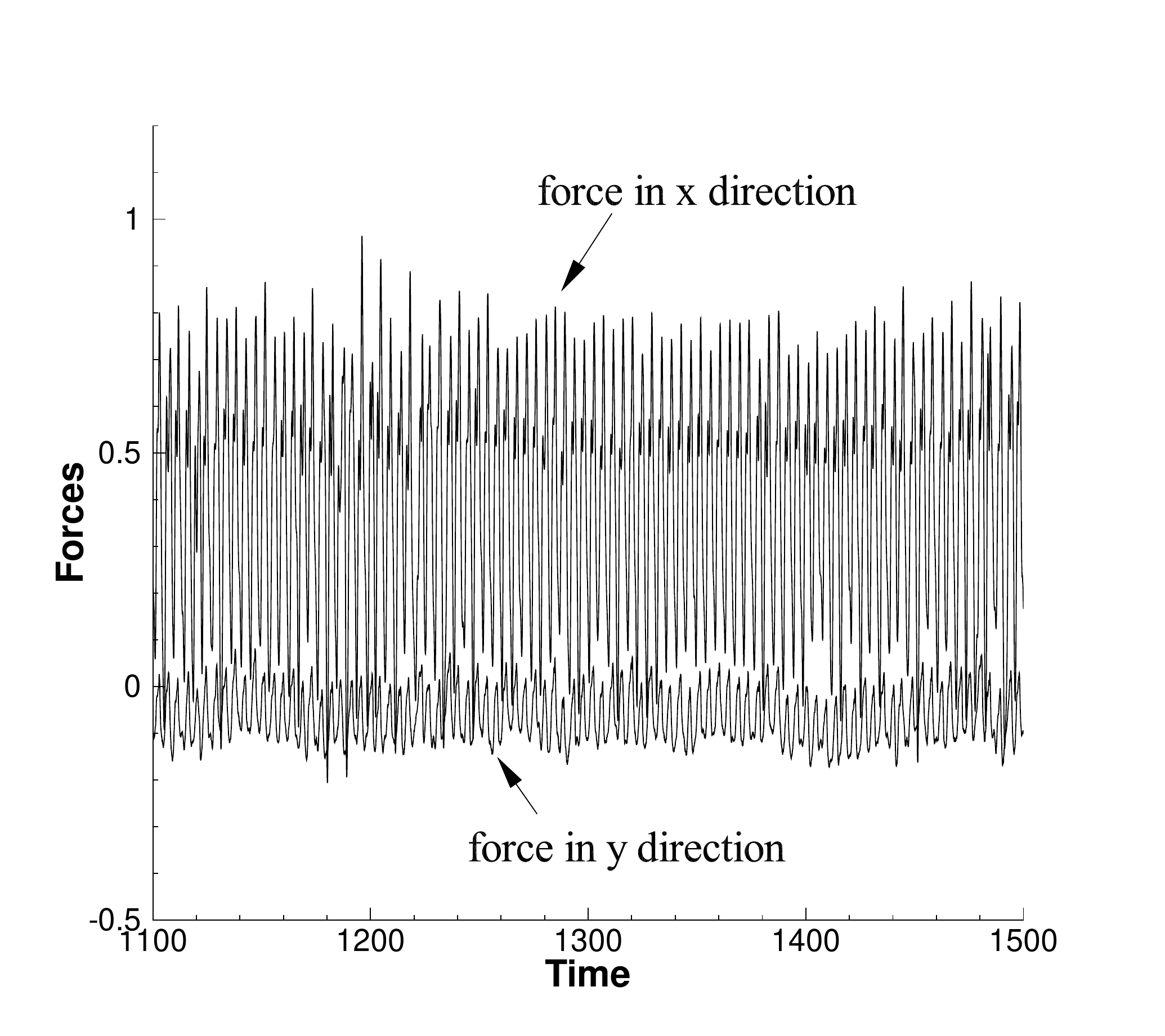}(c)
    \includegraphics[width=3in]{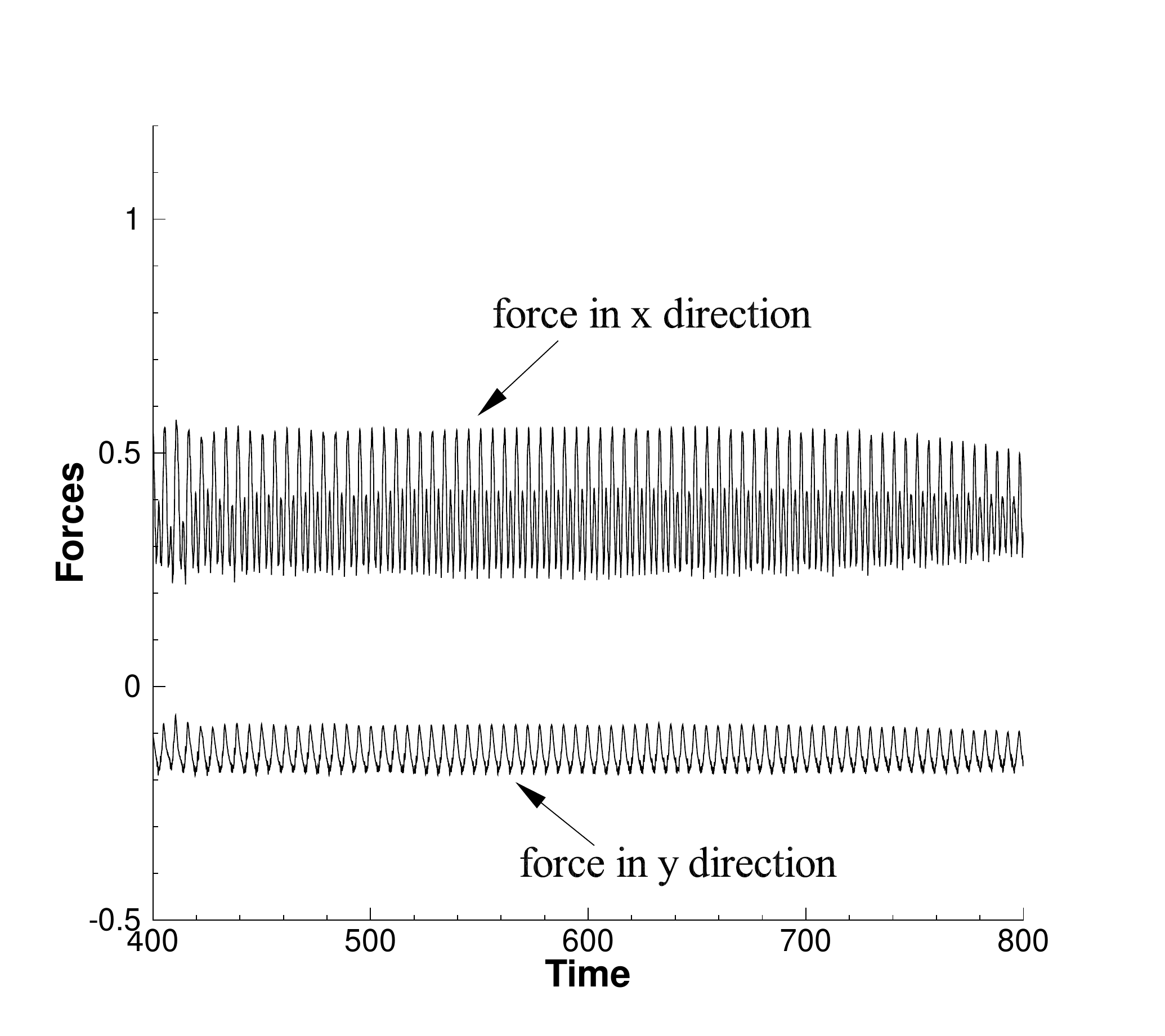}(d)
  }
  \caption{
    Flow past a hemisphere ($\nu=0.001$): time histories of the forces on channel walls
    obtained with different frequency parameter $k_0$ for updating the field $\mbs u_0$.
    (a) $k_0=20$, (b) $k_0=100$, (c) $k_0=200$, (d) $k_0=500$.
  }
  \label{fig:hemis_k0}
\end{figure}

With the current method the field $\mbs u_0$, and
hence the velocity coefficient matrix
(see equation \eqref{equ:def_M1}), is updated
every $k_0$ time steps. We observe that the frequency 
for $\mbs u_0$ update can have an influence on the accuracy of
simulation results.
With too large a $k_0$ value, the simulation  can lose accuracy.
This point is demonstrated by Figure \ref{fig:hemis_k0},
which shows time histories of the forces on walls at $\nu=0.001$
with $\mbs u_0$ updated with different frequencies,
ranging from $k_0=20$ to $k_0=500$.
In this set of simulations we have employed $\Delta t=0.001$,
element order $6$
and $C_0=1000$. It can be observed that the computed forces
have essentially the same characteristics when $\mbs u_0$ is updated
every $20$, $100$, or $200$ time steps.
When $k_0$ increases to $500$, however, 
the computed forces are notably different in terms of
the amplitude, frequency and the overall characteristics.
This indicates that the accuracy  starts to
deteriorate.

\begin{table}[tb]
  \centering
  \begin{tabular}{lllllll}
    \hline
    $\nu$ & $k_0$ & ${\bar f}_x$ & $f'_x$ & ${\bar f}_y$ & $f'_y$ & Driving force \\
    \hline
    0.02 & 10 & 0.393 & 0 & 2.56e-4 & 0 & 0.393  \\
    & 20 & 0.393 & 0 & 8.7e-5 & 0 & 0.393 \\
    & 50 & 0.393 & 0 & 2.61e-4 & 0 & 0.393 \\
    & 100 & 0.393 & 0 & 2.60e-4 & 0 & 0.393 \\
    & 200 & 0.393 & 0 & 2.61e-4 & 0 & 0.393 \\
    & 500 & 0.393 & 0 & 2.59e-4 & 0 & 0.393 \\
    & 1000 & 0.393 & 0 & 2.60e-4 & 0 & 0.393 \\
    \hline
    0.001 & 10 & 0.394 & 0.238 & -0.0668 & 0.0562 & 0.393 \\
    & 20 & 0.394 & 0.238 & -0.0636 & 0.0559 & 0.393 \\
    & 50 & 0.394 & 0.240 & -0.0701 & 0.0567 & 0.393 \\
    & 100 & 0.394 & 0.237 & -0.0670 & 0.0561 & 0.393 \\
    & 200 & 0.394 & 0.243 & -0.0685 & 0.0559 & 0.393 \\
    & 500 & 0.377 & 0.0858 & -0.142 & 0.0300 & 0.393  \\
    & 1000 & 0.382 & 0.184 & -0.105 & 0.0657 & 0.393  \\
    \hline
    0.0002 & 10 & 0.389 & 0.807 & 0.103 & 0.410 & 0.393  \\
    & 20 & 0.388 & 0.812 & 0.102 & 0.401 & 0.393 \\
    & 50 & 0.390 & 0.818 & 0.212 & 0.547 & 0.393  \\
    & 100 & 0.397 & 0.772 & 0.233 & 0.530 & 0.393  \\
    & 200 & 0.395 & 0.554 & 0.182 & 0.772 & 0.393  \\
    & 500 & 0.398 & 0.341 & -0.110 & 0.614 & 0.393  \\
    & 1000 & 0.365 & 0.327 & 0.0396 & 0.818 & 0.393  \\
    \hline
  \end{tabular}
  \caption{
    Flow past a hemisphere:
    Effect of the frequency parameter $k_0$ for $\mbs u_0$ update
    on the computed forces on walls.
  }
  \label{tab:hemis_k0for}
\end{table}


Table \ref{tab:hemis_k0for} provides the mean and rms forces on
the walls at three Reynolds numbers corresponding to
$\nu=0.02$, $0.001$ and $0.0002$ obtained with various $k_0$
(ranging from $10$ to $1000$) for updating the field $\mbs u_0$.
In these tests $\Delta t=0.001$, the element order is $6$, and $C_0=1000$.
These data confirm our observations based on the force
histories. At $\nu=0.02$ the computed forces are basically identical,
irrespective of whether $\mbs u_0$ is updated every $10$ time steps
or every $1000$ time steps.
At $\nu=0.001$ the computed forces are quite close when $k_0=200$ or below.
But their values are notably different with $k_0=500$ and larger.
At $\nu=0.0002$, the computed forces start to show
notable differences when $k_0$ increases to $50$ and larger.
These results suggest that,
when $\mbs u_0$ is updated too rarely, the correction
term $[\mbs N(\mbs u)-\mbs M(\mbs u)]$ in equation \eqref{equ:nse_1}
may become more significant and this can cause larger errors 
in the simulation results.
With higher Reynolds numbers the field $\mbs u_0$ should be
updated more frequently in order to maintain accuracy in
the simulation results.



\begin{figure}
  \centerline{
    \includegraphics[height=2.5in]{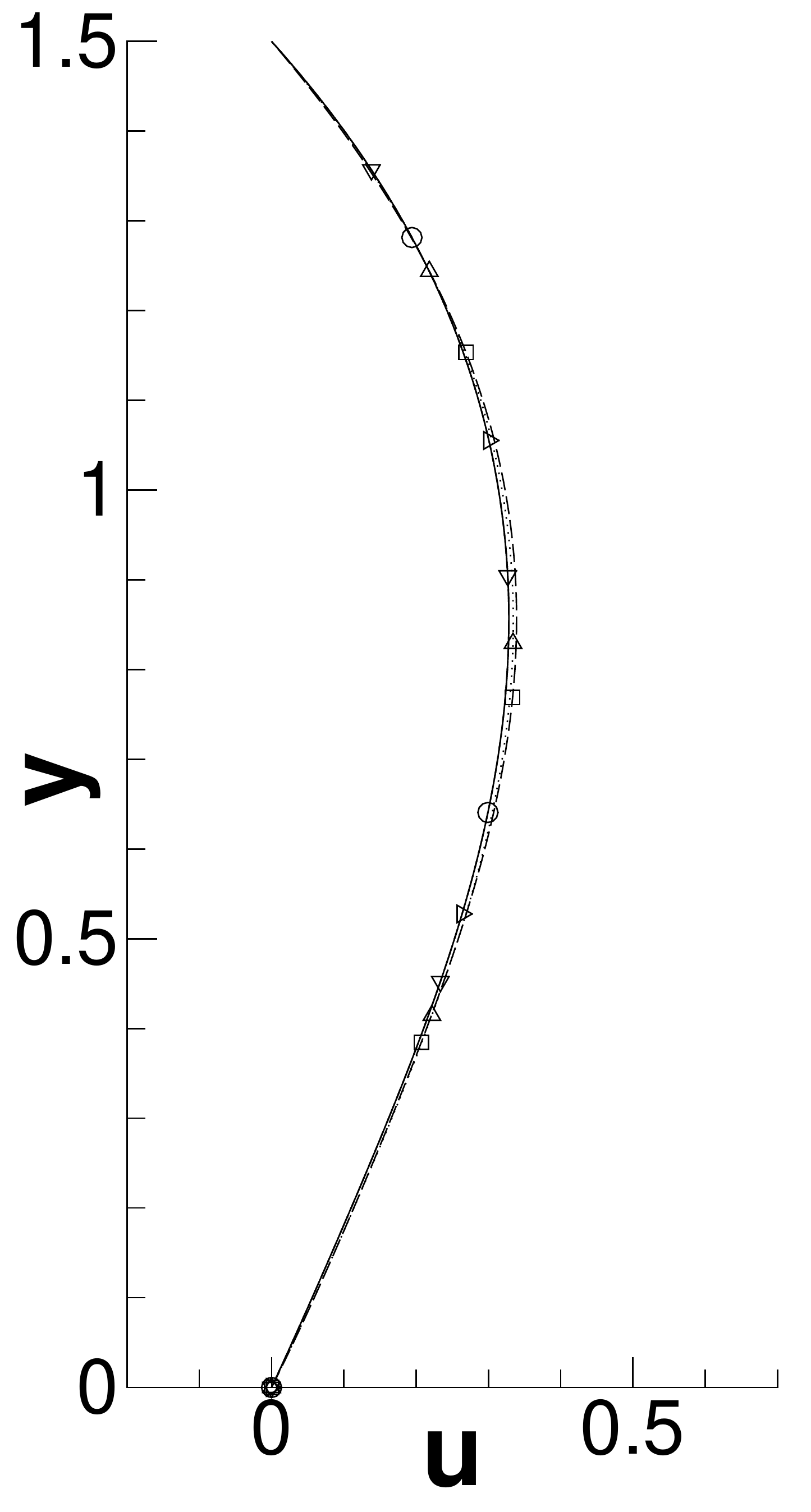}(a)
    \includegraphics[height=2.5in]{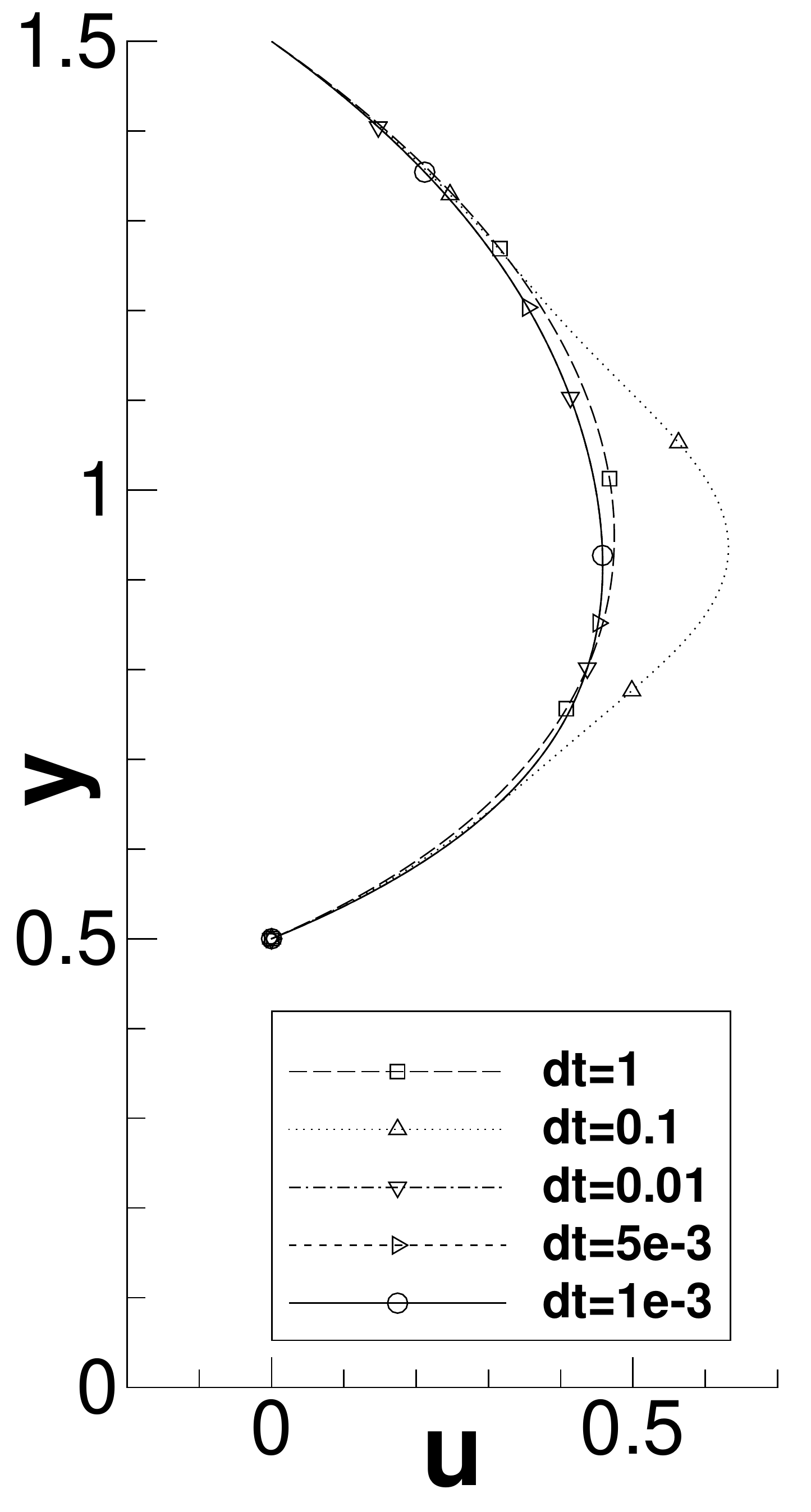}(b)
    \includegraphics[height=2.5in]{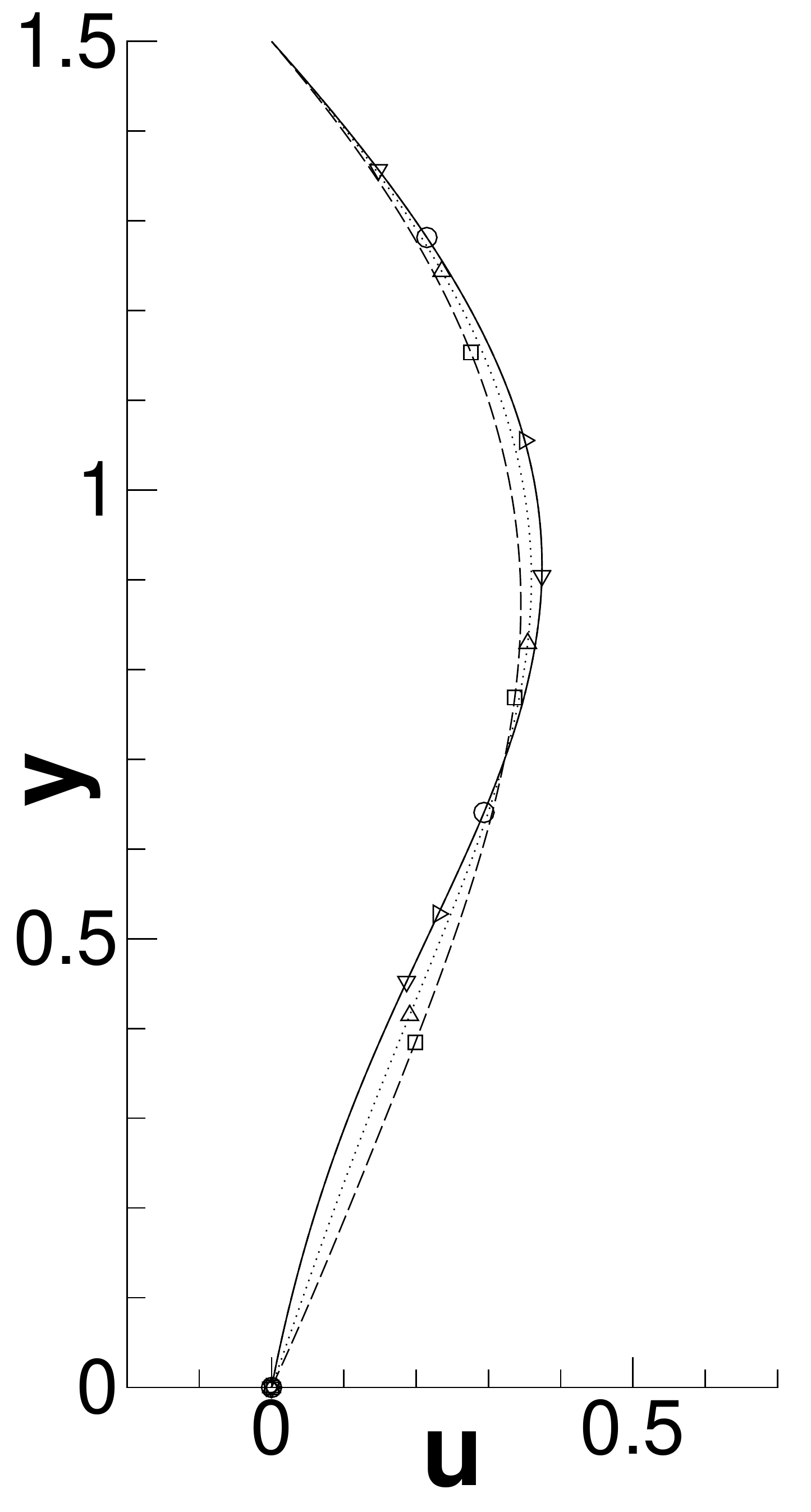}(c)
  }
  \centerline{
    \includegraphics[height=2.5in]{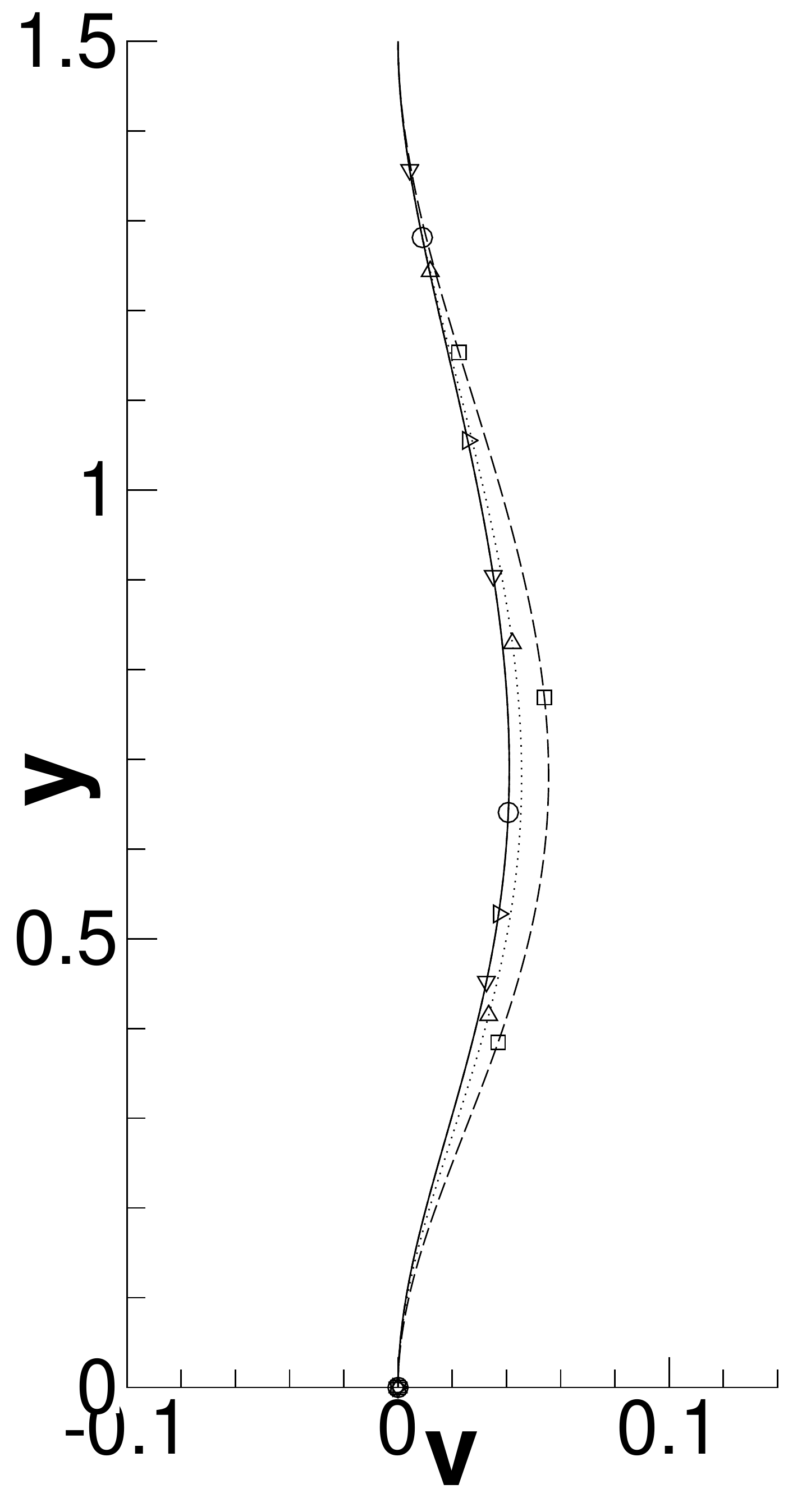}(d)
    \includegraphics[height=2.5in]{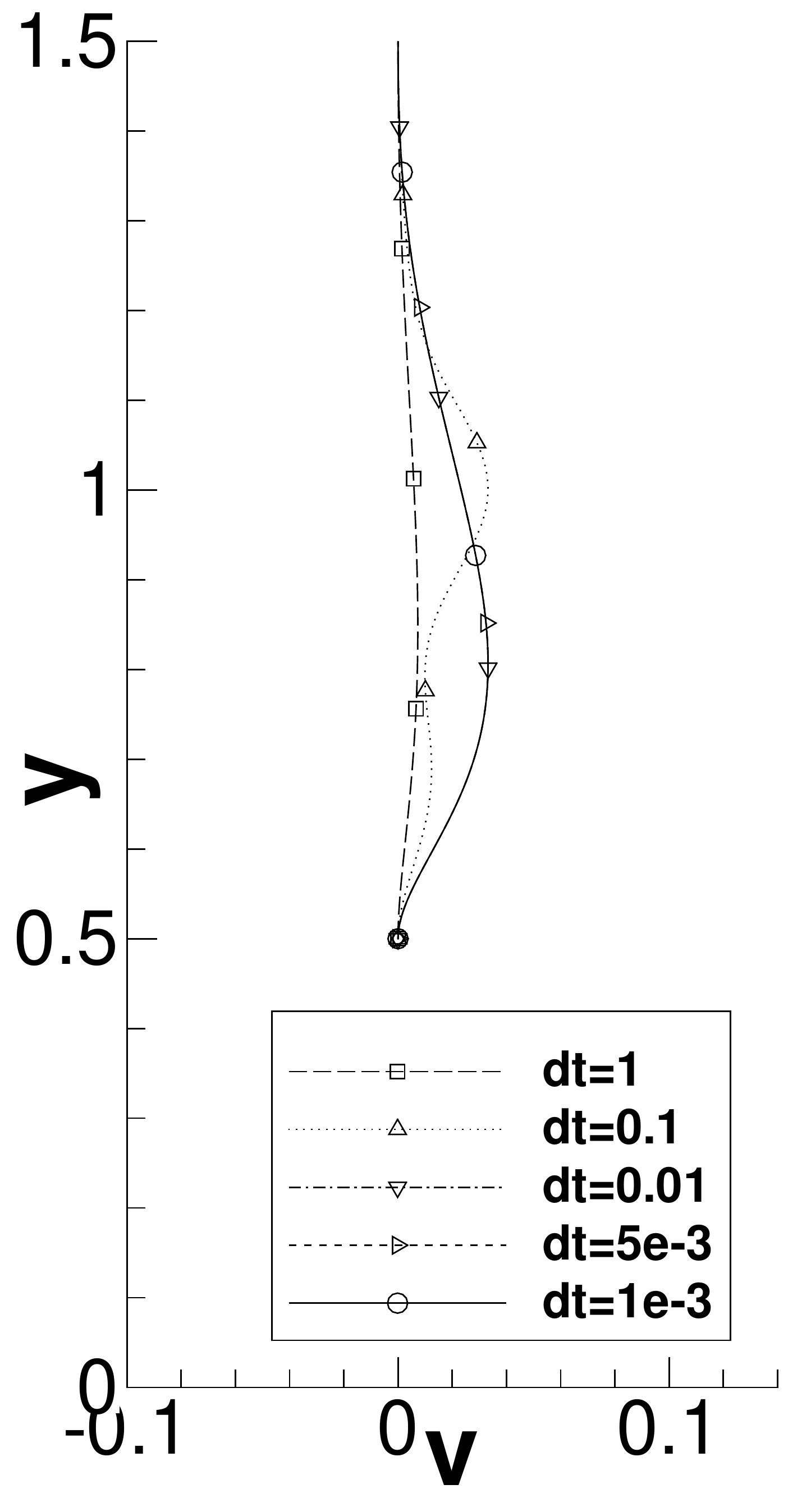}(e)
    \includegraphics[height=2.5in]{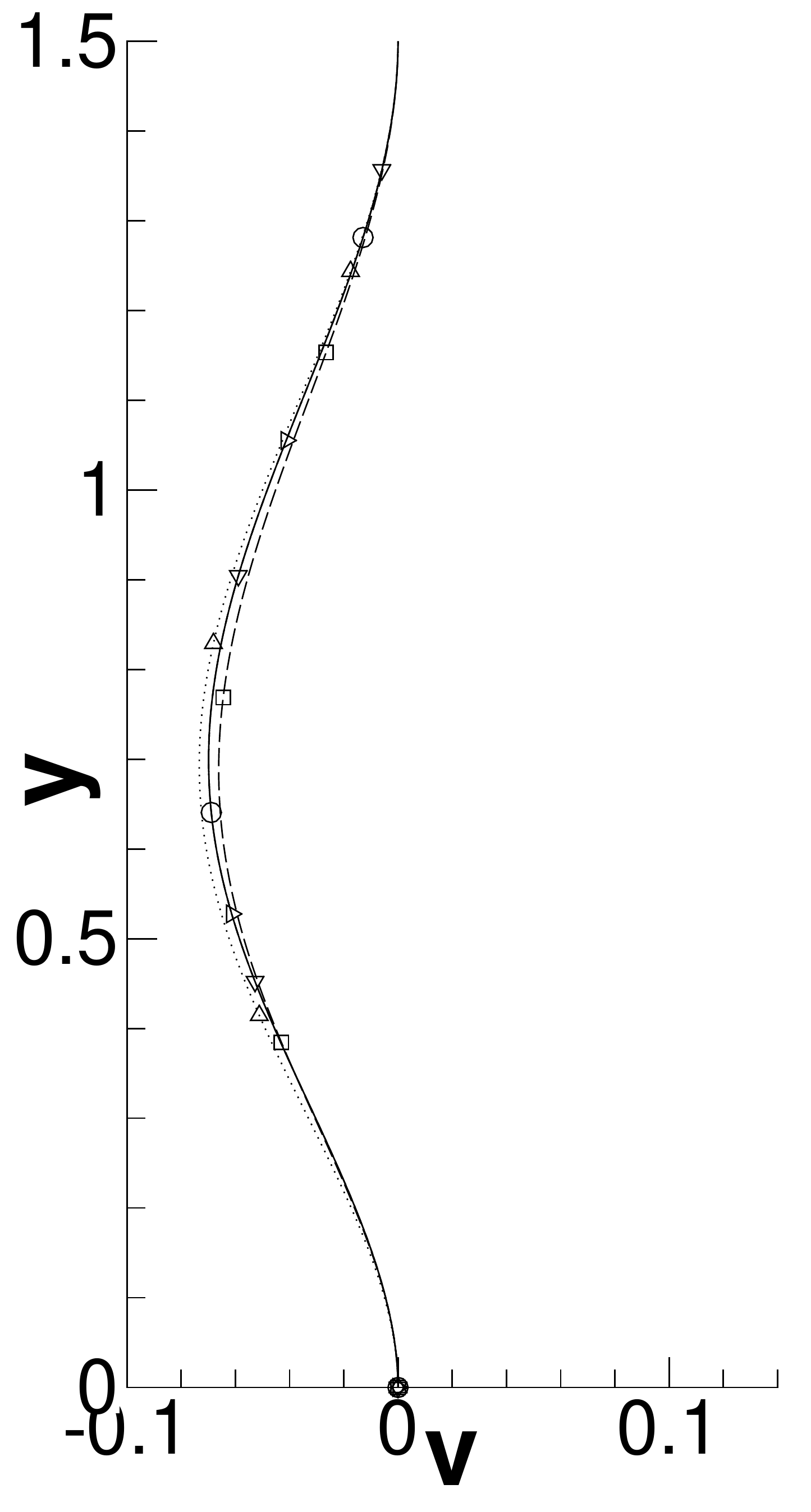}(f)
  }
  \caption{
    Flow past a hemisphere ($\nu=0.02$): Comparison of profiles of
    the streamwise velocity (top row) and vertical velocity (bottom row) at locations
    (a,d) $x/d=-1$, (b,e) $x/d=0$, and (c,f) $x/d=1$ obtained using the
    modified scheme ($\mbs M(\mbs u)=0$)  with
    different time step sizes.
  }
  \label{fig:hemis_Mu0}
\end{figure}


\begin{table}
  \centering
  \begin{tabular}{lllllll}
    \hline
    $\nu$ & $\Delta t$ & ${\bar f}_x$ & $f'_x$ & ${\bar f}_y$ & $f'_y$ & Driving force \\
    \hline
    0.02 & 0.001 & 0.393 & 0 & 1.11e-4 & 0 & 0.393 \\
    & 0.005 & 0.393 & 0 & 2.52e-4 & 0 & 0.393 \\
    & 0.01 & 0.393 & 0 & 2.51e-4 & 0 & 0.393 \\
    & 0.1 & 0.394 & 1.87e-4 & 5.03e-4 & 2.92e-4 & 0.393 \\
    & 1.0 & 0.393 & 1.77e-3 & 8.79e-5 & 8.07e-4 & 0.393 \\
    \hline
    0.001 & 5e-4 & 0.393 & 0.238 & -6.69e-2 & 0.123 & 0.393 \\
    & 0.001 & 0.400 & 0.182 & -2.94e-2 & 0.0954 & 0.393 \\
    & 0.005 & 0.395 & 0.0620 & 8.28e-3 & 8.90e-3 & 0.393 \\
    & 0.01 & 0.394 & 1.17e-4 & 2.86e-2 & 1.59e-4 & 0.393 \\
    & 0.1 & 0.393 & 1.55e-5 & 6.90e-4 & 1.03e-3 & 0.393 \\
    & 1.0 & 0.393 & 3.37e-6 & -4.04e-3 & 1.04e-2 & 0.393 \\
    \hline
  \end{tabular}
  \caption{Flow past a hemisphere:
    Mean and rms forces on the walls attained using
    the modified scheme (with $\mbs M(\mbs u)=0$) with different
    $\Delta t$ for $\nu=0.02$ and $\nu=0.001$.
  }
  \label{tab:hemis_Mu0}
\end{table}

The modified scheme with $\mbs M(\mbs u)=0$ (see Remark~\ref{rem:rem_1})
has also been used to simulate the hemisphere in channel problem.
Figure \ref{fig:hemis_Mu0} shows the steady-state
streamwise and vertical velocity profiles at three downstream locations
 for $\nu=0.02$ obtained using this modified scheme.
These  profiles correspond to
 several time step sizes
ranging from $\Delta t=0.001$ to $\Delta t=1.0$.
$C_0=1000$ and element order $6$ are employed in the simulations.
This figure can be compared with Figure \ref{fig:hemis_vcomp},
which is attained using the current method under identical conditions.
%
While the velocity profiles computed using the modified algorithm
with $\Delta t=0.01$ and smaller
all overlap with one another, those obtained with the larger
$\Delta t=0.1$ and $\Delta t=1.0$ exhibit marked differences
than with smaller $\Delta t$ values. This suggests that
the velocity distributions obtained with the larger $\Delta t$ values
are no longer accurate using the modified scheme.
In contrast, with the current method the velocity profiles computed
with the larger $\Delta t=0.1$ and $\Delta t=1.0$ are identical to
those obtained with the
smaller $\Delta t$ values; see Figure \ref{fig:hemis_vcomp}.
It is further noted that all the velocity profiles obtained
using the current method in Figure \ref{fig:hemis_vcomp} agree with those
profiles obtained using the modified scheme with the smaller $\Delta t$ values.
These data suggest that the current method is superior in accuracy
to the modified scheme.
The current method can produce accurate results at
larger time step sizes where the modified algorithm with
$\mbs M(\mbs u)=0$ ceases to
be accurate.
This is consistent with the observations with the Kovasznay flow
in the previous section.

Table \ref{tab:hemis_Mu0} lists the mean and rms forces
on the walls computed
using the modified scheme ($\mbs M(\mbs u)=0$)
with a number of time step sizes ranging from
$\Delta t=5.0e-4$ to $\Delta t=1.0$ for
$\nu=0.02$ and $\nu=0.001$.
This table can be compared with Table \ref{tab:hemis_ldt},
which is obtained with the current method under identical conditions.
We observe that the modified scheme with $\mbs M(\mbs u)=0$
is more robust for very large time step sizes.
For example, stable simulation results are obtained using
the modified scheme with
$\Delta t=0.1$ and $\Delta t=1.0$ for $\nu=0.001$.
It is already discussed before that
with the current method the BiCGStab linear solver fails
to converge for these two cases.
Note that in the implementation of
the modified scheme with $\mbs M(\mbs u)=0$
the conjugate gradient (CG) solver has been
used to solve the resultant linear systems, because the coefficient
matrix is symmetric positive definite. On the other hand,
in the implementation of the current method, the BiCGStab linear
solver is used for the velocity linear system and
the CG solver is used for the pressure linear system.
In terms of accuracy, the data again indicate that
the current method is superior for
large or fairly large time step sizes.
For $\nu=0.02$, the horizontal and vertical forces ($f_x$, $f_y$)
obtained using the modified scheme with $\Delta t=0.1$
and $\Delta t=1.0$ exhibits slight fluctuations in time,
as shown by the non-zero values of the rms forces corresponding
to these cases in Table \ref{tab:hemis_Mu0}. The current
method, on the other hand, results in a constant force for these
cases. For $\nu=0.001$, the rms forces corresponding to
$\Delta t=0.001 \sim 0.01$ obtained using the modified scheme
exhibit a more pronounced difference when compared with that
corresponding to $\Delta t=5e-4$ (see Table \ref{tab:hemis_Mu0}).
With the current method
there is essentially no difference or this difference is much
smaller (see Table \ref{tab:hemis_ldt}).


\begin{table}
  \centering
  \begin{tabular}{l|c}
    \hline
    Frequency parameter $k_0$ for $\mbs u_0$ update
    & average wall-time/step (seconds)\\ \hline
    $k_0=10$ & 0.0726 \\
    $k_0=20$ & 0.0705 \\
    $k_0=50$ & 0.0693 \\
    $k_0=100$ & 0.0689 \\
    $k_0=200$ & 0.0687 \\
    $k_0=500$ & 0.0686 \\
    $k_0=1000$ & 0.0685 \\
    \hline
  \end{tabular}
  \caption{
    Computational cost of the flow past a hemisphere ($\nu=0.001$):
    average wall-time
    per time step (on two CPU cores)
    for the current method when the coefficient matrix
    is updated once every $k_0$ time steps
    (element order 6, $\Delta t=0.001$).
    With this problem size, it takes 0.109 seconds
    to compute a time step 
    when the coefficient matrix is updated at that particular step, and it takes
    0.0685 seconds to compute a time step
    when the coefficient matrix is not updated at that step.
  }
  \label{tab:hemis_cost}
\end{table}

Let us finally look into the computational cost of the
current method. When the field $\mbs u_0$ is updated
at a time step, the coefficient matrix of the
linear algebraic system for the velocity needs to be re-computed
and re-factorized at that step.
This induces an extra cost, which increases with the problem size
and can become substantial with a fairly large or large element order.
If  $\mbs u_0$ is updated once every $k_0$ time steps,
this extra computational cost is effectively spread over
$k_0$ time steps in the long run. Therefore,
the impact induced by the coefficient matrix update
can be considerably smaller in terms of the average computational cost per
time step. In Table \ref{tab:hemis_cost} we
provide the average wall time per time step (in seconds, using two
CPU cores)
corresponding to different frequency parameter values ($k_0$)
with the current method for the flow past a hemisphere in channel
($\nu=0.001$, element order $6$, $\Delta t=0.001$, $C_0=1000$).
Note that this is the wall time averaged over a number of time steps.
In reality, with this problem size, when $\mbs u_0$ is updated
at a certain time step it takes about $0.109$ seconds (on two CPU cores) to compute
that step. When $\mbs u_0$ is not updated at a time step
it takes about $0.0685$ seconds (on two CPU cores)
to compute that step. The wall-time values
are collected on a Linux cluster in the authors'
institution (Purdue University).
These results indicate that with the current method
if the field $\mbs u_0$ is not updated
very frequently, the impact of
the coefficient-matrix update on the overall computational cost
is not significant.


\section{Concluding Remarks}
\label{sec:summary}

%
%
%

In the current paper we have developed an  energy-stable
scheme for simulating the incompressible Navier-Stokes equations.
The scheme incorporates a pressure-correction type strategy
and the generalized Positive Auxiliary Variable (gPAV) approach.
The salient feature of the algorithm  lies in that
in the gPAV reformulated system the original nonlinear term is
replaced by the sum of a linear term ($\mbs M(\mbs u)$) and a correction term,
and the correction term is put under control by the auxiliary variable.
The scheme satisfies a
discrete energy stability property,
irrespective of the time step sizes. 
Within each time step, the scheme entails the computation of
two copies of the velocity and the pressure, by solving an individual
de-coupled linear
algebraic system for each of these field variables. 
The pressure linear system involves a constant and time-independent
coefficient matrix, which can be pre-computed.
The coefficient matrix for the velocity linear system can be updated
periodically, once every $k_0$ time steps in the current method.
If the linear term is set to zero ($\mbs M(\mbs u)=0$),
the velocity coefficient matrix becomes time-independent and can also
be pre-computed, which corresponds to the modified scheme
suggested in Remark \ref{rem:rem_1}.
The auxiliary variable, on the other hand, is computed by
a well-defined explicit formula, which guarantees the positivity
of its computed values. No nonlinear algebraic solver is involved
in the current method, for either the field variables or
the auxiliary variable.

It is observed that the current method  can produce accurate
results with large (or fairly large) time step sizes for
the incompressible Navier-Stokes equations. The maximum
$\Delta t$ that can lead to accurate simulation results
using the current method is typically
considerably larger than that with the scheme of~\cite{LinYD2019}
or the modified scheme from Remark~\ref{rem:rem_1}.
For example, for the Kovasznay flow (under identical conditions),
the current method can
still produce accurate  results with $\Delta t=0.4$,
and the method from~\cite{LinYD2019} can produce accurate results
with $\Delta t\sim 0.009$ (see Table 1 of \cite{LinYD2019}),
while the modified scheme from Remark~\ref{rem:rem_1}
produces accurate results with only even smaller $\Delta t$ values.
While the current method substantially expands the accuracy range 
for the time step size, it is noted that
when $\Delta t$ increases to a certain level the method
will similarly lose accuracy in the simulation results,
even though the computation may be stable. 
This is similar to those observations in~\cite{LinYD2019,LinLD2019}.

An apparent downside of the current method is the need for periodic update of
the coefficient matrix for the velocity linear algebraic system,
which induces an extra cost when compared with the method from~\cite{LinYD2019}
and the modified scheme from Remark~\ref{rem:rem_1}.
Since this coefficient matrix  is only updated
once every $k_0$ time steps, the extra cost induced by the re-computation
of the coefficient matrix is effectively spread over $k_0$ time
steps. In simulations $k_0$ is typically on the order of several dozen.
So the impact of the coefficient-matrix update
on the overall cost of the
current method is in general quite small, and can be essentially negligible
when $k_0$ is a sizable number.


Another potential drawback of the current method lies in that
the coefficient matrix for the velocity linear system
is non-symmetric due to the $\mbs M(\mbs u)$ term
(but it is positive definite).
In the current implementation we have employed the
BiCGStab  solver
when solving the velocity linear algebraic system.
In numerical simulations
we  observe that when the Reynolds number becomes large
and with very large $\Delta t$
this solver can at times encounter difficulties
for convergence (see Section \ref{sec:hemis}), thus making
the method less robust in these cases.
On the other hand, the modified scheme from
Remark~\ref{rem:rem_1} involves coefficient matrices
that are symmetric positive definite, and the linear systems
are solved using the conjugate gradient (CG) solver
in the current implementation. In numerical experiments
we observe that this method
is very robust with very large $\Delta t$ values
at high Reynolds numbers.
It should be noted that, at those $\Delta t$ values when
the BiCGStab solver starts to encounter difficulty, the simulation
results are already no longer accurate.



\section*{Appendix A. Approximation for the First Time Step}

We summarize the approximation of the flow variables
for the first time step in this Appendix. The scheme below ensures that
the computed values for $\left.R^{n+1}\right|_{n=0}$, $\left.R^{n+1/2}\right|_{n=0}$
and $\left.R^{n+3/2}\right|_{n=0}$
are all positive. The notation here follows that in
the main text.

Given $(\tilde{\mbs u}^0,\mbs u^0, R^0, p^0)$, we compute
the first time step in two substeps.
In the first substep we compute an approximation of
$(\tilde{\mbs u}^1,\mbs u^1, R^1, p^1)$, denoted by
$(\tilde{\mbs u}_a^1,\mbs u_a^1, R_a^1, p_a^1)$.
In the second substep we compute the final
$(\tilde{\mbs u}^1,\mbs u^1, R^1, p^1)$.
These computations are as follows. \\
\noindent\underline{First Substep:}\\
\noindent\underline{For $\tilde{\mbs u}_a^1$:}
\begin{subequations}\label{equ:vel0}
  \begin{align}
    &
    \frac{\tilde{\mbs u}_a^{1}-\mbs u^0}{\Delta t}
    + \mbs M(\tilde{\mbs u}_a^{1}) + \nabla p^0
    -\nu\nabla^2 \tilde{\mbs u}_a^{1}
    + \xi_a\left[\mbs N(\tilde{\mbs u}^{0}) - \mbs M(\tilde{\mbs u}^{0})  \right]
    = \mbs f^{1};
    \label{equ:vel0_1} \\
    &
    \xi_a = \frac{\left(R_a^{1}\right)^2}{E[\bar{\mbs u}_a^{1}]}; \label{equ:vel0_2} \\
    &
    E[\bar{\mbs u}_a^{1}]
    = \int_{\Omega}\frac12\left| \bar{\mbs u}_a^{1}\right|^2d\Omega + C_0;
    \label{equ:vel0_3} \\
    &
    \tilde{\mbs u}_a^{1} = \mbs w^{1}, \quad \text{on} \ \partial\Omega;
    \label{equ:vel0_4}
  \end{align} 
\end{subequations}
\noindent\underline{For $\phi_a^{1}$:}
\begin{align}
  \phi_a^{1} = \nabla\cdot\tilde{\mbs u}_a^{1}; \label{equ:phi0}
\end{align}
\noindent\underline{For $p_a^{1}$ and $\mbs u_a^{1}$:}
\begin{subequations}\label{equ:p0}
  \begin{align}
    &
    \frac{\mbs u_a^{1}-\tilde{\mbs u}_a^{1}}{\Delta t}
    + \nabla\left(p_a^{1}-p^0 + \nu\phi_a^{1}  \right) = 0;
    \label{equ:p0_1} \\
    &
    \nabla\cdot\mbs u_a^{1} = 0; \label{equ:p0_2} \\
    &
    \mbs n\cdot\mbs u_a^{1} = \mbs n\cdot\mbs w^{1}, \quad
    \text{on} \ \partial\Omega; \label{equ:p0_3} \\
    &
    \int_{\Omega} p_a^{1}d\Omega = 0; \label{equ:p0_4}
  \end{align}
\end{subequations}
\noindent\underline{For $R_a^{n+1}$:}
\begin{equation}\label{equ:R0_disp}
  \begin{split}
    &
  \left(R_a^{1} + R^0 \right)
  \frac{R_a^{1} - R^0 }{\Delta t}
  = \int_{\Omega} \tilde{\mbs u}_a^{1}\cdot
  \frac{\mbs u_a^{1}-\mbs u^0}{\Delta t} \\
  &
  +\xi_a \left[
    -\nu\int_{\Omega}\|\nabla\bar{\mbs u}_a^{1} \|^2d\Omega
    + \int_{\Omega}\mbs f^{1}\cdot\bar{\mbs u}_a^{1}d\Omega
    + \int_{\Omega}\left(-\bar{P}_a^{1}\mbs n + \nu\mbs n\cdot\nabla\bar{\mbs u}_a^{1}
    -\frac12 (\mbs n\cdot\mbs w^{1})\mbs w^{1}\right)\cdot\mbs w^{1}d\Omega
    \right] \\
  &
  -\int_{\Omega}\left[
    -\mbs M(\tilde{\mbs u}_a^{1}) - \nabla P_a^{1} +\nu\nabla^2\tilde{\mbs u}_a^{1}
    -\xi_a\left(\mbs N(\tilde{\mbs u}^{0}) - \mbs M(\tilde{\mbs u}^{0})  \right)
    + \mbs f^{1}
    \right]\cdot\tilde{\mbs u}^{n+1} d\Omega \\
  &
  + (1-\xi_a)\left[\left|\int_{\Omega}\mbs f^{1}\cdot\bar{\mbs u}_a^{1}d\Omega \right|
  + \left|
  \int_{\Omega}\left(-\bar{P}_a^{1}\mbs n + \nu\mbs n\cdot\nabla\bar{\mbs u}_a^{1}
    -\frac12 (\mbs n\cdot\mbs w^{1})\mbs w^{1}\right)\cdot\mbs w^{1}d\Omega
  \right|\right].
  \end{split}
\end{equation}
\noindent\underline{Second Substep:}\\
\noindent\underline{For $\tilde{\mbs u}^1$:}
\begin{subequations}\label{equ:vel01}
  \begin{align}
    &
    \frac{\tilde{\mbs u}^{1}-\mbs u^0}{\Delta t}
    + \mbs M(\tilde{\mbs u}^{1}) + \nabla p^0
    -\nu\nabla^2 \tilde{\mbs u}^{1}
    + \xi\left[\mbs N(\tilde{\mbs u}^{0}) - \mbs M(\tilde{\mbs u}^{0})  \right]
    = \mbs f^{1};
    \label{equ:vel01_1} \\
    &
    \xi = \frac{\left(R^{3/2}\right)^2}{E[\bar{\mbs u}^{3/2}]}; \label{equ:vel01_2} \\
    &
    E[\bar{\mbs u}^{3/2}]
    = \int_{\Omega}\frac12\left| \bar{\mbs u}^{3/2}\right|^2d\Omega + C_0;
    \label{equ:vel01_3} \\
    &
    \tilde{\mbs u}^{1} = \mbs w^{1}, \quad \text{on} \ \partial\Omega;
    \label{equ:vel01_4}
  \end{align} 
\end{subequations}
\noindent\underline{For $\phi^{1}$:}
\begin{align}
  \phi^{1} = \nabla\cdot\tilde{\mbs u}^{1}; \label{equ:phi01}
\end{align}
\noindent\underline{For $p^{1}$ and $\mbs u^{1}$:}
\begin{subequations}\label{equ:p01}
  \begin{align}
    &
    \frac{\mbs u^{1}-\tilde{\mbs u}^{1}}{\Delta t}
    + \nabla\left(p^{1}-p^0 + \nu\phi^{1}  \right) = 0;
    \label{equ:p01_1} \\
    &
    \nabla\cdot\mbs u^{1} = 0; \label{equ:p01_2} \\
    &
    \mbs n\cdot\mbs u^{1} = \mbs n\cdot\mbs w^{1}, \quad
    \text{on} \ \partial\Omega; \label{equ:p01_3} \\
    &
    \int_{\Omega} p^{1}d\Omega = 0; \label{equ:p01_4}
  \end{align}
\end{subequations}
\noindent\underline{For $R^{n+1}$:}
\begin{equation}\label{equ:R01_disp}
  \begin{split}
    &
  \left(R^{3/2} + R^{1/2} \right)
  \frac{R^{3/2} - R^{1/2} }{\Delta t}
  = \int_{\Omega} \tilde{\mbs u}^{1}\cdot
  \frac{\mbs u^{1}-\mbs u^0}{\Delta t} \\
  &
  +\xi \left[
    -\nu\int_{\Omega}\|\nabla\bar{\mbs u}^{1} \|^2d\Omega
    + \int_{\Omega}\mbs f^{1}\cdot\bar{\mbs u}^{1}d\Omega
    + \int_{\Omega}\left(-\bar{P}^{1}\mbs n + \nu\mbs n\cdot\nabla\bar{\mbs u}^{1}
    -\frac12 (\mbs n\cdot\mbs w^{1})\mbs w^{1}\right)\cdot\mbs w^{1}d\Omega
    \right] \\
  &
  -\int_{\Omega}\left[
    -\mbs M(\tilde{\mbs u}^{1}) - \nabla P^{1} +\nu\nabla^2\tilde{\mbs u}^{1}
    -\xi\left(\mbs N(\tilde{\mbs u}^{0}) - \mbs M(\tilde{\mbs u}^{0})  \right)
    + \mbs f^{1}
    \right]\cdot\tilde{\mbs u}^{n+1} d\Omega \\
  &
  + (1-\xi)\left[\left|\int_{\Omega}\mbs f^{1}\cdot\bar{\mbs u}^{1}d\Omega \right|
  + \left|
  \int_{\Omega}\left(-\bar{P}^{1}\mbs n + \nu\mbs n\cdot\nabla\bar{\mbs u}^{1}
    -\frac12 (\mbs n\cdot\mbs w^{1})\mbs w^{1}\right)\cdot\mbs w^{1}d\Omega
  \right|\right].
  \end{split}
\end{equation}

The symbols involved in the above equations are explained as follows.
In equations \eqref{equ:vel0_2} and \eqref{equ:R0_disp}
$\bar{\mbs u}_a^1$ is an approximation of $\mbs u_a^1$ and will be specified
later in \eqref{equ:barvar0}. $P_a^1$ and $\bar{P}_a^1$ are given by
\begin{equation}\label{equ:def_P0}
  P_a^1 = p_a^1 + \nu\phi_a^1, \quad
  \bar{P}_a^1 = \bar{p}_a^1 + \nu\bar{\phi}_a^1,
\end{equation}
where $\bar{p}_a^1$ and $\bar{\phi}_a^1$ are approximations of
$p_a^1$ and $\phi_a^1$ to be specified later in \eqref{equ:barvar0}.
In equations \eqref{equ:vel01_2} and \eqref{equ:R01_disp}
$\bar{\mbs u}^{3/2}$, $R^{3/2}$ and $R^{1/2}$ are given by,
\begin{equation}\label{equ:bar32_0}
  \left\{
  \begin{split}
    &
    \bar{\mbs u}^{3/2} = \frac32{\mbs u}_a^1 - \frac12\mbs u^0, \\
    &
    R^{3/2} = \frac32 R^{1} - \frac12 R^0, \\
    &
    R^{1/2} = \frac12\left(R_a^1 + R^0  \right).
  \end{split}
  \right.
\end{equation}
$P^1$ and $\bar{P}^1$ are given by
\begin{equation}\label{equ:def_P01}
  P^1 = p^1 + \nu\phi^1, \quad
  \bar{P}^1 = \bar{p}^1 + \nu\bar{\phi}^1,
\end{equation}
where $\bar{p}^1$ and $\bar{\phi}^1$ are approximations of
$p^1$ and $\phi^1$ to be specified later in \eqref{equ:barvar0}.

A combination of equations \eqref{equ:vel0_1}, \eqref{equ:p0_1}
and \eqref{equ:R0_disp} leads to
\begin{equation}
  \frac{\left(R_a^1 \right)^2-\left(R^0 \right)^2}{\Delta t}
  = \xi_a\left[
    -\nu\int_{\Omega} \left\|\nabla\bar{\mbs u}_a^1  \right\|^2d\Omega
    + B_1 + B_2
    \right]
  + (1-\xi_a)\left(|B_1| + |B_2|  \right),
\end{equation}
where
\begin{equation}
  B_1 = \int_{\Omega}\mbs f\cdot\bar{\mbs u}_a^1d\Omega, \quad
  B_2 = \int_{\Omega}\left(-\bar{P}_a^{1}\mbs n + \nu\mbs n\cdot\nabla\bar{\mbs u}_a^{1}
    -\frac12 (\mbs n\cdot\mbs w^{1})\mbs w^{1}\right)\cdot\mbs w^{1}d\Omega.
\end{equation}
In light of \eqref{equ:vel0_2} we then have
\begin{equation}
  \left\{
  \begin{split}
    &
    \xi_a = \frac{\left(R^0 \right)^2+ (|B_1|+|B_2|)\Delta t}{
      E[\bar{\mbs u}_a^1] + \left[
        \nu\int_{\Omega} \left\| \nabla\bar{\mbs u}_a^1 \right\|^2d\Omega
        + (|B_1|-B_1) + (|B_2|-B_2)
        \right]\Delta t
    }, \\
    &
    R_a^1 = \sqrt{\xi_a E[\bar{\mbs u}_a^1] }.
  \end{split}
  \right.
  \label{equ:cal_xi0}
\end{equation}
Since $R^0>0$ according to \eqref{equ:ic_disc},
we conclude that $\xi_a>0$ and $R_a^1>0$ from
the above equations.
Then based on equation \eqref{equ:bar32_0}
we conclude that $R^{1/2}>0$.

A combination of equations \eqref{equ:vel01_1}, \eqref{equ:p01_1}
and \eqref{equ:R01_disp} leads to
\begin{equation}
  \frac{\left(R^{3/2} \right)^2-\left(R^{1/2} \right)^2}{\Delta t}
  = \xi\left[
    -\nu\int_{\Omega} \left\|\nabla\bar{\mbs u}^1  \right\|^2d\Omega
    + D_1 + D_2
    \right]
  + (1-\xi)\left(|D_1| + |D_2|  \right),
\end{equation}
where
\begin{equation}
  D_1 = \int_{\Omega}\mbs f\cdot\bar{\mbs u}^1d\Omega, \quad
  D_2 = \int_{\Omega}\left(-\bar{P}^{1}\mbs n + \nu\mbs n\cdot\nabla\bar{\mbs u}^{1}
    -\frac12 (\mbs n\cdot\mbs w^{1})\mbs w^{1}\right)\cdot\mbs w^{1}d\Omega.
\end{equation}
Note that $R^{3/2}$ and $R^{1/2}$ are defined by \eqref{equ:bar32_0}.
In light of \eqref{equ:vel01_2} we then have
\begin{equation}
  \left\{
  \begin{split}
    &
    \xi = \frac{\left(R^{1/2} \right)^2+ (|D_1|+|D_2|)\Delta t}{
      E[\bar{\mbs u}^{3/2}] + \left[
        \nu\int_{\Omega} \left\| \nabla\bar{\mbs u}^1 \right\|^2d\Omega
        + (|D_1|-D_1) + (|D_2|-D_2)
        \right]\Delta t
    }, \\
    &
    R^{3/2} = \sqrt{\xi E[\bar{\mbs u}^{3/2}] }, \\
    &
    R^{1} = \frac23 R^{3/2} + \frac13 R^0.
  \end{split}
  \right.
  \label{equ:cal_xi01}
\end{equation}
Since $R^{1/2}>0$, we conclude that $\xi>0$, $R^{3/2}>0$ and $R^1>0$.

In the above formulas
$(\tilde{\mbs u}_a^1, \mbs u_a^1, p_a^1, \phi_a^1)$ and
$(\tilde{\mbs u}^1, \mbs u^1, p^1, \phi^1)$ still need to be determined,
and the variables with overbars need to be specified.
We compute these variables as follows. First define
two sets of field variables $(\tilde{\mbs u}_1^1, \mbs u_1^1, p_1^1, \phi_1^1)$
and $(\tilde{\mbs u}_2^1, \mbs u_2^1, p_2^1, \phi_2^1)$
as solutions to
the following equations: \\
\noindent\underline{For $(\tilde{\mbs u}_1^1, \mbs u_1^1, p_1^1, \phi_1^1)$:}
\begin{equation}\label{equ:til_u1}
  \left\{
  \begin{split}
    &
    \frac{\tilde{\mbs u}_1^1}{\Delta t} + \mbs M(\tilde{\mbs u}_1^1)
    -\nu\nabla^2\tilde{\mbs u}_1^1
    = \mbs f^1 + \frac{{\mbs u}^0}{\Delta t} - \nabla p^0; \\
    &
    \tilde{\mbs u}_1^1 = \mbs w^1, \quad \text{on} \ \partial\Omega;
  \end{split}
  \right.
\end{equation}
\begin{equation}\label{equ:phi1}
  \phi_1^1 = \nabla\cdot\tilde{\mbs u}_1^1;
\end{equation}
\begin{equation}\label{equ:u1p1}
  \left\{
  \begin{split}
    &
    \frac{\mbs u_1^1}{\Delta t} + \nabla p_1^1 = \frac{\tilde{\mbs u}_1^1}{\Delta t}
    - \nabla(-p^0 + \nu\phi_1^1); \\
    &
    \nabla\cdot\mbs u_1^1=0; \\
    &
    \mbs n\cdot\mbs u_1^1 = \mbs n\cdot\mbs w^1, \quad \text{on} \ \partial\Omega; \\
    &
    \int_{\Omega} p_1^1d\Omega = 0.
  \end{split}
  \right.
\end{equation}
\noindent\underline{For $(\tilde{\mbs u}_2^1, \mbs u_2^1, p_2^1, \phi_2^1)$:}
\begin{equation}\label{equ:til_u2}
  \left\{
  \begin{split}
    &
    \frac{\tilde{\mbs u}_2^1}{\Delta t} + \mbs M(\tilde{\mbs u}_2^1)
    -\nu\nabla^2\tilde{\mbs u}_2^1
    = \mbs N(\tilde{\mbs u}^0) - \mbs M(\tilde{\mbs u}^0)
    \\
    &
    \tilde{\mbs u}_2^1 = 0, \quad \text{on} \ \partial\Omega;
  \end{split}
  \right.
\end{equation}
\begin{equation}\label{equ:phi2}
  \phi_2^1 = \nabla\cdot\tilde{\mbs u}_2^1;
\end{equation}
\begin{equation}\label{equ:u2p2}
  \left\{
  \begin{split}
    &
    \frac{\mbs u_2^1}{\Delta t} + \nabla p_2^1 = \frac{\tilde{\mbs u}_2^1}{\Delta t}
    - \nu\phi_2^1; \\
    &
    \nabla\cdot\mbs u_2^1=0; \\
    &
    \mbs n\cdot\mbs u_2^1 = 0, \quad \text{on} \ \partial\Omega; \\
    &
    \int_{\Omega} p_2^1d\Omega = 0.
  \end{split}
  \right.
\end{equation}
It is then straightforward to verify that
the solutions to equations \eqref{equ:vel0_1}--\eqref{equ:R0_disp}
and \eqref{equ:vel01_1}--\eqref{equ:R01_disp}
are, for given $\xi_a$ and $\xi$,
\begin{equation}\label{equ:soln_0}
  \left\{
  \begin{split}
    &
    \tilde{\mbs u}_a^1 = \tilde{\mbs u}_1^1 + \xi_a\tilde{\mbs u}_2^1, \\
    &
    \mbs u_a^1 = \mbs u_1^1 + \xi_a\mbs u_2^1, \\
    &
    \phi_a^1 = \phi_1^1 + \xi_a\phi_2^1, \\
    &
    p_a^1 = p_1^1 + \xi_a p_2^1;
  \end{split}
  \right.
  \qquad\qquad
  \left\{
  \begin{split}
    &
    \tilde{\mbs u}^1 = \tilde{\mbs u}_1^1 + \xi\tilde{\mbs u}_2^1, \\
    &
    \mbs u^1 = \mbs u_1^1 + \xi\mbs u_2^1, \\
    &
    \phi^1 = \phi_1^1 + \xi\phi_2^1, \\
    &
    p^1 = p_1^1 + \xi p_2^1.
  \end{split}
  \right.
\end{equation}
We specify the barred variables as follows,
\begin{equation}\label{equ:barvar0}
  \left\{
  \begin{split}
    &
    \bar{\mbs u}_a^1 = \bar{\mbs u}^1 = \tilde{\mbs u}_1^1 + \tilde{\mbs u}_2^1, \\
    &
    \bar{\phi}_a^1 = \bar{\phi}^1 = \phi_1^1 + \phi_2^1, \\
    &
    \bar{p}_a^1 = \bar{p}^1 = p_1^1 + p_2^1.
  \end{split}
  \right.
\end{equation}
Note that the field equations \eqref{equ:til_u1}--\eqref{equ:u2p2}
can be solved in a way analogous to the
discussions in Section \ref{sec:implementation}.
The details will not be provided here.

Therefore we compute $(\tilde{\mbs u}^1, \mbs u^1, p^1, R^1)$ by
the following procedure:
\begin{itemize}
\item
  Solve equations  \eqref{equ:til_u1}--\eqref{equ:u1p1}
  for $(\tilde{\mbs u}_1^1, \mbs u_1^1, p_1^1, \phi_1^1)$; \\
  Solve \eqref{equ:til_u2}--\eqref{equ:u2p2}
  for $(\tilde{\mbs u}_2^1, \mbs u_2^1, p_2^1, \phi_2^1)$;

\item
  Compute $(\bar{\mbs u}_a^1,\bar{P}_a^1)$ by equations \eqref{equ:barvar0}
  and \eqref{equ:def_P0}; \\
  Compute $\xi_a$ and $R_a^1$ by equation \eqref{equ:cal_xi0}; \\
  Compute $(\tilde{\mbs u}_a^1, \mbs u_a^1, p_a^1)$ by equation \eqref{equ:soln_0};

\item
  Compute $(\bar{\mbs u}^1,\bar{P}^1)$ by equations \eqref{equ:barvar0}
  and \eqref{equ:def_P01}; \\
  Compute $\bar{\mbs u}^{3/2}$ and $R^{1/2}$ by equation \eqref{equ:bar32_0}; \\
  Compute $\xi$ and $R^1$ by equation \eqref{equ:cal_xi01}; \\
  Compute $(\tilde{\mbs u}^1, \mbs u^1, p^1)$ by equation \eqref{equ:soln_0}.
  
\end{itemize}
It is noted that the computed values have the property
$R^{1}>0$, $R^{1/2}>0$ and $R^{3/2}>0$.


\section*{Acknowledgement}
This work was partially supported by
NSF (DMS-1522537).

\bibliographystyle{plain}
\bibliography{obc,mypub,nse,sem,contact_line,interface,multiphase}

\end{document}